\documentclass[10pt]{iopart}

\usepackage[utf8]{inputenc}
\usepackage[T1]{fontenc}
\usepackage{amssymb,amsfonts,textcomp}
\usepackage{graphicx}
\usepackage[dvipsnames]{xcolor}
\usepackage{bm}
\usepackage{hyperref}

\newcommand{\eg}{\textit{e.g.}}
\newcommand{\ie}{\textit{i.e.}}

\newcommand{\Secref}[1]{section~\ref{#1}}
\newcommand{\Figref}[1]{figure~\ref{#1}}
\newcommand{\Eqref}[1]{(\ref{#1})}
\newcommand{\Tabref}[1]{table~\ref{#1}}
\newcommand{\Ocite}[1]{\cite{#1}}

\newcommand{\dipc}{Donostia International Physics Center (DIPC) -- UPV/EHU, E-20018 San Sebasti\'an, Spain}
\newcommand{\ikerbasque}{Ikerbasque, Basque Foundation for Science, E-48013, Bilbao, Spain}
\newcommand{\cinn}{Nanomaterials and Nanotechnology Research Center (CINN), CSIC-UNIOVI-PA, E-33940 El Entrego, Spain}

\begin{document}

\title{Carbon-based nanostructures as a versatile platform for tunable {$\pi$-}magnetism}
\author{Dimas G. de Oteyza$^{1,2}$ and Thomas Frederiksen$^{2,3}$}
\address{$^1$ \cinn}
\address{$^2$ \dipc}
\address{$^3$ \ikerbasque}
\ead{d.g.oteyza@cinn.es}
\ead{thomas\_frederiksen@ehu.eus}

\begin{abstract}
Emergence of $\pi$-magnetism in open-shell nanographenes has been theoretically predicted decades ago but their experimental characterization was elusive due to the strong chemical reactivity that makes their synthesis and stabilization difficult. In recent years, on-surface synthesis under vacuum conditions has provided unprecedented opportunities for atomically precise engineering of nanographenes, which in combination with scanning probe techniques have led to a substantial progress in our capabilities to realize localized electron spin states and to control electron spin interactions at the atomic scale. Here we review the essential concepts and the remarkable advances in the last few years, and outline the versatility of carbon-based $\pi$-magnetic materials as an interesting platform for applications in spintronics and quantum technologies.
\end{abstract}


\maketitle
\ioptwocol


\section{Introduction}

Magnetic materials are key to a wide variety of technological applications. However, to date their practical use is
limited to transition metals with unpaired $d$- or $f$-shell electrons. The same applies even for the so-called
molecular magnetism that has been studied and developed for decades, but still relying on coordination of $d$- or
$f$-block elements to organic ligands \cite{Co.20.Molecularmagnetismchemical,FeVaCa.17.Molecularmagnetismquo}.
Instead, the magnetism associated to $p$-shell electrons in carbon-based
materials has been hardly utilized to date and only recently it is being enthusiastically explored \cite{Ya.10.Emergencemagnetismgraphene,SoSuTe.21.surfacesynthesisgraphene}. There are
multiple reasons for the booming interest that it is raising. On the one hand we find the attractive magnetic
properties offered by light materials like carbon nanostructures, which display weak spin-orbit and hyperfine
couplings, the two main channels responsible for the relaxation and decoherence of electron spins \cite{Ya.10.Emergencemagnetismgraphene,MiHiSi.06.IntrinsicandRashba,Ya.08.HyperfineInteractionsGraphene,SlKeMy.18.Magneticedgestates}. In addition,
they are also expected to display high spin-wave stiffness, which should be mirrored in high Curie temperatures and
long spin correlation lengths \cite{Ya.10.Emergencemagnetismgraphene,YaKa.08.MagneticCorrelationsat}. These materials thus make for a greatly promising platform on which to exploit
spin-polarized currents or the processing of spin-based quantum information \cite{Ya.10.Emergencemagnetismgraphene,SlKeMy.18.Magneticedgestates,LoLoMa.19.Quantumunitstopological,TrBuLo.07.Spinqubitsgraphene,ZhLaRo.18.SinglePhotonEmission}.
On the other hand, carbon-based
materials also display several advantages with regard to their processing, with the potential to allow for a cheap and
scalable production of materials that, beyond their magnetic functionalities, may further include biocompatibility,
light weight or plasticity. Lastly, molecule-based materials profit from the prospects of bottom-up production methods,
which may be among the keys for success if the current miniaturization trend in information technology devices is to be
maintained. 

Until recently, the main drawback faced by carbon-based $\pi$-magnetism was its poor reproducibility. The magnetic properties
or carbon-based nanostructures are extremely susceptible to minute changes in the bonding configuration and at the same
time they are based on unpaired $\pi$-electrons that display great propensity for chemical reactions \cite{StChZe.19.DoDiradicalsBehave}. Indeed, besides information or quantum technology applications, diradicals are also of key importance for the understanding and further development of organic chemistry due to their characteristic reactivity and their inevitable presence, even if often only transiently due to their limited lifetime \cite{StChZe.19.DoDiradicalsBehave}. This is one of the reasons for which diradicals have been extensively studied in the past \cite{StChZe.19.DoDiradicalsBehave,DaWu.15.OpenShellBenzenoid}, but also for the scarce reproducible studies about magnetic carbon-based materials, whose chemical reactivity either hindered their
synthesis by conventional wet chemistry, forcing the addition of protecting groups, or hampered their characterization
due to an excessively short product's lifetime. However, the advent of surface-supported chemistry under vacuum,
typically termed as on-surface synthesis (OSS) \cite{ClOt.19.ControllingChemicalCoupling,WaZh.19.Confinedsurfaceorganic,HeFuSt.17.CovalentBondFormation,DoLiLi.15.SurfaceActivatedCoupling}, has provided unprecedented opportunities for the synthesis and
characterization of this type of materials \cite{SoSuTe.21.surfacesynthesisgraphene,LiFe.20.SyntheticTailoringGraphene}. The inert vacuum environment, in combination with the often stabilizing
effect of the supporting substrate, allows for the synthesis of carbon-based nanostructures and at the same time limits
its possible degradation mechanisms. Furthermore, it places at the researcher's disposal a variety of surface sensitive
characterization techniques with which to characterize the samples, including scanning probe microscopies (SPM) and
spectroscopies (STS) \cite{ClOt.19.ControllingChemicalCoupling,WaZh.19.Confinedsurfaceorganic,HeFuSt.17.CovalentBondFormation}.
The latter allow for a precise determination of the covalent bonding structure as well as for a
simultaneous assessment of the electronic and magnetic properties at the single molecule level. 

These new prospects have caused the research on magnetic carbon-based nanostructures to be booming over the last
years \cite{SoSuTe.21.surfacesynthesisgraphene,LiFe.20.SyntheticTailoringGraphene}.
In turn, the increasing number of researchers that are focusing their efforts into this field have
facilitated an impressive progress in short time. Although the experimental proof of the magnetism associated to carbon
vacancies in graphene (as observed by SPM at the single vacancy level) already dates back to
2007 \cite{UgBrGu.10.MissingAtomas}, the synthesis of iconic molecules with intrinsic magnetic properties that were theoretically devised already in
the 1950s, like triangulene \cite{ClSt.53.AromaticHydrocarbonsLXV}, has only occurred in recent years \cite{PaMiMa.17.Synthesischaracterizationtriangulene}. Along with triangulene, many other highly coveted
carbon-based nanostructures with magnetic properties have been synthesized and studied over the last few years, like
for example Clar's goblet \cite{Cl.72.AromaticSextet,MiBeEi.20.Topologicalfrustrationinduces} and zigzag-edged graphene nanoribbons (GNRs) \cite{RuWaYa.16.surfacesynthesisgraphene}. The growing number of available
systems for experimental study has also brought about an increasing understanding of the properties and of the magnetic
interactions between different magnetic (radical) $\pi$-states. This increased understanding, along with the
continuously improving capabilities of OSS strategies, have further allowed for the generation of
one-and two-dimensional materials with engineered properties as provided by the controlled coupling of appropriately
placed radical $\pi$-states in periodic arrays \cite{GrWaYa.18.Engineeringrobusttopological,RiVeJi.20.Inducingmetallicitygraphene}. That is, the development of the synthetic protocols and the
understanding of the resulting material's properties are going hand in hand and are responsible for the fascinating
advances reached in carbon-based magnetism. 

\begin{figure*}
  \includegraphics[width=\textwidth]{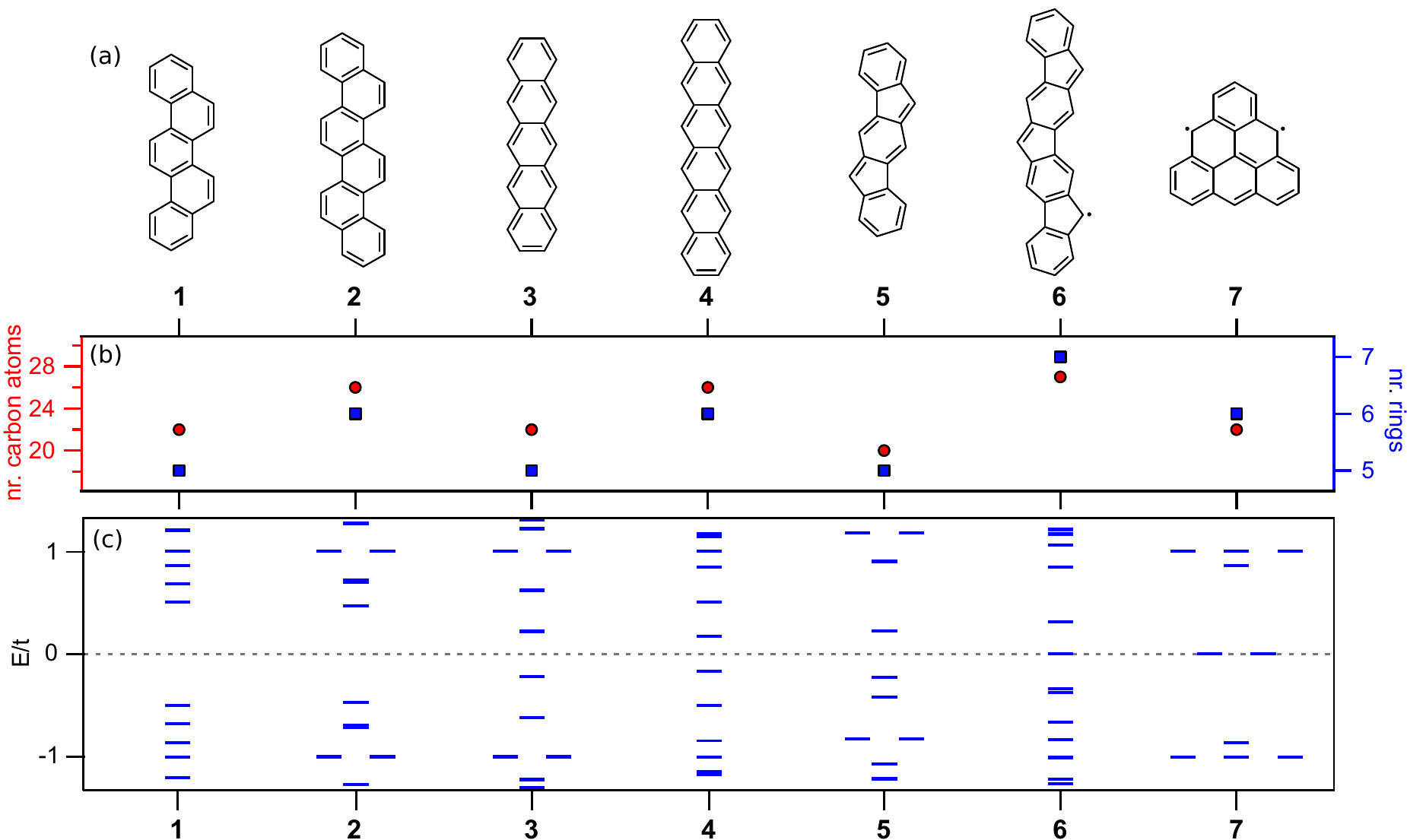}
  \caption{(a) Molecular structures for comparison. (b) Number of carbon atoms and comprised rings. (c) Calculated low-energy spectrum within a tight-binding (TB) model with nearest neighbour hopping $t$.}
  \label{fig:trends}
\end{figure*}

At this point it needs to be remarked that there is a stringent requirement on atomic precision for the synthesis of the carbon nanostructures. Minor changes in the bonding structure may result in dramatic changes to the optoelectronic and magnetic properties, as exemplified in \Figref{fig:trends} with seven different molecular structures. Although they all feature a comparable number of carbon atoms and associated rings, the energy of their calculated frontier orbitals, which is displayed in the bottom part, varies enormously, with decreasing HOMO--LUMO gaps and the appearance of zero-energy states when moving from left to right. As explained later on, low HOMO--LUMO gaps (and thus even more zero-energy states) may not warrant but at least facilitate the appearance of magnetic properties in carbon nanostructures (indeed when going from left to right there is a gradual change from a purely closed-shell character, through gradually increasing open-shell character and all the way to a high-spin molecule). From the calculated orbital energies in \Figref{fig:trends}c we can therefore already extract some qualitative trends for parameters that, although again being no guarantee for magnetism, generally promote its appearance. One is increasing molecular size, as can be inferred from comparison of the isomorphic structures (\textbf{1} \textit{vs.}~\textbf{2}, \textbf{3} \textit{vs.}~\textbf{4} and \textbf{5} \textit{vs.}~\textbf{6}), which show a smaller gap for the larger structures. Another one is the presence of zigzag edges, as can be inferred from comparing \textbf{1} and \textbf{2} with \textbf{3} and \textbf{4}, or also with structure \textbf{7}, which displays zigzag edges all over its perimeter and has two zero-energy states. Lastly, also the presence of non-benzenoid rings facilitates the appearance of magnetism, as may be inferred from comparing \textbf{1} and \textbf{2} with \textbf{5} and \textbf{6}, the latter of which even displays a zero-energy state. Also a comparison of \textbf{3} with \textbf{5} reveals a comparable HOMO--LUMO gap in spite of the lower number of carbon atoms in the latter. The reasons underlying the effects of each of these parameters on the material's magnetic properties will be discussed and explained in more detail in the following sections.

This review is organized as follows. We begin with an overview in \Secref{sec:computational} of commonly used computational approaches to describe the electronic structure of graphene nanostructures and relationships between band topology, emergence of spin-polarization and many-body spin states. In \Secref{sec:counting} we present well-established counting rules to infer the relationship between nanographene geometry and its ground-state spin properties. After a review in \Secref{sec:experimental} of the experimental techniques for nanographene synthesis and scanning probe characterization, we proceed to discuss the different mechanisms to induce magnetism. Section \ref{sec:interactions} is devoted to interactions among spin states and \Secref{sec:reactivity} to the chemical reactivity associated with the open-shell character of the involved spin states. Finally, an outlook is presented in \Secref{sec:outlook}.

\section{Computational approaches}
\label{sec:computational}
Carbon has four valence electrons in the second shell which favors the formation of four single bonds.
In the case of planar carbon structures like graphene, the robust trigonal structure arises from a $sp^2$
hybridization of one $2s$ and two $2p$ carbon orbitals to form doubly-occupied $\sigma$ bonds with its three neighbours.
The remaining electron, associated with the out-of-plane orbital $2p_z$, forms $\pi$-bonding with its neighbours,
resulting in a half-filled $\pi$ band at the origin of interesting electronic properties of graphene,
such as ``relativistic'' massless Dirac fermions and valley degeneracy \cite{CaGuPe.09.electronicpropertiesof,MeSoBa.16.Physicalpropertieslow}, 
emergence of $\pi$-magnetism at edges and defects \cite{Ya.10.Emergencemagnetismgraphene}, and 
exceptional transport phenomena \cite{ToLuRo.20.IntroductionGrapheneBased}.
The electronic structure of $sp^2$ carbon systems can be addressed computationally at different levels as outlined in the following sections.

\subsection{Density functional theory}
In condensed matter physics, density functional theory (DFT) is probably the most popular approach to describe the electronic structure of materials \cite{Ma.04.Electronicstructurebasic}.
It is theoretically founded in the Hohenberg--Kohn theorems, establishing that the electron \emph{density} that minimizes the total energy corresponds to that of the ground state of the system. In principle, this implies a drastic reduction of complexity since the ground state of a many-body problem can be described by the electron density instead of the full wave function.
The Kohn--Sham (KS) equations (KS-DFT) make this theory tractable in practice by introducing an effective non-interacting Schrödinger equation for one-electron KS orbitals that lead to the same electron density as for the interacting system. The KS wave function of the $N$-electron system is then written as a Slater determinant of these single-particle orbitals.
The only unknown is the so-called exchange-correlation (XC) functional, which leads to typical schemes such as the local-density approximation (LDA), the generalized-gradient approximation (GGA), etc.
Having fixed the XC functional, the problem is then solved iteratively for the self-consistent solution to the KS equations.

The reason for the popularity of DFT nowadays is the tractability of systems of thousands of atoms, without any system-specific parameters, with high accuracy for ground state properties such as lattice constants and phonons for solids as well as geometry and vibrations for molecules.
A recent Special Topic Issue \cite{ShMaMa.20.Electronicstructuresoftware} on electronic structure software provides a broad overview of many quantum chemistry codes based on DFT and beyond.

DFT is also successful in describing many aspects of carbon $\pi$-magnetism as is evident from the vast body of literature covered in this review. It allows, for instance, to compute the spin density for adsorbed atoms, molecules and extended 2D structures on metal surfaces and interfaces.
Spin in KS-DFT \cite{JaRe.12.Spindensityfunctional} is typically described in the symmetry-broken form, where the spin density of the non-interacting electrons are targeting an accurate description of the energetics of low-spin states at the expense of matching the exact spin state of the interacting problem.
This occurs because the KS orbitals for the two spin components are allowed to differ in the spin-unrestricted formulation.
This leads to wave functions that are not true eigenstates of the total spin operator $\hat S^2$ (only of the $\hat S_z$ projection), \ie, they are in principle spin-contaminated.
This limitation, which applies more generally to all mean-field descriptions, is important to keep in mind for applications to open-shell molecules.
Wave-function based methods from computational chemistry (configuration-interaction, coupled cluster, etc.) are therefore more suitable for accurately describing spin states, but these are limited to small systems.

\subsection{Tight binding}
\label{sec:tb}
\nocite{AsMe.76.Solidstatephysics}

\begin{figure*}
	\centering
	\includegraphics[height=8cm]{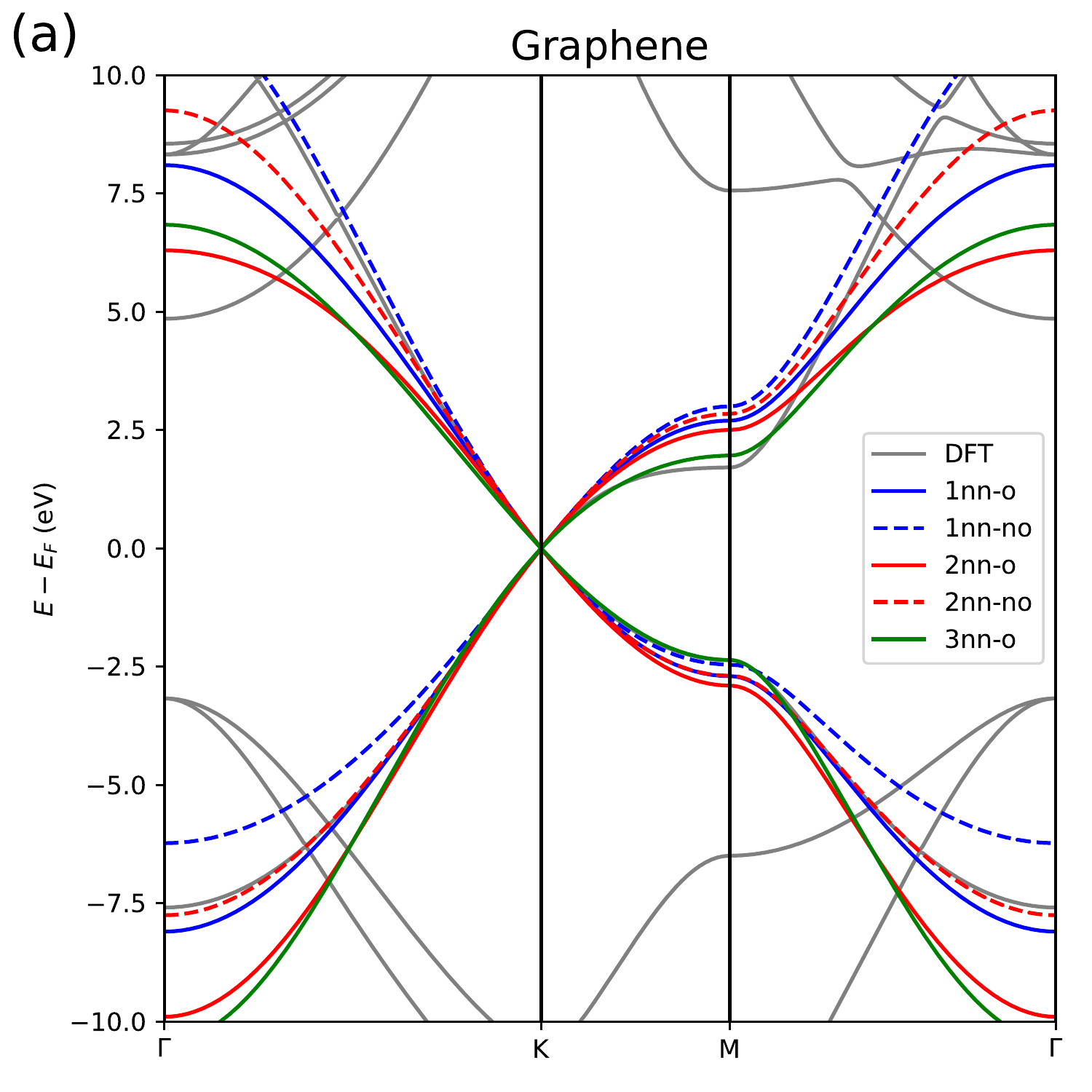}
	\includegraphics[height=8cm]{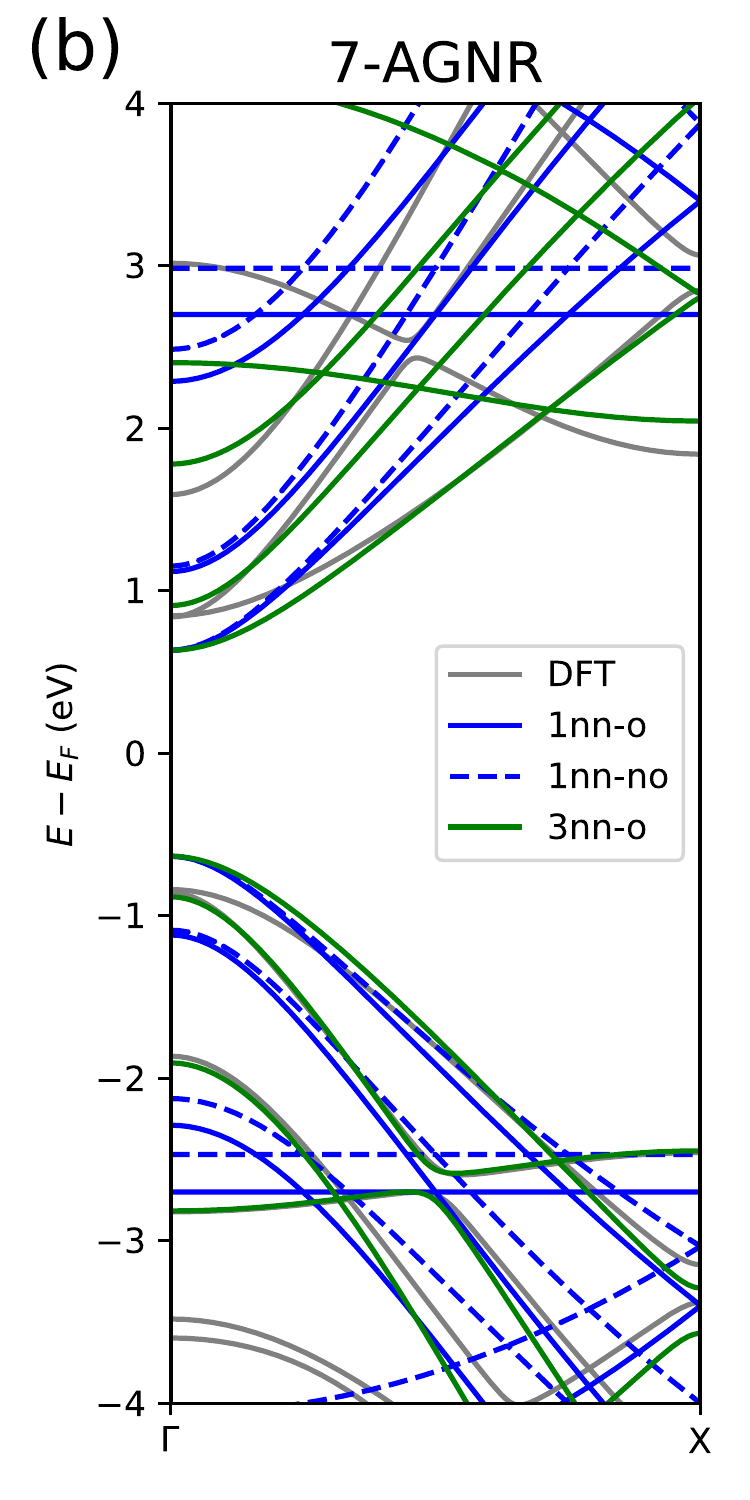}
	\caption{Comparison of electronic band structures from different TB models and DFT for
	(a) graphene and (b) a hydrogen-terminated 7-AGNR.
	The nearest neighbour (1nn, blue) model includes $t_1=-2.7$ eV,
	the next-nearest neighbour (2nn, red) model also $t_2=-0.2$ eV, and
	the third-nearest neighbour (3nn, green) model additionally $t_3=-0.18$ eV.
	Both orthogonal (-o, full lines) and nonorthogonal (-no, dashed lines) basis variants are considered,
	the latter with a nearest-neighbor overlap set to $s_1=0.1$.
	The gray bands correspond to SIESTA \cite{SoArGa.02.SIESTAmethodab} DFT-GGA calculations
	using a TZTP (DZP) basis set and lattice constant $a=2.460$ ($4.260$) {\AA} for graphene (7AGNR).}
	\label{fig:tb-bands}
\end{figure*}

Due to the orthogonality of the $\sigma$ and $\pi$ electrons in planar structures, it is possible to study them separately. The advantages of such effective descriptions involving only the latter are simplicity and numerical efficiency, which may also help to reach a better understanding of the mechanisms at play.

In the tight-binding (TB) approximation one describes electrons in a solid
in terms of the atomic orbitals of isolated atoms and the corrections that arise due to their overlaps \cite{AsMe.76.Solidstatephysics}. The electrons are also treated as independent (noninteracting) particles. The (multi-band) TB Hamiltonian can be written as
\begin{equation}
\hat H^0 = \sum_{ij\sigma} H^0_{ij} \hat c_{i\sigma}^\dagger \hat c_{j\sigma}^{\phantom\dagger},
\end{equation}
where $\hat c^\dagger_{i\sigma}$ is the fermionic creation operator of an electron with spin $\sigma=\{\uparrow,\downarrow\}$ in basis orbital $i$.
The operators satisfy the usual anticommutation relations
$\{\hat c_{\mu}, \hat c_{\nu}\} = \{\hat c^\dagger_{\mu}, \hat c^\dagger_{\nu}\} = 0$ 
and 
$\{\hat c_{\mu}^{\phantom\dagger}, \hat c^\dagger_{\nu}\} = \delta_{\mu\nu}$.
The matrix elements $H^0_{ij}=\langle i |\hat H^0 | j\rangle$
describe the onsite energies along the diagonal 
and the strength of electron hopping between orbitals $i$ and $j$ in the off-diagonals.
The matrix elements are real and satisfy $H^0_{ij}=H^0_{ji}$.
In general, the basis orbitals may be nonorthogonal represented by the overlap integrals $S_{ij}=\langle i|j\rangle$. For simplicity, one may assume orthogonality and write $S_{ij}=\delta_{ij}$.
For $sp^2$ carbon the simplest description includes just the 2$p_z$ orbital
at each carbon atom with nearest-neighbour hopping matrix elements $t=-2.7$ eV \cite{CaGuPe.09.electronicpropertiesof}.

The time-independent Schr\"odinger equation for the TB Hamiltonian can be written as as a generalized eigenvalue problem
\begin{eqnarray}
\mathbf {H}^0 \bm{\psi}_\alpha = E_\alpha \mathbf{S} \bm{\psi}_\alpha
\end{eqnarray}
where $E_\alpha$ is the eigenenergy corresponding to state $\bm{\psi}_\alpha = \sum_i|i\rangle\langle i|\alpha \rangle$.

Different TB approximations for the $\pi$-bands of graphene are shown in \Figref{fig:tb-bands}a, constructed with the free \textsc{sisl} software \cite{zerothi_sisl}
using typical parameters from the literature \cite{CaGuPe.09.electronicpropertiesof,ReMaTh.02.Tightbindingdescription}.
In the simplest description with only nearest neighbour interactions 
$t_1$ and orthogonal basis states the bands are perfectly symmetric with respect to the Fermi level 
(full line blue bands, \texttt{1nn-o}).
When a finite overlap is introduced (blue dashed bands, \texttt{1nn-no})
an asymmetry between empty and filled states develops.
Introduction of second-nearest neighbour interactions $t_2$ breaks the sublattice symmetry
(red bands, \texttt{2nn}) and lowers the bands at $\Gamma$.
Further, inclusion of third-nearest neighbour interactions $t_3$ (green bands, \texttt{3nn-o}), which represent atomic distances 
only marginally larger than the next-nearest pairs, enables a lowering of the empty band around $M$.
For comparison, we also show the corresponding bands from DFT calculations (gray bands) \cite{SoArGa.02.SIESTAmethodab}.

Figure \ref{fig:tb-bands}b concerns a similar comparison between TB and DFT calculations for a different system, namely the $\pi$-band states of a 7-atom wide armchair graphene nanoribbon (7-AGNR). The third-nearest neighbour parametrization from \Ocite{HaUpSa.10.Generalizedtightbinding} (green bands, \texttt{3nn-o}) provides the most accurate description.
Overall, TB provides a good description of the carbon $\pi$-electron system.

\subsection{The Hubbard model}
While the TB method is a well-established starting point for describing the electronic structure of carbon-based systems, the method itself cannot account for the emergence of $\pi$-magnetism due to the neglect of electron interactions. One way to include them is called the Hubbard model.

The Hubbard model was introduced in 1963 as an approximate description of electron correlation effects in narrow energy bands of solids, taking into account the atomistic nature in a minimal fashion \cite{Hu.63.Electroncorrelationsnarrow}. It was initially proposed to understand the properties of transition and rare-earth metals with partly filled $d$- and $f$-bands. However, despite its apparent simplicity it contains very rich physics, and has been useful to describe phenomena such as correlated metal-insulator transitions, magnetism, and superconductivity. 

In the (fermionic) Hubbard model one adds to the TB Hamiltonian $\hat H_0$ a term that accounts for short-range electron interactions. The model is defined by the Hamiltonian \cite{Hu.63.Electroncorrelationsnarrow}
\begin{equation}
\hat H = \hat H^0 + U_0 \sum_{i} \, \hat n_{i\uparrow} \hat n_{i\downarrow},
\label{eq:Hubbard-model}
\end{equation}
where $\hat n_{i\sigma} = \hat c_{i\sigma}^\dagger \hat c_{i\sigma}$ is the number operator for site $i$ and $U_0>0$ is the onsite Coulomb repulsion energy. The model thus ignores long-range Coulomb effects by only considering the interaction energy when two electrons occupy the same orbital.

Solving the Hubbard model is in general a difficult problem due to the quartic structure of the operators appearing in \Eqref{eq:Hubbard-model} and the exponential growth of Fock-space with system size, thus sharing challenges with full many-body descriptions (\Secref{sec:manybody}).
In order to solve the Hubbard model for moderately large nanographenes,
methods based on configuration interaction in the so-called complete active space (CAS) have been developed
\cite{DuLaPa.08.Electronelectroninteractions,OrFe.20.Probinglocalmoments}.
This approximation consists of splitting the single-particle spectrum into ``frozen'' low- and high-energy sectors and an active space which is 
handled numerically exact to capture the low-energy states accurately.
For example, CAS(6,6) formed by six electrons in six single-particle states was used to describe the low-energy states of triangulene dimers \cite{MiBeEi.20.CollectiveAllCarbon} and trimers \cite{MiCaWu.21.Observationfractionaledge}.
The CAS-Hubbard approach to zero-modes in nanographenes has been useful to establish exchange rules for diradicals \cite{OrBoGa.19.ExchangeRulesDiradical}.

\subsubsection{Example: half-filled Hubbard dimer}
\label{sec:Hubbard-dimer}

\begin{figure}
	\includegraphics[width=\columnwidth]{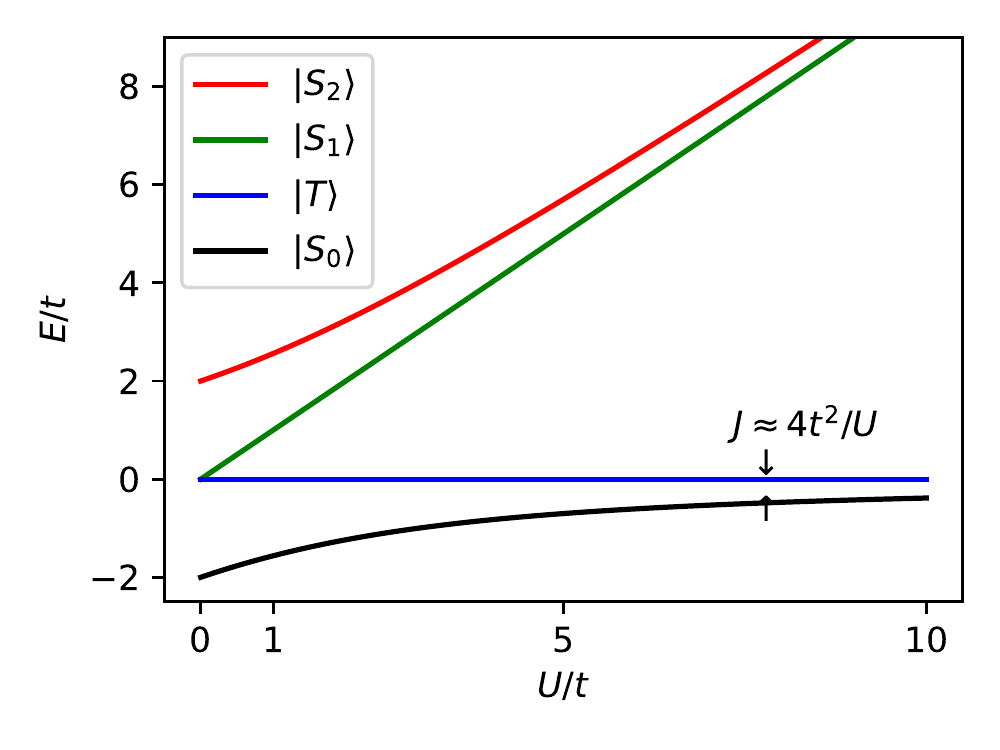}
	\caption{Energy spectrum of the half-filled Hubbard dimer. In absence of a magnetic field the ground state is a singlet $|S_0\rangle$ and the first excited state a triplet $|T\rangle$. In the limit of large $U/t$ the low-energy spectrum can be described by the Heisenberg spin dimer with antiferromagnetic exchange coupling $J=4t^2/U$.}
	\label{fig:hubbard-dimer}
\end{figure}

A simple, illustrative example of analytic solution of the Hubbard model is the case of a half-filled dimer, \ie, a system composed of two electronic orbitals and occupied by two electrons. We define it as
\begin{eqnarray}
\hat H &=& \sum_{\alpha=L,R}\sum_\sigma E_\alpha \hat n_{\alpha\sigma} 
-t \sum_\sigma (\hat c^\dagger_{L\sigma} \hat c_{R\sigma}^{\phantom\dagger} + \mathrm{h.c.}) \nonumber\\
&& + \sum_{\alpha=L,R} U_\alpha\; \hat n_{\alpha\uparrow} \hat n_{\alpha\downarrow},
\end{eqnarray}
where $E_\alpha$ the single-particle energy of orbital $\alpha \in \{L,R\}$, $t$ the effective hopping matrix element between the orbitals, and
\begin{eqnarray}
U_\alpha = U_0 \int d\mathbf{r} \, |\psi_\alpha(\mathbf{r})|^4 
\end{eqnarray}
the effective Coulomb repulsion for each orbital.
Projecting this Hamiltonian on the following basis of two-electron states
\begin{eqnarray}
\{\hat c^\dagger_{L\uparrow} \hat c^\dagger_{R\downarrow} |0\rangle,\;
\hat c^\dagger_{L\downarrow} \hat c^\dagger_{R\uparrow} |0\rangle,\;
\hat c^\dagger_{L\uparrow} \hat c^\dagger_{L\downarrow} |0\rangle,\;
\hat c^\dagger_{R\uparrow} \hat c^\dagger_{R\downarrow} |0\rangle \},
\end{eqnarray}
where $|0\rangle$ represents the vacuum state,  we arrive at the simple matrix form
\begin{eqnarray}
\hat H &=&
\left(
\begin{array}{cccc}
\varepsilon_L + \varepsilon_R & 0 & -t & -t \\
0 & \varepsilon_L + \varepsilon_R & t & t \\
-t & t & 2 \varepsilon_L + U_L & 0 \\
-t & t & 0 & 2 \varepsilon_R + U_R
\end{array}\right),\nonumber\\
\end{eqnarray}
which is readily diagonalized. In the symmetric limit $\varepsilon_L=\varepsilon_R=0$ and $U_L=U_R=U$ the eigenenergies are
\numparts
\begin{eqnarray}
E_{S_0} &=& U/2-\sqrt{(2t)^2 + (U/2)^2},\\
E_{T} &=& 0,\\
E_{S_1} &=& U,\\
E_{S_2} &=& U/2+\sqrt{(2t)^2 + (U/2)^2},
\end{eqnarray}
\endnumparts
and the corresponding eigenstates
\numparts
\begin{eqnarray}
|S_0\rangle &\propto& \big(\alpha [\hat c^\dagger_{L\uparrow} \hat c^\dagger_{R\downarrow} 
- \hat c^\dagger_{L\downarrow}\hat c^\dagger_{R\uparrow} ]
+ \hat c^\dagger_{L\uparrow} \hat c^\dagger_{L\downarrow}
+ \hat c^\dagger_{R\uparrow} \hat c^\dagger_{R\downarrow}\big)|0\rangle, \label{eq:S0}\\
|T\rangle &=& \frac 1{\sqrt{2}}(\hat c^\dagger_{L\uparrow} \hat c^\dagger_{R\downarrow} 
+ \hat c^\dagger_{L\downarrow}\hat c^\dagger_{R\uparrow} )|0\rangle, \\
|S_1\rangle &=& \frac 1{\sqrt{2}}(\hat c^\dagger_{L\uparrow} \hat c^\dagger_{L\downarrow} 
- \hat c^\dagger_{R\uparrow} \hat c^\dagger_{R\downarrow})|0\rangle, \\
|S_2\rangle &\propto& \big(\beta [\hat c^\dagger_{L\uparrow} \hat c^\dagger_{R\downarrow} 
- \hat c^\dagger_{L\downarrow}\hat c^\dagger_{R\uparrow} ] 
+ \hat c^\dagger_{L\uparrow} \hat c^\dagger_{L\downarrow}
+ \hat c^\dagger_{R\uparrow} \hat c^\dagger_{R\downarrow}\big)|0\rangle, \quad\quad 
\end{eqnarray}
\endnumparts
where $\alpha = -2 t/ E_{S_0}$ and $\beta = -2t/E_{S_2}$. Here $|S_i\rangle$ can be identified as the singlet states while $|T\rangle$ is the $S_z=0$ triplet state. An illustration of the energy spectrum of the Hubbard dimer is shown in \Figref{fig:hubbard-dimer}.

In the limit $U/t\rightarrow 0$, where $\alpha\rightarrow 1$, we see that the ground state is an equal mix of the open- and closed-shell configurations. Conversely, in the limit $U/t\rightarrow \infty$, where $\alpha\rightarrow\infty$, only the open-shell configuration remains. Here the ground state energy can be expanded in $t$ to read
\begin{eqnarray}
E_{S_0} &\rightarrow& -4t^2/U.
\end{eqnarray}
This allows to map the low-energy spectrum in this limit onto the antiferromagnetic Heisenberg model for a spin-1/2 dimer
\begin{eqnarray}
\hat H_\mathrm{eff} &=& J \; \Big( \vec{\mathbf{S}}_L\cdot \vec{\mathbf{S}}_R - \frac 14 \Big),
\end{eqnarray}
where $\vec{\mathbf{S}}_i$ is the spin-1/2 operator for site $i$ and the exchange coupling can identified as $J=4t^2/U$. From the relations 
$2 \vec{\mathbf{S}}_L\cdot \vec{\mathbf{S}}_R = (\vec{\mathbf{S}}_L + \vec{\mathbf{S}}_R)^2 - \vec{\mathbf{S}}_L^2 - \vec{\mathbf{S}}_R^2$ and $\vec{\mathbf{S}}^2 = S(S+1)$,
it is easy to see that the eigenenergies of $\hat H_\mathrm{eff}$ correspond to $-J$ for the singlet ($S=0$) and $0$ for the triplets ($S=1$), exactly as derived above for the Hubbard dimer.

\subsection{Mean-field Hubbard approximation}
The Hubbard model can in principle be solved by numerically exact diagonalization, but the Hilbert space grows exponentially with the number of lattice sites. This makes a brute-force approach to even small nanographenes impractical.

\begin{figure}
	\includegraphics[height=8cm]{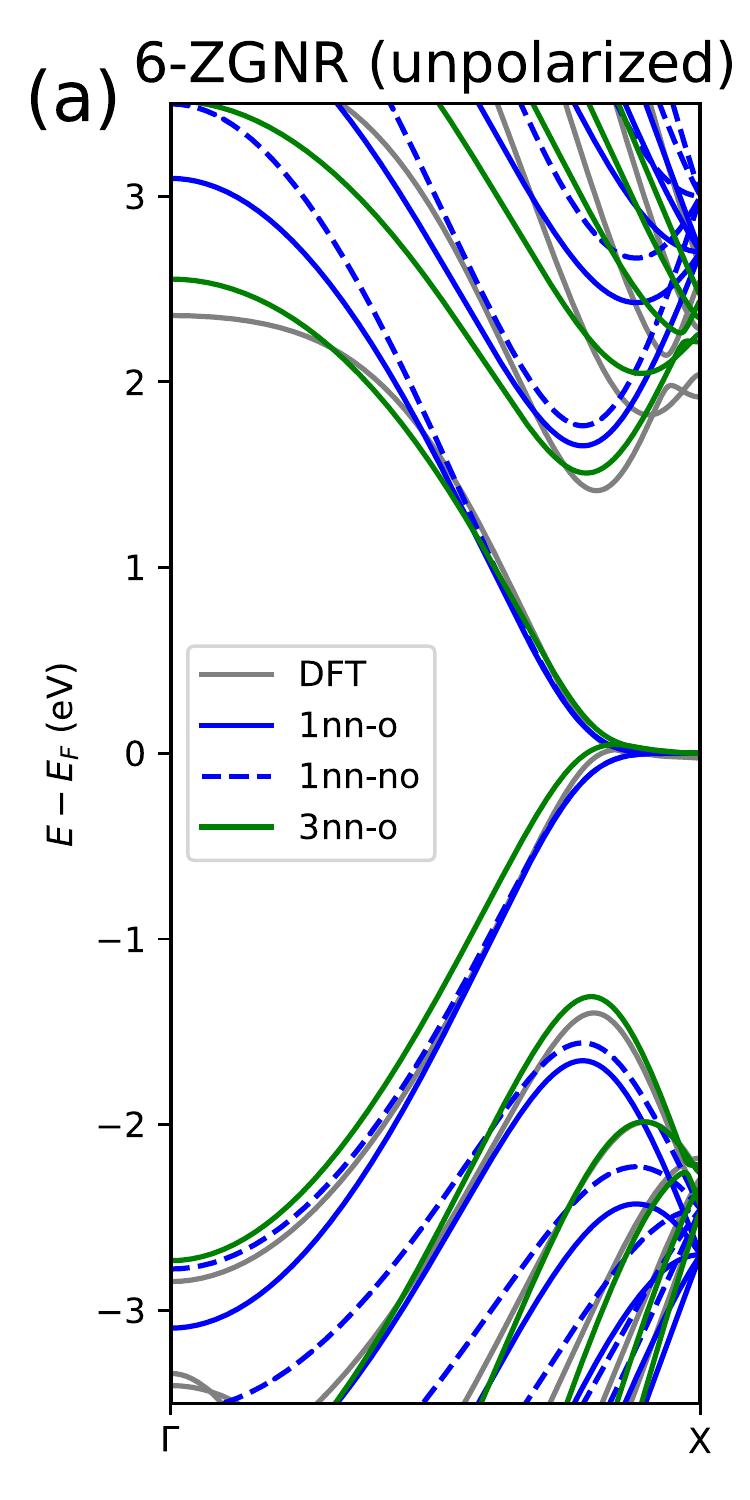}
	\includegraphics[height=8cm]{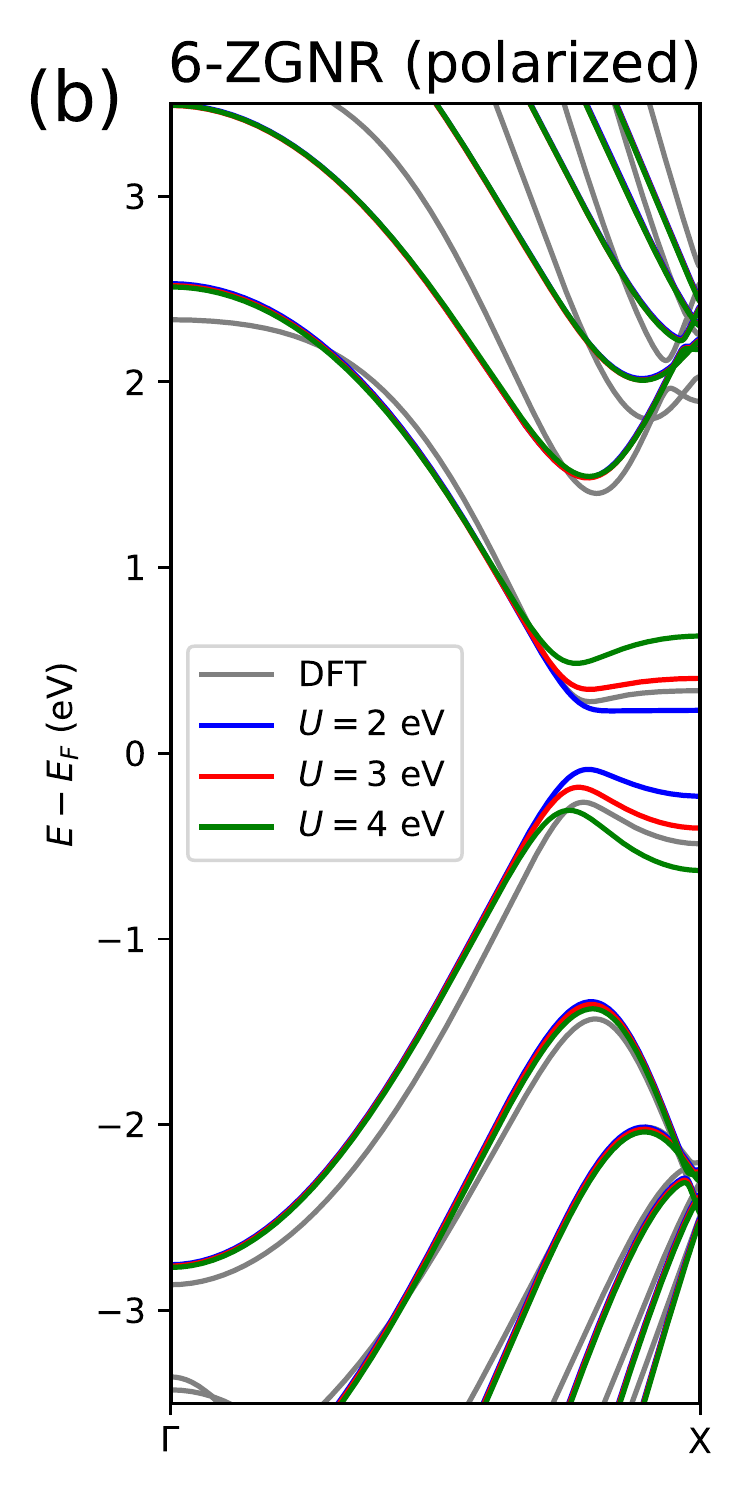}
	\includegraphics[width=0.9\columnwidth]{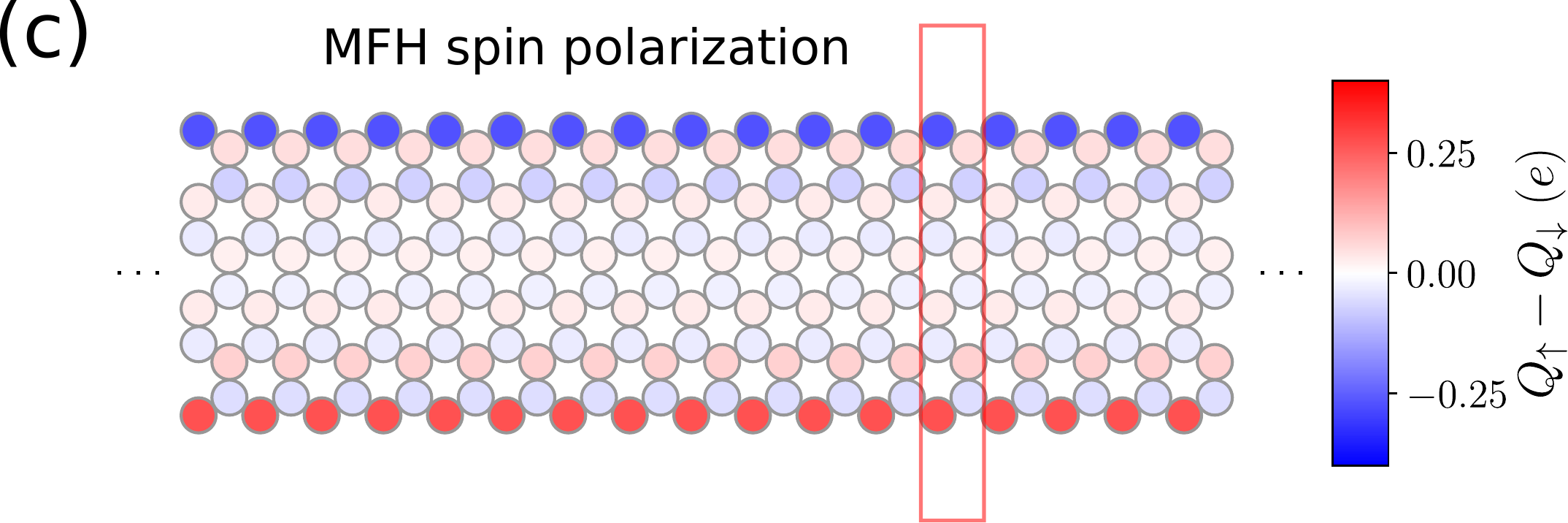}
	\caption{Comparison of electronic band structures from MFH and DFT for a hydrogen-terminated 6-ZGNR.
	(a) For spin-degenerate (unpolarized) calculations the ZGNR is metallic and displays a flat band towards $X$. The TB parameters are the same as those
	of \Figref{fig:tb-bands}. (b) Spin-polarized
	calculations leads to symmetry-broken solutions with edge-polarization \cite{FuWaNa.96.PeculiarLocalizedState,SoCoLo.06.EnergyGapsin} and the opening of a correlation gap. The MFH model here is based on the third-nearest neighbour model (\texttt{3nn-o}) and $U$ is varied in the range 2-4 eV (coloured bands).
	In both panels the gray bands correspond to SIESTA \cite{SoArGa.02.SIESTAmethodab} DFT-GGA calculations
	using a DZP basis set and lattice constant of $a=2.460$ {\AA}.
	(c) Spin polarization from MFH calculations (3nn-o, $U=3$ eV). The unit cell is indicated by the red box.}
	\label{fig:zgnr-bands}
\end{figure}

Instead a dramatic simplification is achieved if one resorts to the mean-field approximation. By defining the occupation operators $\hat n_{i\sigma} = \langle \hat n_{i\sigma}\rangle +\delta\hat n_{i\sigma}$ in terms of the fluctuations around their average values, one can write the product of operators as
\begin{eqnarray}
\hat n_{i\uparrow} \hat n_{i\downarrow} 
&=& (\langle \hat n_{i\uparrow}\rangle + \delta \hat n_{i\uparrow})
	(\langle \hat n_{i\downarrow}\rangle + \delta \hat n_{i\downarrow})\\
&=& \hat n_{i\uparrow} \langle\hat n_{i\downarrow}\rangle
+\langle\hat n_{i\uparrow}\rangle \hat n_{i\downarrow} 
-\langle\hat n_{i\uparrow}\rangle \langle\hat n_{i\downarrow}\rangle 
+\delta\hat n_{i\uparrow} \delta\hat n_{i\downarrow}\nonumber\\
&\approx& \hat n_{i\uparrow} \langle\hat n_{i\downarrow}\rangle
+\langle\hat n_{i\uparrow}\rangle \hat n_{i\downarrow} 
-\langle\hat n_{i\uparrow}\rangle \langle\hat n_{i\downarrow}\rangle, \nonumber
\end{eqnarray}
where the fluctuations (last term) has been neglected. This leads to the mean-field Hubbard (MFH) model of the form
\begin{equation}
\hat H_\mathrm{MFH} = \hat H_0 + U \sum_{i} (\hat n_{i\uparrow} \langle\hat n_{i\downarrow}\rangle
+\langle\hat n_{i\uparrow}\rangle \hat n_{i\downarrow} 
-\langle\hat n_{i\uparrow}\rangle \langle\hat n_{i\downarrow}\rangle).\nonumber\\
\end{equation}
An example of MFH results for the band structure of a 6-atom wide zigzag GNR (ZGNR) is shown in \Figref{fig:zgnr-bands} (colored bands). It is well-established that symmetry-broken solutions (\Figref{fig:zgnr-bands}b) are in very good agreement with DFT-GGA calculations (gray bands) when using $U\sim 3$ eV on top of the third-nearest neighbour parametrization \cite{Ya.10.Emergencemagnetismgraphene,HaUpSa.10.Generalizedtightbinding}. 
The MFH model has been shown as a useful model to understand the emergence of magnetism in $sp^2$ carbon systems \cite{Ya.10.Emergencemagnetismgraphene,MiBeEi.20.Topologicalfrustrationinduces,HaUpSa.10.Generalizedtightbinding,FuWaNa.96.PeculiarLocalizedState,FePa.07.MagnetismGrapheneNanoislands,Ya.08.MagnetismDisorderedGraphene,SoMuFe.10.Hydrogenatedgraphenenanoribbons, SoTuDu.15.Spintransporthydrogenated,OrLaMe.16.Engineeringspinexchange,GrPoJa.17.Nanostructuredgraphenespintronics,LiSaCo.19.Singlespinlocalization}.
A free implementation of MFH in Python is available \cite{hubbard}.

\subsection{Wave-function based methods and many-body theory}
\label{sec:manybody}

The problem of electron correlation in open-shell nanographenes can also be addressed with full {ab initio} quantum chemistry approaches.
This is particularly pertinent when strong correlation effects are at play.
However, the computational cost of traditional wave-function based methods (configuration interaction, coupled cluster, multi-configurational self-consistent fields etc.) typically scale very unfavorable with the system size and are thus limited to rather small molecules. 
Examples of quantum chemistry calculations for graphene nanoislands include CAS-SCF and multireference NEVPT2 methods \cite{CuEvLe.18.ElectronicStructureGraphene}, restricted active space configuration interaction (RASCI)
\cite{SaCaCa.19.Triangulargraphenenanofragments},
and multi-reference averaged quadratic coupled cluster method (MR-AQCC) \cite{PlPaGe.13.MultiradicalCharacterOne}.
Other approaches are based on
quantum chemical density matrix renormalization group
(DMRG) that allows to address the full $\pi$-orbital space of large nanographenes \cite{HaDoAv.07.radicalcharacteracenes, ChSh.11.DensityMatrixRenormalization,
MiKuYa.13.MorePiElectrons,
SaUrVe.21.UnravellingOpenShell,BrBrKo.21.Massivelyparallelquantum}.
Many-body diagrammatic techniques based on the GW approximation
was also applied to the Hubbard model to describe correlation effects in GNR heterojunctions \cite{JoJaBo.19.CorrelatedTopologicalStates}.

Another important aspect beyond the properties of \emph{isolated} nanographenes is the impact of the interaction with a supporting surface. This can for instance give rise to Kondo-like phenomena on metal surfaces arising from the coupling of a magnetic system with the conduction electrons in the substrate.
This situation has been described theoretically based on the multi-orbital Anderson model in the non-crossing approximation \cite{MiCaWu.21.Observationfractionaledge,KoLo.11.Multiorbitalnon, JaOrFe.21.Renormalizationspinexcitations}.

\subsection{Topological band analysis}
\label{sec:top-band-analysis}

Another useful way to analyze, understand or predict radical states in $sp^2$-carbon structures is through topology \cite{GrWaYa.18.Engineeringrobusttopological,CaZhLo.17.TopologicalPhasesGraphene,LiCh.18.TopologicalPropertiesGapped,RiVeCa.18.Topologicalbandengineering,CiSaTo.20.Tailoringtopologicalorder,FrBrLi.20.MagnetismTopologicalBoundary,ZhCaLo.21.TopologicalPhasesGraphene}.
The topological classes are characterized by a $\mathbb{Z}_2$ invariant that takes on the values 0 or 1 for trivial and
nontrivial characters, respectively. The relevant quantity to characterize the topological class of one-dimensional (1D) systems is the Zak phase
$\gamma_n$ of its $n$ occupied bands, which is obtained from an integral of the Berry connection across the 1D Brillouin
zone \cite{Za.89.Berrysphaseenergy,Re.00.ManifestationsBerrysphase}
\begin{equation}
\label{eq:zak}
\gamma _n=i \frac{2\pi } a  \int _{-\pi /a}^{\pi /a} dk
\big\langle u_{nk} \big| \frac{\partial}{\partial k} \big | u_{nk} \big\rangle,
\end{equation}
where $k$ is the wave vector, $a$ is the unit cell size, and $u_{nk}$ is the periodic part of the electron Bloch wave function in
band $n$. Note that for systems with band-crossings or degenerate states there is an arbitrariness to the definition of the Zak phase for the individual bands. However, the total phase $\gamma = \sum_{n\in\mathrm{occ}}\gamma_n$ (\ie, summed over occupied bands) is invariant and unaffected by this arbitrariness as long as the filled and empty bands do not intersect \cite{Re.00.ManifestationsBerrysphase}.

The Zak phase $\gamma$ depends on the choice of shape and origin of the unit cell and can in general take any value. However, if the unit cell possesses inversion or mirror symmetry, it is possible to choose the origin such that $\gamma$ (mod $2\pi$) is quantized to either 0 or $\pi$. The corresponding $\mathbb{Z}_2$ invariant is then obtained from
\begin{equation}
\label{eq:z2}
\left(-1\right)^{\mathbb{Z}_2}=e^{i \gamma}.
\end{equation}
The $\mathbb{Z}_2$ invariant is also related to the parity of the Bloch wave functions at the time-reversal invariant
momentum points ${\Gamma}$ and ${\pi}/a$ by \cite{CaZhLo.17.TopologicalPhasesGraphene,CiSaTo.20.Tailoringtopologicalorder,ZhCaLo.21.TopologicalPhasesGraphene}
\begin{equation}
\label{eq:parity}
\left(-1\right)^{\mathbb{Z}_2}
= \prod_{n\in\mathrm{occ}} 
\langle \psi _{n\Gamma } |\widehat  P |\psi_{n\Gamma} \rangle 
\langle \psi _{n\pi/a} | \widehat  P |\psi_{n\pi/a} \rangle,
\end{equation}
where  $\widehat  P$  is the parity operator (applying to either inversion or mirror symmetry) and  $\psi
_{\mathit{nk}}$  is the wave function of the $n$-th occupied band at momentum $k$. 

According to the bulk-boundary correspondence, at the
interface between materials belonging to different topology classes one (or an odd number) of localized zero-energy modes will emerge. Such in-gap states in $sp^2$ carbon structures are also manifestations of $\pi$-radicals \cite{GrWaYa.18.Engineeringrobusttopological,CaZhLo.17.TopologicalPhasesGraphene,LiCh.18.TopologicalPropertiesGapped,RiVeCa.18.Topologicalbandengineering,CiSaTo.20.Tailoringtopologicalorder,FrBrLi.20.MagnetismTopologicalBoundary,ZhCaLo.21.TopologicalPhasesGraphene}.

\begin{figure}
	\includegraphics[width=\columnwidth]{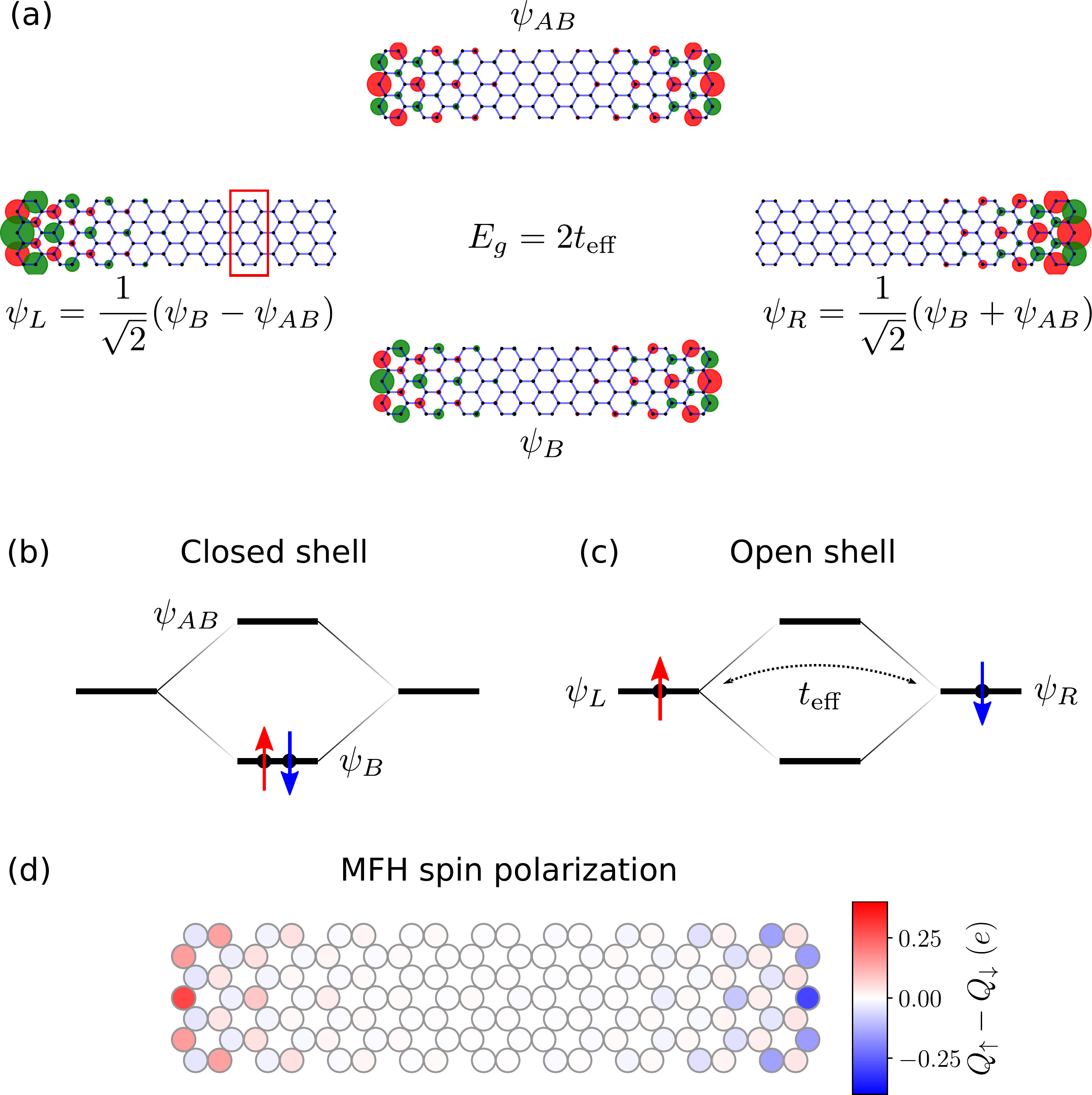}
	\caption{Illustration of the emergence of zero-modes in a finite 7-AGNR that belongs to the topology class $\mathbb{Z}_2=1$. The corresponding unit cell is indicated by the red box. (a) The bonding and antibonding single-particle orbitals $\psi_B$ and $\psi_{AB}$ (center column) fall inside the energy gap of the infinite ribbon and originate from the hybridization of two zero-modes $\psi_L$ and $\psi_R$ (outer columns) through the effective tunnel coupling $t_\mathrm{eff}$.
	(b) Simple picture of the closed-shell configuration for short 7-AGNRs ($t_\mathrm{eff}\gg U_B$), and (c) open-shell configuration for long 7-AGNRs ($t_\mathrm{eff}\ll U_B$).
	(d) Calculated spin polarization with MFH using $U=3$ eV.}
	\label{fig:hybridization}
\end{figure}

As an example, applying these concepts to the band structure of a 7-AGNR described
by the unit cell shown by the red box in \Figref{fig:hybridization}a, one concludes that this
ribbon belongs to the topological class $\mathbb{Z}_2=1$ \cite{CaZhLo.17.TopologicalPhasesGraphene}. As vacuum is trivial $\mathbb{Z}_2=0$, one zero-mode edge state is expected at each zigzag terminus. 
To illustrate this, the low-energy wave functions for a finite 7-AGNR are 
superimposed on the structures in \Figref{fig:hybridization}a. Two states, $\psi_B$ and $\psi_{AB}$ are found inside the energy gap of the infinite 7-AGNR (\Figref{fig:tb-bands}b), corresponding to the bonding and antibonding combinations of the topological zero-modes $\psi_L$ and $\psi_R$ coupled through the effective tunnel matrix element $t_\mathrm{eff}$. The end states are readily identified by the proper linear combinations.
As sketched in \mbox{\Figref{fig:hybridization}b-c} this system may adopt either a closed- or open-shell configuration depending on the strength of the Coulomb repulsion energy of the bonding orbital $U_B= U\int |\psi_B|^4 dr$ with respect to the hybridization energy $|2t_\mathrm{eff}|$. A crossover occurs naturally as the length of the GNR is increased due to the exponentially decaying matrix element $|t_\mathrm{eff}|$.
The characteristic spin polarization map computed from the local difference between up and down spin populations with MFH is shown in \Figref{fig:hybridization}d for $U=3$ eV.

\section{Counting rules}
\label{sec:counting}

\subsection{Ovchinnikov's rule and Lieb's theorem}
\label{sec:lieb}
Graphene is a so-called bipartite system in which $sp^2$-hybridized carbon atoms are arranged in two alternating hexagonal
sublattices. That is, carbon atoms of one sublattice are covalently bound only to carbons of the opposite sublattice.
The low-energy electronic states that prevalently determine most physico-chemical processes of graphene and its
nanostructured derivatives are $\pi$-orbitals that arise from the hybridization of the carbon $p_z$ electrons. Each
$sp^2$-hybridized carbon atom has one $p_z$ electron. Indeed, as described in the previous sections, a one-orbital approximation with the TB or
MFH models has been extremely successful to describe the properties of graphene and its
nanostructured derivatives \cite{Ya.10.Emergencemagnetismgraphene,Ya.13.GuideDesignElectronic}.

In a bipartite structure, each $p_z$ electron can only form $\pi$-bonds with electrons belonging to atoms of the
opposite sublattice. It automatically follows that, if the two sublattices have a different number of atoms, the
electrons will not be able to bind pairwise. The electrons that remain unpaired become $\pi$-radicals with an
associated spin $S = 1/2$. This is an intuitive basis for Lieb's theorem \cite{Li.89.TwotheoremsHubbard}, which in turn follows from
Ovchinnikov's rule \cite{Ov.78.Multiplicitygroundstate}. The theorem states that bipartite systems with repulsive electron-electron interactions ($U > 0$), as is the case of graphene and carbon-based nanostructures, will display a ground state with net
spin according to 
\begin{equation}
S = \frac 12 |N_A-N_B|,
\end{equation}
where $N_A$ and $N_B$ are the number of atoms belonging to each sublattice. A molecular system with an unbalanced amount of
atoms in each sublattice will thus necessarily display a net spin. 

\begin{figure}
\includegraphics[width=\columnwidth]{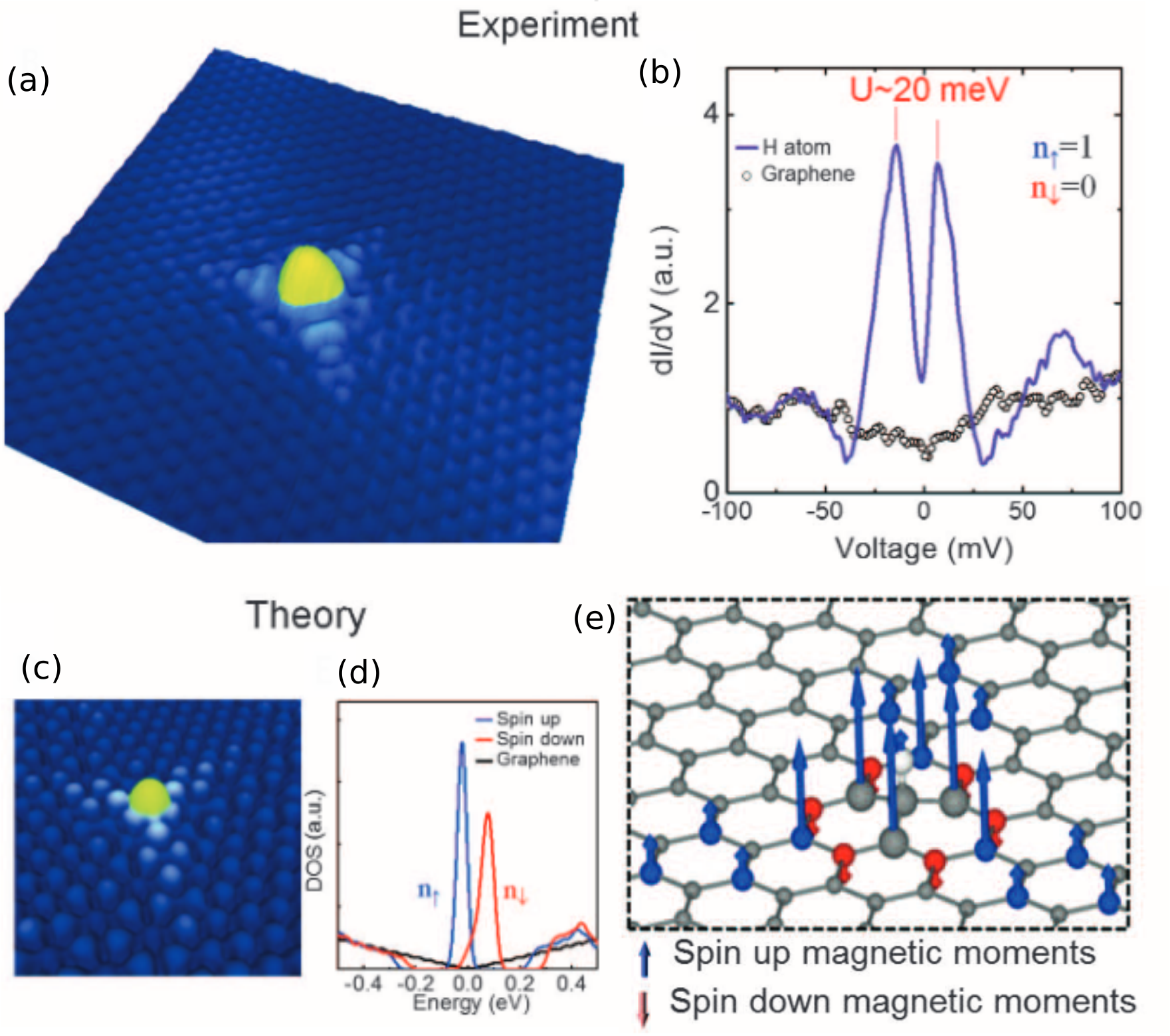}
\caption{(a) STM topography of a single H atom chemisorbed on neutral graphene. (b) $dI/dV$ spectrum measured on the H atom, showing a fully polarized peak at $E_F$, and measured on bare graphene far from the H atom. (c) DFT-simulated STM image and (d) density of states of an H atom chemisorbed on neutral graphene. (e) Calculated magnetic moments induced by H chemisorption (the lengths of the arrows signify the relative magnitudes of the magnetic moments). From \Ocite{GoGoMa.16.Atomicscalecontrol}. Reprinted with permission from AAAS.}
\label{fig:graphene}
\end{figure}

A very intuitive example of such a system is defective graphene, in which the local removal of one $p_z$ electron endows
the defect with magnetic properties \cite{PaFeBr.08.Vacancyinducedmagnetism}. Experimentally, the removal of a $p_z$ electron from graphene has been obtained by
different means, like forming atomic vacancies by low-energy ion bombardment in combination with an annealing
treatment \cite{UgBrGu.10.MissingAtomas}, or by $sp^3$ rehybridization of a C atom upon hydrogenation when exposed to atomic hydrogen \cite{GoGoMa.16.Atomicscalecontrol}.
A scanning tunneling microscopy (STM) image for the latter case is shown in \Figref{fig:graphene}, and a spectroscopic analysis on top of it clearly 
revealed the Coulomb gap of the magnetic state (\Figref{fig:graphene}b). Theoretical modelling of such a hydrogenated defect not
only nicely reproduces its contrast in STM images (\Figref{fig:graphene}c) but also the DOS in correlation to the STS spectra (\Figref{fig:graphene}d) \cite{GoGoMa.16.Atomicscalecontrol}.
The spin density distribution is further displayed in \Figref{fig:graphene}e. In agreement with the experimental and
calculated STM images, the magnetic state and therefore the net spin extends over several nm, but almost exclusively on
carbon atoms of the sublattice not hosting the extra hydrogen (\Figref{fig:graphene}e). Only a residual oppositely oriented spin
density is observed on the sublattice of the hydrogenated atom, which is a manifestation of the spin polarization
caused by the exchange interaction of the unpaired electron with the fully populated states \cite{Ya.10.Emergencemagnetismgraphene}.
For cases in which two
such defects are close to one another, DFT calculations and experiments reveal a clear magnetic signal and the spin's
ferromagnetic alignment (total spin $S = 1$) for defect pairs on the same sublattice, and a totally quenched magnetic
signal when they are on opposite sublattices \cite{GoGoMa.16.Atomicscalecontrol}. These findings are in perfect agreement with Lieb's theorem, which
predicts a ground state with $S=0$ if the defects are on opposite sublattices, and with $S=1$ if they are located on the same
sublattice. It is important to keep in mind, however, that Ovchinnikov's rule and Lieb's theorem no longer apply for doped systems \cite{CaWaLe.14.Edgemagnetizationlocal}.

\subsection{Nullity}

\begin{figure}
	\includegraphics[width=\columnwidth]{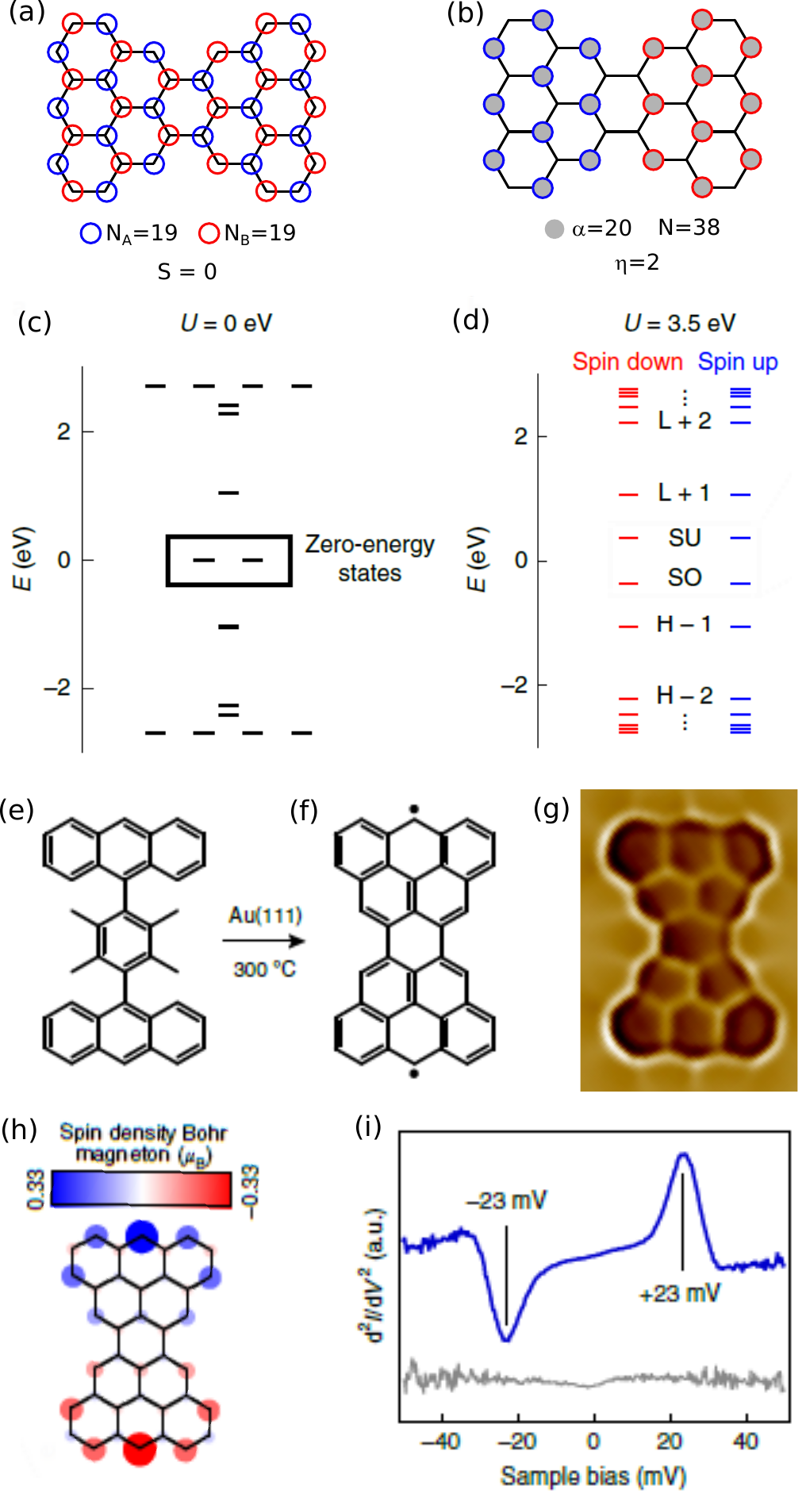}
	\caption{(a) Structure of Clar's goblet with carbon atoms belonging to sublattice A (marked in blue) and to sublattice B (marked in red). (b) Atoms that maximize the number of sites that are not nearest neighbor to each other, marking with the colored perimeter the sublattice they belong to. (c) Orbital energy spectrum of Clar's goblet calculated with TB and (d) with the MFH model. (e) Reactant utilized for the first successful synthesis of the Clar's goblet. (f) Chemical structure and (g) measured bonding structure of the product. (h) Calculated spin density distribution according to the MFH model. (i) $d^2/dV^2$ spectra acquired on Clar's goblet in the vicinity of the Fermi level, revealing inelastic spin excitation at $\pm 23$ mV.
	Panels c-i reprinted by permission from Springer Nature: Nature Nanotechnology \Ocite{MiBeEi.20.Topologicalfrustrationinduces}. Copyright (2020)}
	\label{fig:goblet}
\end{figure}

Another simple and very useful counting rule to predict magnetic properties in electronically neutral carbon-based nanostructures is the
analysis of the structure's so-called ``nullity'' ($\eta$). Nullity represents the number of zero-energy states
that result from solving the nearest-neighbor TB Hamiltonian for a specific nanostructure, dating back to early work on the spectral properties of unsaturated hydrocarbons \cite{Lo.50.SomeStudiesMolecular}. Zero-energy
states fulfill the Stoner criterion \cite{Ya.10.Emergencemagnetismgraphene} and thus undergo spin-polarization for any $U>0$ to avoid the
instability associated with low energy electrons, since alternative mechanisms like a Peierls distortion or charge
ordering have been shown to be inefficient in the case of graphene nanostructures \cite{Ya.10.Emergencemagnetismgraphene,PiChMo.07.Electronicstructuremagnetic}.
As a consequence, nullity can be
used to predict the number of $\pi$-radicals in a system, which is given by 
\begin{equation}
\eta = 2\alpha - N ,
\end{equation}
where $\alpha$ is the maximum number of sites that are not nearest neighbors to each other and $N$ is the total number
of sites. Whereas for bipartite structures $\alpha$ often coincides with the number of atoms belonging to the
dominant sublattice (or of either sublattice if they display the same number of sites), that is not always the case.
This is shown by way of example in \Figref{fig:goblet} with the iconic ``Clar's goblet'' molecule. 

We note that the nullity can be understood in terms of two contributions $\eta=\eta_\mathrm{pre}+\eta_\mathrm{sup}$, where $\eta_\mathrm{pre}=|N_A-N_B|$ is the \emph{predictable} number of zero modes from sublattice imbalance and $\eta_\mathrm{sup}\in2\mathbb{Z}$ is the remaining \emph{supernumerary} zero modes (of topological origin) that come in pairs \cite{WeScBe.16.GraphenevacanciesSupernumerary}.

Figure \ref{fig:goblet}a shows the molecular structure of Clar's goblet, differentiating with blue and red colors the sites belonging to either
sublattice, $N_A$ and $N_B$. Following Lieb's theorem, the total spin of the molecule corresponds to $S = 0$.
However, counting the maximum number of non-adjacent sites $\alpha$  as displayed in \Figref{fig:goblet}b, we obtain $\alpha = 20$. Given the total number of atoms $N = 38$, the system's
nullity amounts to $\eta =\eta_\mathrm{sup} = 2$. This is corroborated with TB calculations, which indeed predict the
presence of two zero-energy states (\Figref{fig:goblet}c) that become spin-polarized as Coulomb interactions are included with the
Hubbard model (\Figref{fig:goblet}d). This is not in contradiction with the prediction of Lieb's theorem, but
implies that the two $\pi$-radicals are antiferromagnetically oriented. This antiferromagnetic spin orientation is
also confirmed with theoretical calculations as displayed in \Figref{fig:goblet}h, in which the blue and red colors now denote the
spin density with up and down orientation, respectively. The antiferromagnetic alignment of the two radical states may
seem in contradiction with Hund's rule, according to which the spins should align ferromagnetically. The explanation of
this apparent conflict is that the two spins can be considered as belonging to independent systems, each comprising one
sublattice. Indeed, the stabilizing bonding interaction between nearest neighbor C atoms is promoted by the
antiferromagnetic alignment of their spins (as opposed to the destabilizing effect of equal spins on adjacent sites) \cite{RaGhGh.19.queststabletriplet},
which in bipartite structures leads to two sublattices displaying opposite spin orientations. The predictions from
Lieb's theorem can therefore be understood as following Hund's rule separately on the
two sublattices, and with the spins of each sublattice aligned antiferromagnetically due to a superexchange mechanism \cite{Ya.10.Emergencemagnetismgraphene}.
Indeed, as can be observed from \Figref{fig:goblet}h, the up and down spin densities on Clar's goblet are each distributed over
different sublattices \cite{Ya.10.Emergencemagnetismgraphene,MiBeEi.20.Topologicalfrustrationinduces}.

Although Clar's goblet was theoretically devised already in 1972 \cite{Cl.72.AromaticSextet} it was not until 2020 that such molecule has been
synthesized and characterized \cite{MiBeEi.20.Topologicalfrustrationinduces}. Starting from the reactant displayed in \Figref{fig:goblet}e, its deposition and subsequent
annealing on a Au(111) surface led to the formation of Clar's goblet (\Figref{fig:goblet}f), which could ultimately be imaged
with bond-resolving microscopy as shown in \Figref{fig:goblet}g \cite{MiBeEi.20.Topologicalfrustrationinduces}. The magnetism associated to the ${\pi}$-radicals was proved by
the characterization of inelastic singlet-to-triplet spin-flip excitations (\Figref{fig:goblet}i) that furthermore revealed a
remarkably large exchange coupling of 23 mV. 

\subsection{Clar sextets}
\label{sec:Clar-sextets}

It is well-established that aromatization of conjugated structures increases their band gap
and overall stability. An easy way to quantify the aromaticity with simple counting rules is analyzing the number of
Clar sextets (a six-membered carbon ring comprising three conjugated $\pi$-bonds) contained within a molecular
structure \cite{Cl.72.AromaticSextet,So.13.FortyyearsClars,KoKu.15.OrganicChemistryGraphene}. The energetically favored and therefore dominant resonance structure of polyaromatic hydrocarbons is
supposed to be the structure displaying the largest number of Clar sextets, which is termed as the ``Clar formula''.
The aromatic stabilization energy per Clar sextet is estimated to be in the order of $\approx 90$ kJ/mol \cite{SlLi.01.EnergeticsAromaticHydrocarbons,KoHiNa.10.SynthesisCharacterizationTeranthene}, which
happens to be about one third of the energy of a C--C $\pi$-bond ($\approx 270$ kJ/mol), as assessed from the
rotational barrier of ethylene \cite{KoHiNa.10.SynthesisCharacterizationTeranthene,RaLo.55.NitricOxideCatalyzed}. From this comparison one may expect that the breaking of a $\pi$-bond, along
with the associated generation of two $\pi$-radicals, may be favored if three or more additional Clar sextets can
be gained \cite{KoKu.15.OrganicChemistryGraphene,SuLePa.13.DibenzoheptazethreneIsomersDifferent}. As will be shown in later sections with a number of examples, the ratio of three sextets per broken $\pi$-bond (that
is, three sextets per pair of radicals) indeed marks an approximate threshold at which the molecules adopt an open-shell configuration. The larger the molecular structures are, the more possibilities there are to generate a sufficient
number of Clar sextets to compensate for the generation of radicals, hinting at the stronger tendency of larger
molecules to display radicals. 

It must be kept in mind, however, that this proposed threshold is only indicative and may vary from structure to structure. Some of the factors affecting this threshold will be elaborated on in later sections. 

\section{Experimental approaches}
\label{sec:experimental}

\subsection{On-surface synthesis}

A key part in the booming development of carbon-based magnetism has been clearly the progress in the synthetic capabilities. As will be discussed later in more detail, carbon nanostructures with magnetic properties are normally very reactive, which complicates their synthesis and characterization. In this respect, OSS performed under vacuum conditions offers opportunities that were previously unavailable to more conventional approaches. The inert vacuum environment, along with the generally stabilizing effect of the supporting substrate surface, allow for the synthesis of carbon nanostructures that under different conditions would either degrade extremely fast or can not be directly achieved. In spite of being a relatively young research field, OSS has been reviewed extensively and we refer the reader to more detailed descriptions on the virtues and possibilities offered by this approach \cite{ClOt.19.ControllingChemicalCoupling,WaZh.19.Confinedsurfaceorganic,HeFuSt.17.CovalentBondFormation,DoLiLi.15.SurfaceActivatedCoupling}.

\begin{figure}
 	\includegraphics[width=\columnwidth]{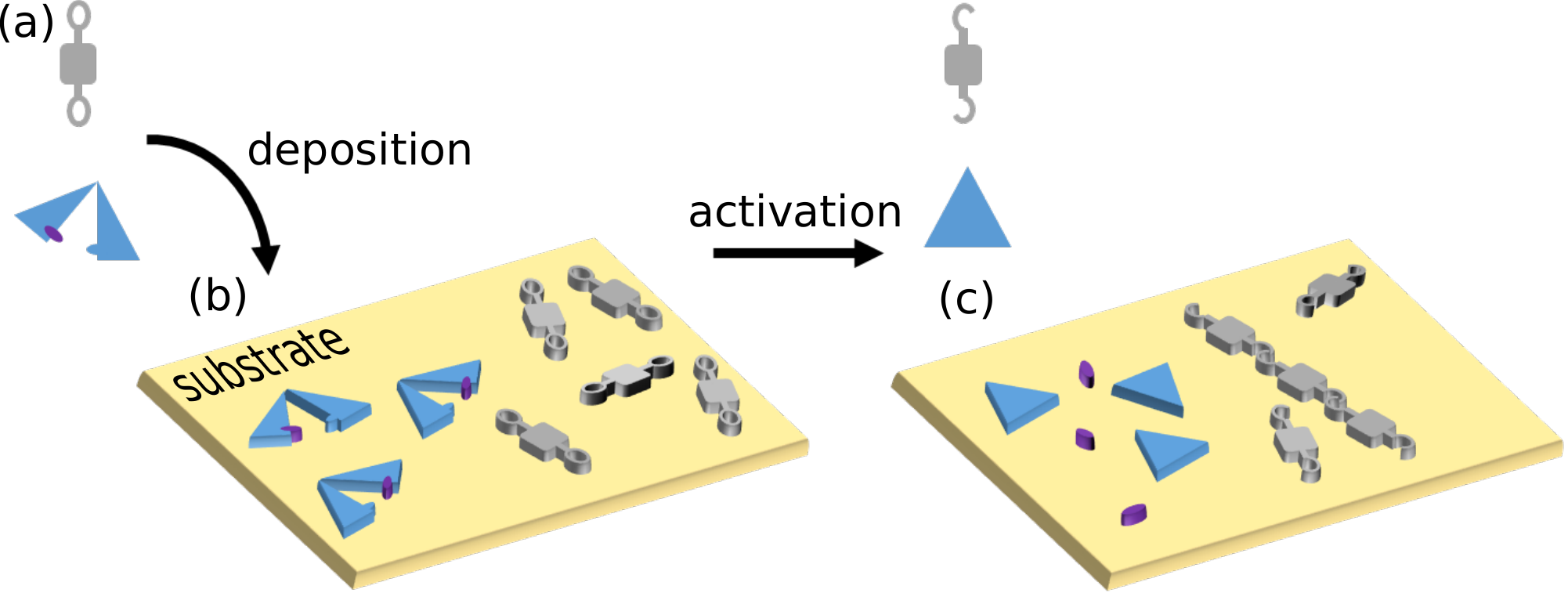}
 	\caption{(a) Schematic representation of two reactant molecules. (b) Reactant molecules deposited onto a substrate. (c) Sample after the application of stimuli to activate the reactions, which occur at the intramolecular level for the blue reactants (along with the generation of byproducts represented by the purple molecular segments), and promote intermolecular coupling reactions for the grey reactants.}
 	\label{fig:oss}
\end{figure}

In short, the OSS approach consists in the deposition of appropriately designed molecular reactants (\Figref{fig:oss}a) onto a substrate surface (\Figref{fig:oss}b), on top of which the molecules react and remain adsorbed (\Figref{fig:oss}c).
Within this frame, there is huge space for variations. Most of all with regard to the starting reactants, but also for the deposition step and the activation method. The starting reactants are designed so as to form the target product either by intramolecular reactions (as represented with the blue reactants in \Figref{fig:oss}), by intermolecular coupling processes (as represented with the grey reactants in \Figref{fig:oss}), or by combinations thereof. The utilized reactions (and the associated functional groups required on the precursors) are manifold \cite{ClOt.19.ControllingChemicalCoupling,WaZh.19.Confinedsurfaceorganic,HeFuSt.17.CovalentBondFormation,DoLiLi.15.SurfaceActivatedCoupling}, but the most common ones applied to date in the synthesis of magnetic carbon-nanostructures are cyclodehydrogenation reactions and dehalogenative aryl--aryl coupling, which is commonly referred to as Ullmann coupling. When it comes to the  deposition step, the most common approach for surfaces under vacuum is sublimation from Knudsen cells. However, there are also alternatives like the deposition from the gas phase through leak-valves \cite{VaFaSi.16.QuasiOneDimensional}, or from solution by controlled injection sources \cite{HoCoLa.21.LargeStarpheneComprising,RaRoGu.17.TwoDimensionalFolding,HiMePa.18.ElectrosprayDeposition}. Finally, for the activation step the vast majority of reported works utilize heat. Nevertheless, local activation by scanning probes have also been proved successful and provide exquisite control over the reaction process \cite{HlBaMe.01.InducingAllStepsChemicalReaction,KaScLo.19.SpHybridizedMolecularCarbon,ZhIhAh.21.ConstructingCovalentOrganic}. Alternatively, photoactivation is another option that has also been proved successful and, although hardly exploited to date in OSS, bears great potential \cite{PaLoCh.19.PhotochemistrySurfaceSynthesis}. 

The fact that the products obtained by OSS remain adsorbed on the surface is probably the most important difference with respect to heterogeneous catalysis \cite{DoLiLi.15.SurfaceActivatedCoupling}, for which chemical reactions are equally supported/catalyzed by solid substrates, but the products desorb again after reacting. Although this difference may at first seem minor, it has important implications. On the one side, it severely limits the scalability of OSS as a production method, since it can be considered as heterogeneous catalysis suffering from poisoning and the transfer of products onto different surfaces or environments requires relatively complicated protocols \cite{ClOt.19.ControllingChemicalCoupling}. Besides, also reaction byproducts may remain on the surface (see for example the purple parts of the blue reactant in (\Figref{fig:oss}), which may in turn be detrimental for the characterization or for the functionality of the system or following reactions. However, on the positive side, it allows using a battery of surface science techniques (specially for OSS performed under vacuum) that can result in a delightfully detailed characterization of the products even at the single molecule level by, \eg, SPM.        
Although the OSS approach can be applied in different environments, it is most commonly performed under vacuum, which is certainly the case for the synthesis of magnetic carbon nanostructures due to their mentioned lack of stability. The progress achieved within the last few years with the OSS of carbon nanostructures displaying magnetic properties or at least with zigzag edges (which as mentioned earlier generally promote the appearance of magnetism in the nanostructures) has been specifically summarized in recent review articles \cite{SoSuTe.21.surfacesynthesisgraphene,LiFe.20.SyntheticTailoringGraphene}.

\subsection{Characterization techniques}

As mentioned above, most of the progress achieved in the field has been thanks to the new synthetic capabilities offered by OSS under vacuum. As occurs with conventional chemistry, also in OSS the yield with which the target product is obtained varies largely from system to system. However, because target as well as byproducts remain on the surface and no means have been developed yet for purification processes, ensemble averaging measurement techniques may be hampered depending on the nature and amount of byproducts contributing to it. This is one of the reasons for which SPM are the most popular characterization techniques applied in the wider field of OSS, as well as more specifically to the study of magnetic carbon nanostructures. Nevertheless, SPM techniques specifically developed to directly measure the magnetism with spatial resolution like spin-polarized STM (SP-STM) \cite{Wi.09.SpinMappingNanoscale} or electron-spin-resonance STM (ESR-STM) \cite{BaNiMa.12.ElectronSpinResonance} have not yet been applied successfully to carbon nanostructures. 
Only ensemble ESR measurements have provided direct measurement of magnetism and the coherent manipulation of edge states on functionalized GNRs \cite{SlKeMy.18.Magneticedgestates}.
A magnetic signal that has been experimentally measured with spatial resolution by means of SPM is the Zeeman splitting under magnetic fields \cite{LiSaCa.20.UncoveringTripletGround,ZhLiZh.20.EngineeringMagneticCoupling}. Other than that, most evidences for magnetism in carbon nanostructures by SPM-based measurements with single molecule resolution have been obtained by indirect means. 

\begin{figure*}
 	\includegraphics[width=\textwidth]{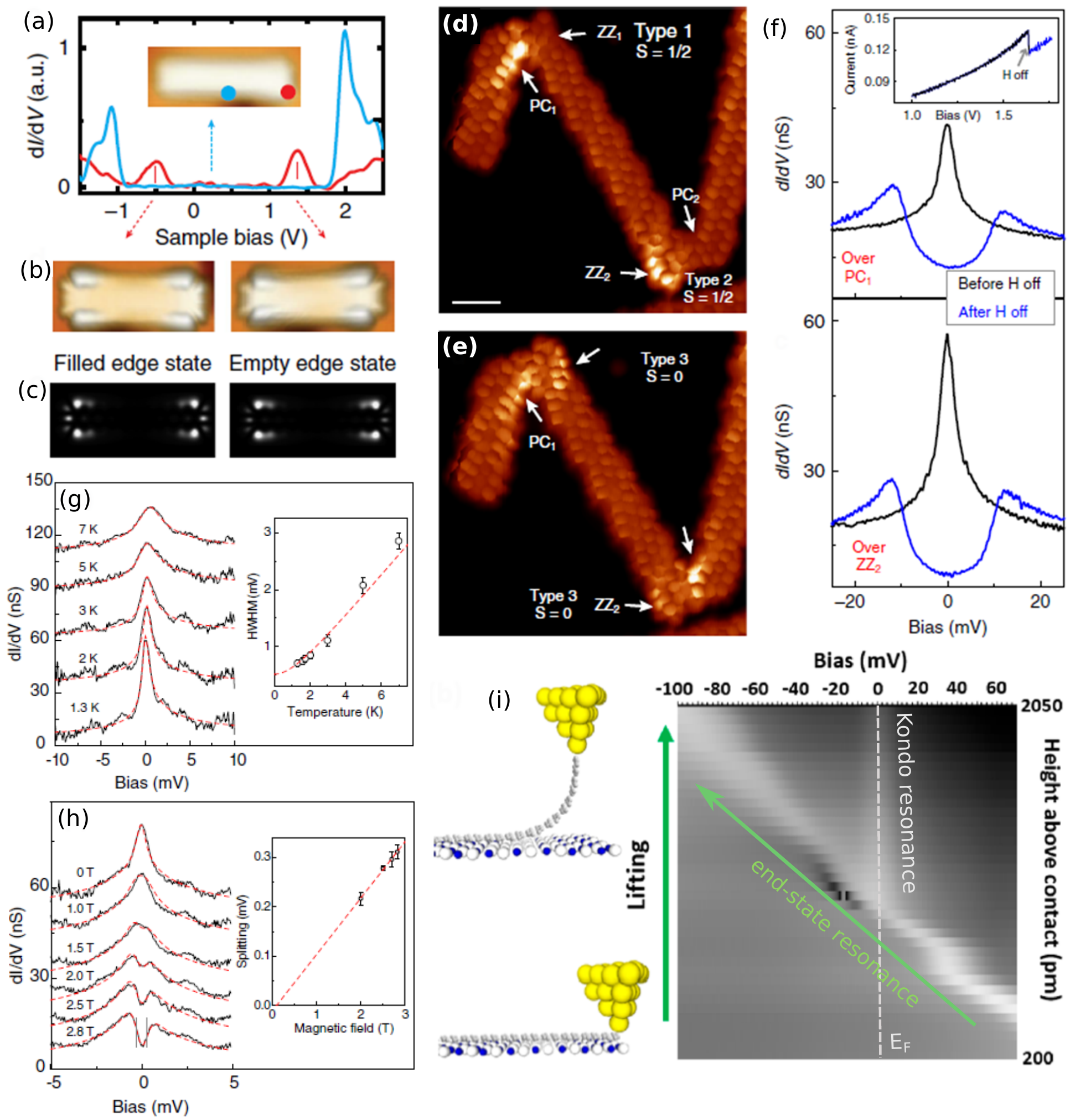}
 	\caption{(a) Conductance point spectra in the center (blue) and towards the ends (red) of a zigzag-terminated 7-AGNR on NaCl decoupling layers on Au(111). The exact location for each of the spectra is shown in the inset. In the central part the signal reveals the valence and conduction band onsets, whereas the spectra at the reveal in-gap end states. (b) Conductance maps of the two in-gap states. (c) Calculated density of states for the end states at 4 \r{A} above the carbon backbone. (d) Constant height STM image of two GNR heterojunctions, each with with one extra H atom that passivates a radical. (e) Similar STM image after tip-induced removal of the extra H atoms. (f) Conductance point spectra at each of the two junctions before (black line) and after (blue line) removal of the H atom. The inset shows the current trace during removal of one of the H atoms. (g) Temperature dependence of the Kondo resonance for an extended triangulene molecule. The inset plots the half width at half maximum (HWHM) at each temperature extracted by fitting a Frota function (red dashed lines) and corrected for the thermal broadening of the tip. The plot includes the fit to the empirical Fermi liquid model expression. (h) Magnetic field dependence of the Kondo resonance with the field strength indicated in the figure, measured at $T = 1.3$ K. The red dashed lines show the simulated curves using a model for a $S=1$ system. The inset shows the dependence of Zeeman splitting of the Kondo resonance with magnetic fields, determined from the bias position of steepest slope (indicated on the spectrum at 2.8 T). The dashed line fits the Zeeman splitting with a g factor of $1.98 \pm 0.07$. (i) Stacked conductance spectra as a function of the lifting height (as measured with respect to the contact point) of a 22 unit cell long 5-AGNR on top a NaCl decoupling layer on Au(111). Panels a-c reproduced with permission from \Ocite{WaTaPi.16.Giantedgestate}. https://creativecommons.org/licenses/by/4.0/ 
 	 Panels d-f reproduced with permission from \Ocite{LiSaCo.19.Singlespinlocalization}. https://creativecommons.org/licenses/by/4.0/
 	 Panels g-h reprinted with permission from \Ocite{LiSaCa.20.UncoveringTripletGround}. Copyright (2020) by the American Physical Society. 
 	 Panel i adapted with permission from \Ocite{LaBrBe.20.ProbingMagnetismTopological}. Copyright (2021) American Chemical Society.}
 	\label{fig:methods}
\end{figure*}

The conceptually simplest signal utilized to indirectly prove the single occupation of molecular orbitals and the associated magnetism of molecular nanostructures is the presence of a Coulomb gap. An example thereof is shown in \Figref{fig:methods}a-c \cite{WaTaPi.16.Giantedgestate}. Figure \ref{fig:methods}a displays STS data revealing the onsets of the valence and conduction bands (blue curve) of 7-AGNRs and, at the zigzag ends (red curve), two resonances within the ribbon's band gap below and above the Fermi level, respectively. These resonances are associated with magnetic GNR end states, whose origin will be discussed in later sections. The resonance below the Fermi level (negative sample bias) corresponds to the singly occupied state as probed when extracting the electron by the scanning probe, whereas the one above the Fermi level (positive sample bias) is the same state as probed when injecting a second electron into the magnetic state. The second electron suffers the Coulomb repulsion from the first electron and the resonance thus appears at an energy $U$ above the negative resonance. The similar spatial distribution of the two resonances (\Figref{fig:methods}b) supports their common origin from the same singly occupied  radical state, and final proof is given by the good match with the calculated density of states (DOS) predicted for those magnetic end states (\Figref{fig:methods}c) \cite{WaTaPi.16.Giantedgestate}. The magnitude of the Coulomb gap depends on several parameters, one of which is the screening potential of the state's environment. Magnetic states in molecules adsorbed directly on metals will, \eg, display much smaller Coulomb gaps than if adsorbed on a thin dielectric layer \cite{WaTaPi.16.Giantedgestate}. Another important parameter is the electronic state's extension. That is, the more extended the orbital is that hosts the unpaired electron, the lower the Coulomb repulsion will be if filled with a second electron \cite{WaSaCa.22.MagneticInteractions}. All in all, the magnitudes of reported Coulomb gaps can vary from the meV range, as found for example for point defects in graphene on SiC substrates \cite{GoGoMa.16.Atomicscalecontrol}, to the eV range in more localized states of molecules adsorbed on metals \cite{MiBeEi.20.Topologicalfrustrationinduces,MiBeEi.20.CollectiveAllCarbon}, which can grow to several eV by simply intercalating some thin dielectric layer between molecule and metal substrate \cite{PaMiMa.17.Synthesischaracterizationtriangulene,WaTaPi.16.Giantedgestate}.

An alternative signal that is often used as proof for magnetism is a zero-bias resonance, namely a so-called Kondo resonance, that can appear when magnetic states interact with a metallic substrate \cite{Ko.68.EffectOrdinaryScattering,TeHeSc.09.SpectroscopicmanifestationsKondo,KoGl.01.RevivalKondoEffect}. A magnetic state with energy $\varepsilon_0$ hosts an unpaired electron with spin $S=1/2$ whose $z$-component displays up or down orientation. Quantum mechanically there is a limited time of the order of $h/|\varepsilon_0|$ or $h/|\varepsilon_0+U|$ (typically some femtoseconds) during which the magnetic state can be empty or doubly occupied, respectively \cite{Ko.68.EffectOrdinaryScattering,TeHeSc.09.SpectroscopicmanifestationsKondo,KoGl.01.RevivalKondoEffect}. The former requires the electron in the magnetic state to populate an empty state of the substrate, whereas the latter requires the population of the magnetic state by a second electron from the metal. Either case can subsequently relax leaving the magnetic state with a single electron of reversed spin. That is, the impurity's spin state is constantly fluctuating and a combination of all these spin exchange processes results in the appearance of an additional Kondo resonance at the Fermi energy that can thus be directly related to the magnetic moment of the screened state \cite{Ko.68.EffectOrdinaryScattering,TeHeSc.09.SpectroscopicmanifestationsKondo,KoGl.01.RevivalKondoEffect}.

An example of such a Kondo resonance in carbon nanostructures is shown in \Figref{fig:methods}d-f, in particular at the junctions of chiral GNRs \cite{LiSaCo.19.Singlespinlocalization}. Such junctions in principle display two radical states. However, one often appears passivated by hydrogen, which leaves a single magnetic state with $S=1/2$ per junction. Conductance spectra at those positions display a clear zero-energy resonance that corresponds to a Kondo peak and therefore proves the presence of the local spin (black curves in \Figref{fig:methods}f). 

However, an unambiguous discrimination of the Kondo origin normally requires differentiating it from other possible sources of zero-bias features like tip-effects or low-energy inelastic excitations \cite{TeHeSc.09.SpectroscopicmanifestationsKondo}. There are multiple ways to do so. In scanning probe spectroscopy experiments, this is most commonly performed analyzing the temperature-dependent width of the resonance \cite{LiSaCa.20.UncoveringTripletGround,MiBeBe.20.TopologicalDefectInduced}, which displays an anomalous broadening at a much faster pace than expected from a conventional thermal broadening, as shown, \eg, in \Figref{fig:methods}g with measurements obtained on extended triangulene molecules \cite{LiSaCa.20.UncoveringTripletGround}. Furthermore, the width and its temperature dependence determine the radical's Kondo temperature, which is in turn like a measure for the interaction of the magnetic impurity with the supporting substrate and its electron reservoir \cite{TeHeSc.09.SpectroscopicmanifestationsKondo}. Alternatively, the same information can also be extracted from the temperature-dependent resonance intensity \cite{FeFrPa.08.VibrationalKondoEffect}.

Proof for the magnetic origin of the zero-bias resonance can also be obtained analyzing its response to a magnetic field, which causes a Zeeman splitting of the resonance that becomes observable at sufficiently strong fields that depend on its width or Kondo temperature \cite{LiSaCa.20.UncoveringTripletGround,ZhLiZh.20.EngineeringMagneticCoupling}. An example thereof is shown in \Figref{fig:methods}h, corresponding to the field-dependent conductance spectra on the same extended triangulene molecules as the temperature-dependent data of \Figref{fig:methods}g \cite{LiSaCa.20.UncoveringTripletGround}.

Another approach is ramping the associated state's energy, which can in turn  modify its occupation. This approach is often utilized in devices \cite{GoGoKa.98.FromKondoRegime,ScHu.17.GateControlledKondo} but not so often in SPM because in its most obvious application mode it requires a gateable system \cite{JiLoMa.18.InducingKondoScreening}. However, a similar ``gating'' effect can be obtained, \eg, when gradually lifting a molecule with the scanning probe. The fading electrostatic influence of the substrate causes the molecular states to gradually realign. This can be seen for example in \Figref{fig:methods}i, which displays conductance spectra while lifting a 5-atom wide AGNR (5-AGNR) \cite{LaBrBe.20.ProbingMagnetismTopological}. These ribbons feature magnetic end states that, when adsorbed on a Au(111) surface, become unoccupied due to charge transfer to such high work-function substrate. However, while lifting the ribbon from one of its ends, the associated end state realigns, approaching the Fermi level and becoming singly occupied again. From that moment on, the Kondo resonance appears in the conductance spectra \cite{LaBrBe.20.ProbingMagnetismTopological}.
Other transport phenomena with partially lifted GNRs have also been reported \cite{LiSaCo.19.Singlespinlocalization,FrBrLi.20.MagnetismTopologicalBoundary,KoAmJo.12.Voltagedependentconductance,ChAfSc.18.BrightElectroluminescenceSingle,LiFrMe.19.ElectricallyAddressingSpin, MaFiCr.22.Chargetransporttopological}.

A different type of signal that has been also used to demonstrate the magnetism of carbon nanostructures by SPM and spectroscopy is the presence of inelastic spin-flip excitations. A beautiful example thereof is shown again in \Figref{fig:methods}d-f with the GNR junctions \cite{LiSaCo.19.Singlespinlocalization}. By a controlled tip-induced dehydrogenation of the passivated carbon site, a second radical can be recovered. The proximity of the two radicals, along with the respective extension of the associated spin-densities, allows for a notable exchange interaction between the two spins. The ground state is a singlet, but the tunneling electrons can excite the electrons into a triplet configuration with ferromagnetically aligned spins, which appears as clear steps in the conductance spectra at $V \approx \pm 10$ mV that corresponds to the excitation energy. The steps therefore not only prove the presence of spins within the carbon nanostructure, but additionally provide information on their exchange interactions.

\section{Origin of magnetism in carbon-based materials}

\subsection{Sublattice imbalance}

\begin{figure}
	\includegraphics[width=\columnwidth]{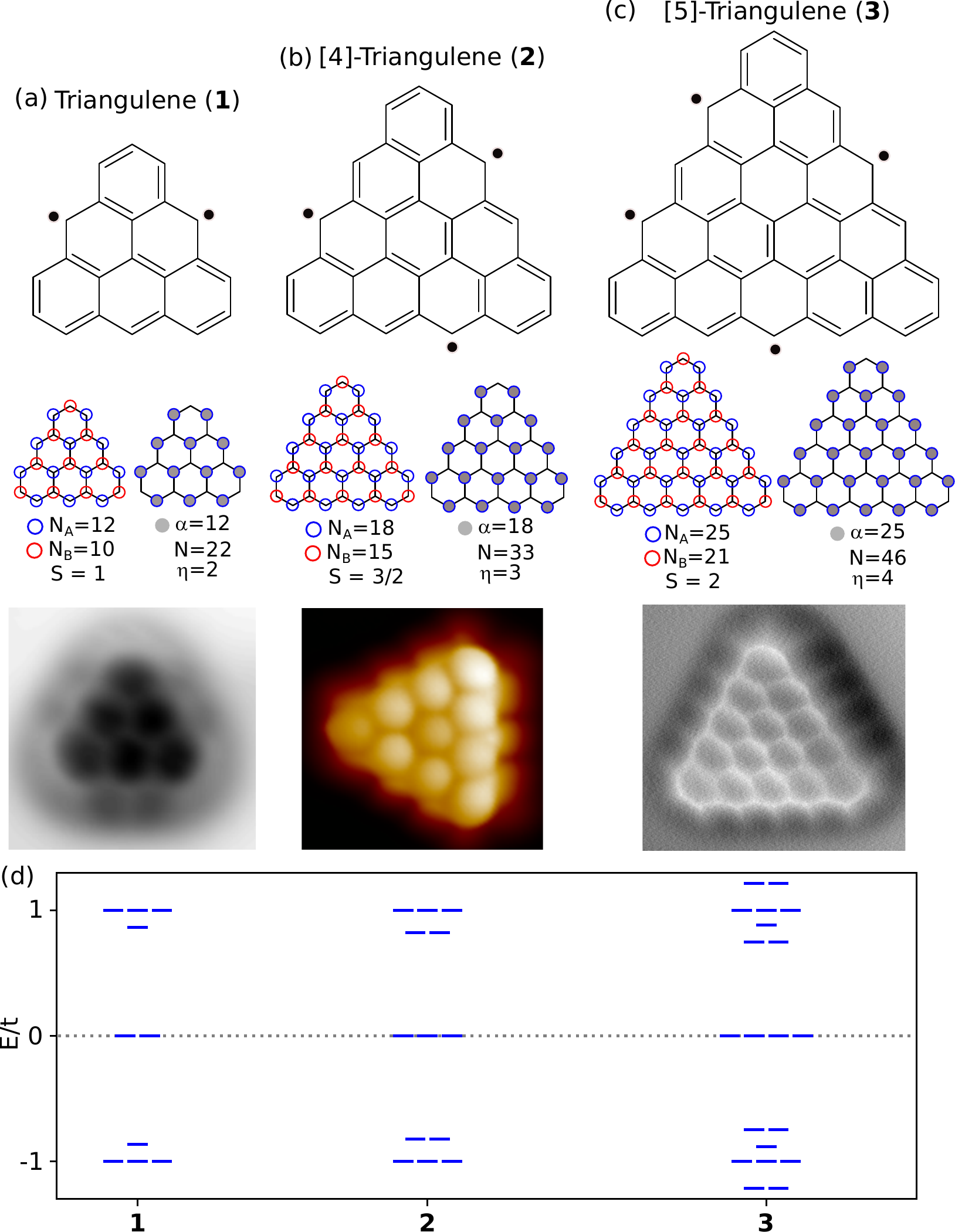}
	\caption{Molecular structure, atoms belonging to each sublattice marked in blue ($N_A$) and red ($N_B$), atoms making for the largest number of non-adjacent sites marked in grey ($\alpha$), and the bond-resolving SPM images of the associated structures for (a) triangulene, (b) [4]-triangulene, (c) [5]-triangulene, (d) Calculated low-energy spectrum for each of the structures within a TB model with nearest neighbour hopping $t$.
	Experimental image in panel a reprinted by permission from Springer Nature: Nature Nanotechnology \Ocite{PaMiMa.17.Synthesischaracterizationtriangulene}. Copyright (2017). 
	Experimental image in panel b reprinted with permission from \Ocite{MiBeEi.19.SynthesisCharacterization}. Copyright (2019) American Chemical Society.
	Experimental image in panel c reproduced with permission from \Ocite{SuTeHu.19.Atomicallyprecisebottom}}
	\label{fig:sublattice_imbalance}
\end{figure}

As already explained above in \Secref{sec:lieb}, molecules formed by alternant structures but with a different number of atoms in each
sublattice, $N_A$ and $N_B$, display a net spin of magnitude $S=|N_A-N_B|/2$ \cite{Li.89.TwotheoremsHubbard}.
An intuitive example thereof, shown earlier
in \Figref{fig:graphene}, are defects in graphene in which certain $p_z$ electrons are removed from the network either by creating C
vacancies \cite{UgBrGu.10.MissingAtomas} or by $sp^3$ rehybridization upon hydrogenation \cite{GoGoMa.16.Atomicscalecontrol}. However, there is also a large variety of finite molecular
structures that display sublattice imbalance. Some iconic examples are triangular molecules with their sides made up by
zigzag edges (\Figref{fig:sublattice_imbalance}), so called [$n$]-triangulenes, where $n$ is the number of hexagons along each molecular edge. In
particular, the cases with $n=2$ and $n=3$ are also called phenalenyl and triangulene, respectively. The latter is pictured
in \Figref{fig:sublattice_imbalance}a. Whereas 12 of its atoms belong to sublattice A, only 10 atoms belong to sublattice B, which according to
Lieb's theorem \cite{Li.89.TwotheoremsHubbard} implies a net spin $S = 1$. On the other hand, the largest number of atoms that are not nearest
neighbors to each other is 12, coinciding with the atoms of sublattice A. Given that the total number of atoms is 22,
the nullity $\eta$ amounts to 2. That is, the counting rules predict for triangulene two ${\pi}$-radicals and $S=1$, implying a ferromagnetic alignment of the two radical states. Indeed, it can be immediately realized that no
Kekul\'e structure can be drawn for triangulene. Although different resonance structures may be drawn, all of them
require the addition of two $\pi$-radicals, which furthermore are located in the same sublattice. It is
energetically most favorable for spins in adjacent sites to be antiferromagnetically coupled, so as to allow for
bonding interactions between them \cite{RaGhGh.19.queststabletriplet}. In bipartite lattices this generic assumption causes atoms on the same lattice to
share the same spin, while being antiferromagnetically aligned to atoms of the opposite sublattice. The presence of two
radicals located on atoms of the same sublattice is thus in perfect agreement with Lieb's theorem and the nullity
parameter. 

Although triangulene was already hypothesized by Erich Clar in 1953 \cite{ClSt.53.AromaticHydrocarbonsLXV}, its synthesis by conventional chemistry has only been achieved recently, and only with the help of bulky edge substituents to sterically protect the radical sites at its zigzag edges \cite{ArShSh.21.SynthesisandIsolation,VaMaKa.22.TrimesityltrianguleneaPersistent}.
Instead, OSS under UHV conditions allowed its first synthesis without further substituents in 2017 (\Figref{fig:sublattice_imbalance}a) \cite{PaMiMa.17.Synthesischaracterizationtriangulene}.
Its magnetism was
indirectly confirmed by the characterization of its singly occupied molecular orbitals. Soon after, larger members of
the triangulene family have been synthesized and characterized under vacuum as well, as is the case of [4]-triangulene
(\Figref{fig:sublattice_imbalance}b) \cite{MiBeEi.19.SynthesisCharacterization}, [5]-triangulene (\Figref{fig:sublattice_imbalance}c) \cite{SuTeHu.19.Atomicallyprecisebottom}, [7]-triangulene with \cite{SuFaMu.21.SurfaceSynthesisCharacterization} and without \cite{MiXuEi.21.Synthesischaracterization7triangulene} a well-defined hole in its center, or
other extended triangulene derivatives \cite{LiSaCa.20.UncoveringTripletGround}. As the size is increased, also the nullity and the sublattice imbalance
become greater (\Figref{fig:sublattice_imbalance}b,c), which translates into high-spin molecules whose magnetic properties have been further
confirmed with calculations and experiments \cite{LiSaCa.20.UncoveringTripletGround,MiBeEi.19.SynthesisCharacterization,SuTeHu.19.Atomicallyprecisebottom,SuFaMu.21.SurfaceSynthesisCharacterization,MiXuEi.21.Synthesischaracterization7triangulene}. By way of example, the calculated low-energy spectrum for triangulene, [4]-triangulene and [5]-triangulene are displayed in \Figref{fig:sublattice_imbalance}d, confirming the increasing number of zero-energy states in each for the structures from two to three and four, which correspond also to the number of radicals on the molecule. Indeed, that is exactly the number of radicals that need to be drawn on the resonance forms to satisfy the valence four of its carbon atoms. 

Further examples with sublattice imbalance are, \eg, heptauthrene, in which $N_A-N_B = 2$ and the net spin is therefore $S = 1$ \cite{SuLiDu.20.AtomicallyPreciseSynthesis}, or rhombus-shaped molecules whose symmetry is broken by a missing C atom, causing a sublattice imbalance of 1 and
thus $S = 1/2$ \cite{ZhLiXu.20.Designerspinorder}. As explained above for the graphene case, the hydrogenation of finite molecules can equally cause a
rehybridization of C atoms from $sp^2$ to $sp^3$, which in terms of the $p_z$ electron counting rules is equivalent to the
removal of that atomic site. Therefore, under a scanning probe microscope, a controlled tip-induced dehydrogenation of
$sp^3$ hybridized C atoms can be used to effectively add $p_z$ orbitals and thereby change the magnetic properties of the
molecules in a controlled manner \cite{MiBeEi.20.Topologicalfrustrationinduces,LiSaCo.19.Singlespinlocalization,LiSaCa.20.UncoveringTripletGround,SuLiDu.20.AtomicallyPreciseSynthesis}.
An example thereof was already shown in \Figref{fig:methods}d-f, in which doubly
hydrogenated edge atoms in GNR junctions were dehydrogenated, creating additional radical states whose
interaction with already present radicals could be assessed from inelastic spin-flip excitations in STS measurements \cite{LiSaCo.19.Singlespinlocalization}.
The graphene $sp^2$-to-$sp^3$ rehybridization by hydrogen or other adsorbates/defects, and their implications for $\pi$-magnetism, has been studied extensively by calculations \cite{SoMuFe.10.Hydrogenatedgraphenenanoribbons,SoTuDu.15.Spintransporthydrogenated,PaFeBr.08.Vacancyinducedmagnetism,LeFoAy.03.MagneticPropertiesDiffusion, YaHe.07.Defectinducedmagnetism, BoKaLi.08.HydrogengrapheneElectronic, LeSoRo.11.MagnetismDependentTransport, SoLeOr.11.MagnetoresistanceMagneticOrdering,SaAySa.12.Universalmagneticproperties, 
PiKa.22.HydrogenAtomsZigzag}.

\subsection{Topological frustration}

\begin{figure*}
  \includegraphics[width=\textwidth]{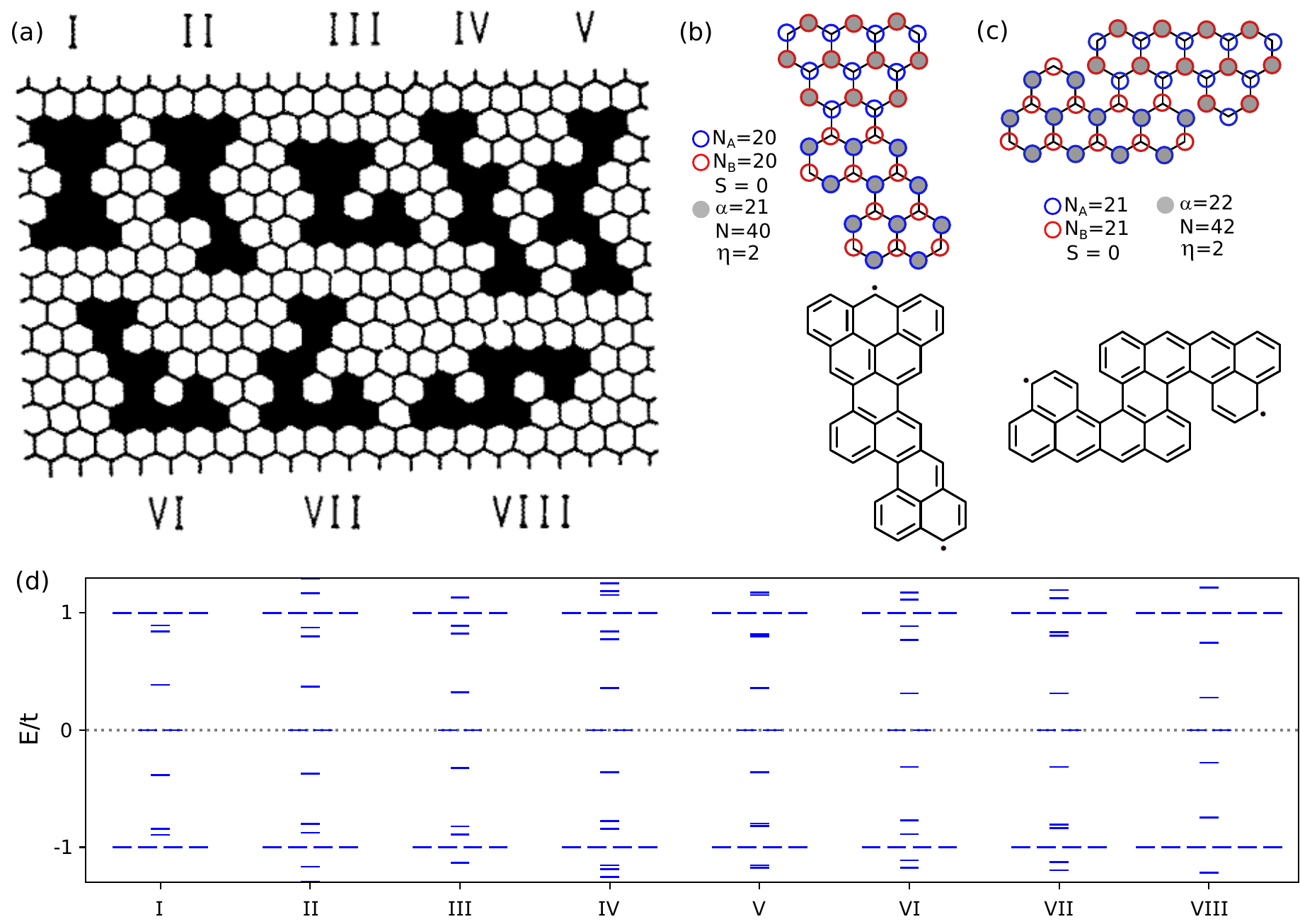}
  	\caption{(a) The eight smallest structures comprising eleven hexagons that display topological frustration in spite of their sublattice balance. (b) Chemical structure and marked atoms belonging to each sublattice as marked by blue and red circles, respectively, as well as the atoms making for the largest number of non-adjacent sites (filled in grey) for structure II. (c) Same for structure VIII. (d) Calculated low-energy spectrum within a TB model with nearest neighbour hopping $t$ for structures I to VIII. Panel a reprinted by permission from Springer Nature: J. Math. Chem. \Ocite{CyBrCy.90.huntconcealednon}. Copyright (1990)}
	\label{fig:topfrustration}
\end{figure*}

By topological frustration it is meant that there is no possible way to draw a Kekul\'e structure for the molecules. All
the molecular structures discussed in the previous section displaying sublattice imbalance suffer topological
frustration. 
Chemical structure drawings require the addition of explicit radicals to satisfy the valence four of
carbon atoms. This is easily understood with the qualitative arguments outlined in the description of the counting
rules for bipartite lattices as in nanographenes. Each carbon atom has one $p_z$ electron that can only bind to electrons
of the opposite sublattice. It automatically follows that, if the two sublattices have a different number of atoms, the
electrons will not be able to bind pairwise, leaving some unpaired electron as a radical. 
A similar scenario can also apply to non-benzenoid structures in which the sublattices are no longer well-defined. By way of example, the presence of odd-membered rings may result in structures with an odd number of $p_z$ electrons, which automatically prevents their pairwise coupling. An example thereof can be found in structure \textbf{6} of \Figref{fig:trends}.
However, topological
frustration can also occur for benzenoid carbon nanostructures displaying sublattice balance, which were formerly termed as
``concealed non-Kekul\'ean'' structures \cite{CyBrCy.90.huntconcealednon}.
The smallest of such structures (as counted by the number of comprised
hexagons) consist of eleven hexagons, which can be arranged in eight different configurations that display topological
frustration (see structures I to VIII in \Figref{fig:topfrustration}a). The number of possible ``concealed non-Kekul\'ean
structures'' rapidly grows with increasing molecular size \cite{CyBrCy.90.huntconcealednon}.

As illustrated in \Figref{fig:topfrustration}b and \Figref{fig:topfrustration}c with structures II and VIII (as well as with the Clar's goblet I in \Figref{fig:goblet}), each of
the ``concealed non-Kekul\'ean
structures'' in \Figref{fig:topfrustration} displays sublattice balance and therefore singlet character according to Lieb's theorem. However,
all of them display $\eta = 2$, as confirmed also with the TB calculations displayed in \Figref{fig:topfrustration}d, which reveal two zero-energy states for each of the structures. This can be reconciled with the singlet character by an antiferromagnetic
orientation of the two radicals' spins. The antiferromagnetic alignment can also be also expected from the location of the
radicals, which may be drawn in the chemical structures at various positions, but with the constraint of having always
one on each sublattice (\Figref{fig:topfrustration}b and \Figref{fig:topfrustration}c). Structures that display sublattice balance and topological frustration at the same time and that have been synthesized
and characterized experimentally are for example the Clar's goblet shown in \Figref{fig:goblet} \cite{MiBeEi.20.Topologicalfrustrationinduces} (structure I in \Figref{fig:topfrustration}a) or
triangulene dimers linked at their vertices \cite{MiBeEi.20.CollectiveAllCarbon}.

\begin{table*}
\caption{Examples of structure classification.}
\label{tab:classifications}
\begin{indented}
\item[]\begin{tabular}{l  c  c}
		\br
		& Kekulé & non-Kekulé \\
		\mr
		Sublattice balance & ZGNRs, chiral GNRs\cite{LiSaMe.21.TopologicalPhaseTransition}, rhombenes \cite{MiYaCh.21.Largemagneticexchange} & Clar's goblet\cite{MiBeEi.20.Topologicalfrustrationinduces} (\Figref{fig:topfrustration})\\
		($N_A=N_B$) & ($S=0$) & ($S=0$)\\
		\mr
		Sublattice imbalance & \emph{none} & Triangulene family (\Figref{fig:sublattice_imbalance}a-d), \\
		($N_A\neq N_B$)  && Extended triangulene (ETRI)\cite{LiSaCa.20.UncoveringTripletGround} (\Figref{fig:spin-interactions}a-c) \\
		& & ($S>0$)\\
		\mr
		Mechanism & Coulomb repulsion vs.~hybridization gap & Topological frustration (zero-energy states)\\
		\br
	\end{tabular}
\end{indented}
\end{table*}

\subsection{Polarization of low-energy states}

The carbon-based nanostructures discussed in previous sections (see also \Tabref{tab:classifications}) with sublattice imbalance and topological frustration all
feature zero-energy states (as predicted by the nullity parameter $\eta$), for which any repulsive Coulomb
interaction triggers a spin-polarization that reduces the density of states around the Fermi level and thereby avoids
the associated instability.
However, the same scenario can happen also with molecular states that are not strictly at
zero energy ($\eta=0$), requiring a larger Coulomb repulsion the farther the orbitals are from the Fermi level.
At this point it is worth remarking that the Coulomb repulsion discussed in this context is not the on-site Coulomb
repulsion included in the Hubbard-model, but the effective Coulomb repulsion felt by an electron upon double population
of a molecular orbital. The latter is proportional to the former, but depends on additional parameters like the
molecular state's extension. That is, a more extended $\pi$-radical state will normally display a lower Coulomb
repulsion than a more localized one \cite{WaSaCa.22.MagneticInteractions}. In turn, the extension of the radical state depends on its hybridization with the
following fully occupied (unoccupied) low-energy electronic states \cite{Ca.17.ParaQuinodimethanesUnified}. The lower the carbon nanostructure's electronic
band gap, the closer the HOMO and LUMO will be to the in-gap radical states, promoting their hybridization and thereby
the radical's delocalization \cite{CiSaTo.20.Tailoringtopologicalorder,WaSuGr.19.surfacesynthesischaracterization}.

In a simplified two-electron picture, the polarization of low-energy states takes place when the Coulomb energy of
putting a second electron into one orbital is larger than the energy cost of creating two spatially separated radical states (with reduced Coulomb repulsion) that can be understood as a linear combination of the
frontier occupied and unoccupied states (\Figref{fig:hybridization}) \cite{StChZe.19.DoDiradicalsBehave,TrMa.18.PredictingOpenShell,YeCh.16.RoleKekuleNon,YaDaYa.16.Naturegroundelectronic}.
A similar argumentation also holds for more extended systems with
more electrons, as for example the  case of ZGNRs, for which one-third of its valence band forms
a nearly zero-energy flat band that consequently undergoes spin polarization and forms the so-called GNR
edge-states \cite{Ya.10.Emergencemagnetismgraphene,RuWaYa.16.surfacesynthesisgraphene,FuWaNa.96.PeculiarLocalizedState,SoCoLo.06.EnergyGapsin,SoCoLo.06.Halfmetallicgraphene,MaJiHa.14.Roomtemperaturemagnetic}.

There are multiple ways to qualitatively understand the behavior of molecular nanostructures undergoing (or not) such
spin polarization, which often shows a notable size-dependence with relatively well-defined thresholds. Although with some exceptions \cite{KoXeSc.14.SignatureDiraccone,EiKueGa.20.DodecaceneGeneratedSurface}, in conjugated materials the band gap typically shrinks with increasing size \cite{GuPe.13.piElectronConjugation}.
Recalling that the closer the electronic states are to
the Fermi level, the lower the Coulomb repulsion is needed to favor the formation of spatially separated spin-polarized
states \cite{Ya.10.Emergencemagnetismgraphene}, larger structures with lower band gaps will thus facilitate the spin-polarization. 

A complementary explanation to the size-dependent spin-polarization is that two singly occupied radical states hybridize
into a fully occupied HOMO and a fully unoccupied LUMO as they are brought sufficiently close to one another in carbon
nanostructures below a certain size threshold. Such threshold is again dependent on the radical's delocalization
described above, and thus strongly dependent on the particular bonding structure. By way of example, finite GNRs with longitudinal armchair-shaped edges and 5 carbon atoms across their width are open-shell structures
with radical states at their zigzag ends \cite{LaBrBe.20.ProbingMagnetismTopological}. However, as the ribbon's length drops to 14 unit cells or less (which
basically comprise 28 rows or less of zigzag C atoms from end to end), the two end-states hybridize, conferring the
ribbon a predominantly closed-shell character \cite{LaBrBe.20.ProbingMagnetismTopological}. For comparison we now focus on differently shaped GNRs, namely chiral (3,1) ribbons, whose longitudinal edges are formed by a regular alternation of three zigzag
and one armchair unit \cite{OtDiGa.16.SubstrateIndependentGrowth,MeMoCa.20.Transferringaxialmolecular}. This type of ribbons hosts spin-polarized edge-states. However, as their width drops to 4
or less rows of zigzag C atom from side to side, the edge-states hybridize and result in a conventional closed-shell
semiconducting structure \cite{SuOs.15.EnergeticsElectronStates,MeLiGa.18.UnravelingElectronicStructure}.

\subsubsection{Chemical counting rules to predict spin polarization of low-energy states}
\label{sec:Clar-threshold}

The perhaps most intuitive way to qualitatively understand the spin polarization of low-energy states is based on
chemical counting rules. As already explained in \Secref{sec:Clar-sextets}, the stabilization energy of three new Clar sextets approximately compensates the energetic cost of forming a pair of radicals by breaking a $\pi$-bond. Thus, as we will see with a number of examples, if three or more sextets can be formed for each new pair of radicals, this will normally be favorable, marking an approximate threshold at which polarization of low-energy
states sets in. The larger are the molecular structures, the more possibilities there are to generate a sufficient
number of Clar sextets, supporting again the stronger tendency of larger
molecules to produce radicals by polarization of low-energy states. 

\begin{figure*}
	\includegraphics[width=\textwidth]{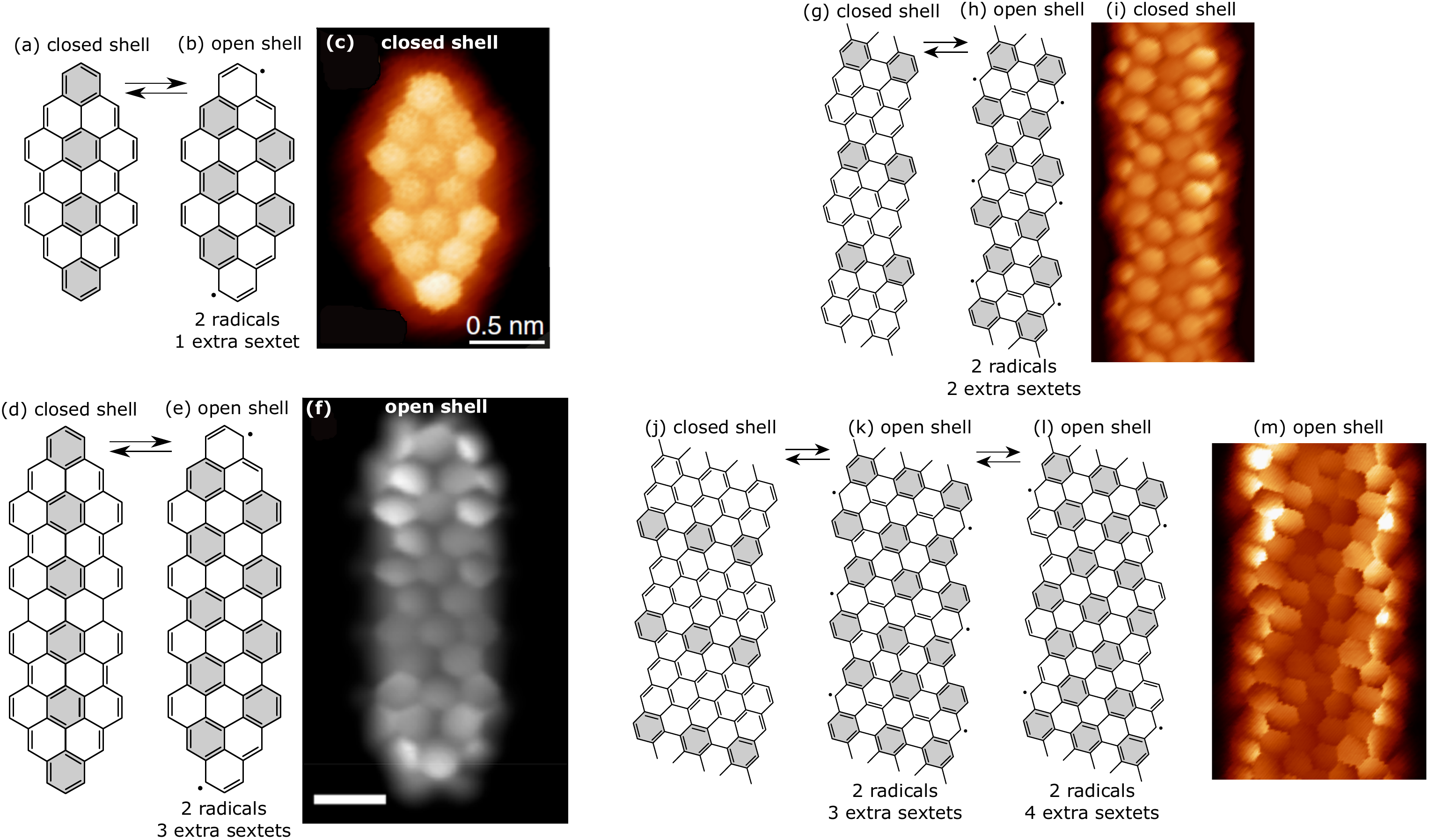}
	\caption{Closed-shell (a) and open-shell (b) chemical structure drawings for a triangulene dimer, along with an experimental bond-resolving STM image of the structure on Au(111) (c). Closed-shell (d) and open-shell (e) chemical structure drawings for an isomorphic structure with an extended carbon backbone in the central part, along with an experimental bond-resolving STM image of the structure on Au(111) (f). Closed-shell (g) and open-shell (h) chemical structure drawings for (3,1) chiral GNRs with four zigzag lines across the width of their unit cell, along with an experimental bond-resolving STM image of the structure on Au(111) (i). Closed-shell (j), open-shell (k) and another open-shell structure with a larger supercell (l) for (3,1) chiral GNRs with six zigzag lines across the width of their unit cell, along with an experimental bond-resolving STM image of the structure on Au(111) (m). For each chemical structure, the Clar sextets are highlighted in grey.
	Panel c  reprinted with permission from \Ocite{LiSaCa.20.UncoveringTripletGround}. Copyright (2020) by the American Physical Society. 
	Panel f reprinted with permission from \Ocite{MiBeBe.20.TopologicalDefectInduced}. Copyright (2020) American Chemical Society. 
	Panels i and m reprinted with permission from \Ocite{LiSaMe.21.TopologicalPhaseTransition}. https://creativecommons.org/licenses/by/4.0/}
	\label{fig:sextets}
\end{figure*}

In a first example, we analyze the number of Clar sextets (four) in the Clar formula of a molecule consisting of two
fused triangulene moieties (\Figref{fig:sextets}a). As shown in \Figref{fig:sextets}b, the number of Clar sextets can be increased from four to
five if two $\pi$-radicals are included into the structure. This molecule has been synthesized under vacuum
supported on a Au(111) surface (\Figref{fig:sextets}c) and its characterization revealed a closed-shell ground state, as expected
from the sub-threshold amount of new Clar sextets per pair of radicals \cite{LiSaCa.20.UncoveringTripletGround} On the same substrate, a similar molecular
structure with an extended molecular backbone as shown in \Figref{fig:sextets}d-f has also been studied \cite{MiBeBe.20.TopologicalDefectInduced}. Figure \ref{fig:sextets}d shows its Clar
formula, characterized by six Clar sextets. A non-Kekul\'ean structure including two radicals increases the number of
sextets to nine (\Figref{fig:sextets}e). That is, this larger molecule reaches the threshold of three additional sextets per pair of
radicals and is thus expected to undergo spin polarization. Its detailed analysis indeed confirmed the open-shell
nature of its ground state, as can be readily inferred indirectly from the STM image in \Figref{fig:sextets}f that shows the
characteristically enhanced contrast associated with low-energy states around the two molecular ends superposed to the
resolved molecular backbone structure \cite{MiBeBe.20.TopologicalDefectInduced}.

In the following we describe another example of more extended structures, namely chiral GNRs of varying
width. Chiral GNRs display an alternating number of zigzag and armchair segments. They are predicted to
be open-shell structures with so-called edge states consisting of $\pi$-radicals \cite{YaCaLo.11.Theorymagneticedge}. There is, however, a threshold
width below which the ribbons become conventional closed-shell structures \cite{CaWaLe.14.Edgemagnetizationlocal,LiSaMe.21.TopologicalPhaseTransition,SuOs.15.EnergeticsElectronStates,JiSo.15.Bandgaposcillation}. This threshold width
depends on the chirality, being lower the closer the ribbon's edge orientation is to a pure zigzag direction \cite{CaWaLe.14.Edgemagnetizationlocal,SuOs.15.EnergeticsElectronStates,JiSo.15.Bandgaposcillation}. The chirality dependence can also be explained with Clar sextet counting rules, but in the following we
will focus on the width dependence for a fixed chirality. In particular, we focus on (3,1)-chiral GNRs, which is the
only type of atomically precise chiral GNRs that have been synthesized to date with varying width \cite{LiSaMe.21.TopologicalPhaseTransition,OtDiGa.16.SubstrateIndependentGrowth,MeMoCa.20.Transferringaxialmolecular}. These
ribbons have three zigzag units per unit cell, followed by an armchair segment. Calculations on ribbons with four zigzag lines across the width of its unit cell predict no
magnetization, whereas increasing the unit cell width by only two zigzag lines more already results in an open
shell-structure with magnetic edge states \cite{LiSaMe.21.TopologicalPhaseTransition,SuOs.15.EnergeticsElectronStates}. For the narrow ribbons, the 
Clar formula displays two sextets per unit cell (\Figref{fig:sextets}g), which can be increased to four upon creation of a pair of
radicals (\Figref{fig:sextets}h). In line with the calculations, this is below the proposed threshold and these ribbons are thus not
expected to present magnetic edge states. The synthesis and characterization of this type of ribbons (\Figref{fig:sextets}i) has
indeed proved the absence of edge states, interpreted as a rehybridization of the edge states from either side with one
another as they are brought excessively close \cite{MeLiGa.18.UnravelingElectronicStructure}. For the wider ribbons, the Clar formula displays three sextets per
unit cell (\Figref{fig:sextets}j), which can be increased to six (three extra sextets) in an open shell structure with two radicals
per unit cell (\Figref{fig:sextets}k). In fact, considering the extended nature of the ribbons that allows creating larger
supercells, the ratio of newly created sextets per radical pair can be increased up to four (\Figref{fig:sextets}l). In any case it
is beyond the proposed threshold and is thus expected to favor the open-shell structure, in agreement with the
theoretical predictions \cite{SuOs.15.EnergeticsElectronStates}. Ribbons of this width have also been synthesized and characterized (\Figref{fig:sextets}m) \cite{LiSaMe.21.TopologicalPhaseTransition}.
The results confirm the presence of the edge states, as may already be inferred from the obvious density of states
around the Fermi level that provides the characteristic enhanced contrast at the GNR edges in the bond-resolving STM
images at low bias (\Figref{fig:sextets}m) \cite{LiSaMe.21.TopologicalPhaseTransition}.

The same threshold also explains the transition from closed-shell to open-shell in other molecules as a function of
their size, like in the case of periacenes \cite{SaUrVe.21.UnravellingOpenShell,BiUrMu.22.SynthesisCharacterizationPeriHeptacene, MiLoPi.18.TailoringBondTopologies}. A non-Kekul\'ean structure with two additional radicals allows for the
creation of two extra Clar sextets in the case of bisanthene. This is sub-threshold and bisanthene is therefore
expected to show a predominantly closed-shell character. However, as the periacene's size is increased, \eg, to
peripentacene or periheptacene, the non-Kekul\'ean structure with two radicals already allows for the creation of three extra Clar
sextets, whereby the threshold is reached and an open-shell structure becomes in principle favorable. SPM experiments on those structures have indeed confirmed the open-shell character of peripentacene and periheptacene and the
closed-shell character of bisanthene \cite{SaUrVe.21.UnravellingOpenShell, BiUrMu.22.SynthesisCharacterizationPeriHeptacene}. Similar conclusions and the same threshold can be extracted comparing, \eg,
rhombene molecules of different sizes \cite{MiYaCh.21.Largemagneticexchange}.

\subsubsection{Alternative factors influencing the counting rule thresholds}
\label{sec:threshold exceptions}

It should be kept in mind, however, that these counting rules are only a guidance and not at all rigorous. Additional
factors may affect the energies of open- and closed-shell structures and thereby shift the proposed threshold value of
three Clar sextets per radical pair.  

\begin{figure}
	\includegraphics[width=\columnwidth]{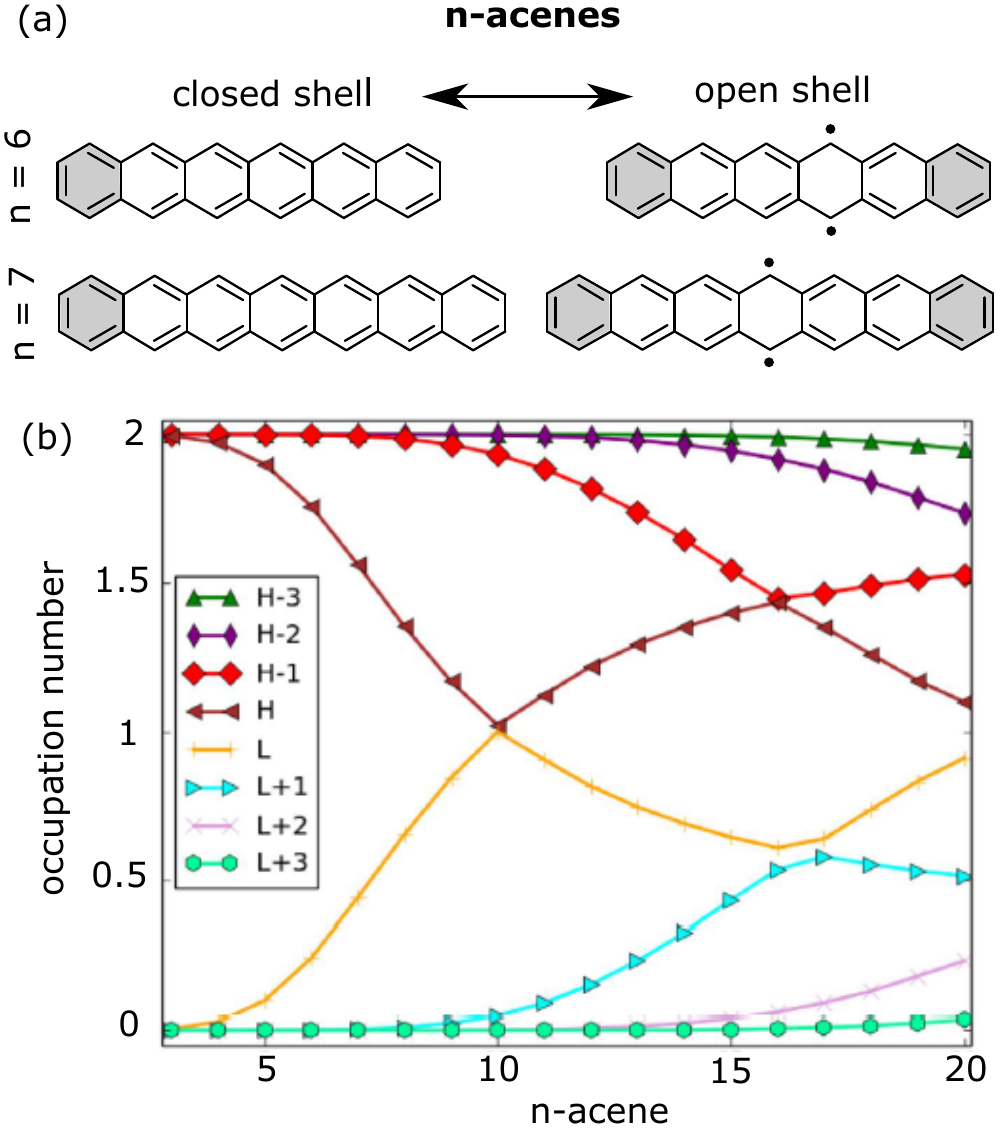}
	\caption{(a) Closed-shell and open-shell structure drawings for two acenes, namely hexacene and heptacene. (b) Active orbital occupation numbers of the low-energy orbitals for the lowest singlet states of n-acenes as a function of the acene length n, as calculated with DFT. 
	Panel b reproduced with permission from \Ocite{YeCh.16.RoleKekuleNon}.}
	\label{fig:acenes}
\end{figure}

An example thereof are acenes, which consist of linearly fused six-membered rings. A characteristic of this family of
molecules is that they only display one single Clar sextet independently of their size (\Figref{fig:acenes}a). The Clar sextet can
furthermore migrate freely to any of the six-membered rings, diluting even more the effect of the sextet's aromatic
stabilization. Indeed, calculations reveal that the homodesmic stabilization energy per $\pi$-electron scales
inversely with the acene size \cite{SlLi.01.EnergeticsAromaticHydrocarbons}. All this endows the acenes with a particularly low-energy gap as compared to similarly
sized molecules of different topologies \cite{ToeBe.21.PushingLimitsAcene,KoKu.17.BenzenoidQuinodimethanes}. As a result, above a certain size threshold that is typically assumed to
be around hexacene \cite{HaDoAv.07.radicalcharacteracenes,BeDuSt.04.OligoacenesTheoreticalPrediction}, the molecules develop an increasing open-shell character that can be understood as a growing
occupation of the LUMO orbital at the expense of the HOMO occupation
(\Figref{fig:acenes}b) \cite{HaDoAv.07.radicalcharacteracenes,YeCh.16.RoleKekuleNon} since the radical states actually
correspond to linear combinations of the occupied and unoccupied molecular orbitals \cite{StChZe.19.DoDiradicalsBehave,YaDaYa.16.Naturegroundelectronic}. An occupation of HOMO and
LUMO, each by one electron, would correspond to a full diradical, whereas the gradual occupancy of higher orbitals
(\eg,~\mbox{LUMO+1}) at the expense of lower lying orbitals (\eg,~\mbox{HOMO--1}) marks the onset of polyradical character, as has
been predicted to occur for longer acenes (\Figref{fig:acenes}b). Indeed, as a ``rule-of-thumb'' one can roughly associate an
initiating radical pair generation with every six rings \cite{HaDoAv.07.radicalcharacteracenes}. This is also the reason for the strongly compromised
stability for higher acenes with six or more rings \cite{ToeBe.21.PushingLimitsAcene,ZaBe.12.Reactivityacenesmechanisms}, which correlates well with the acene size at which a certain
diradical character starts to set in. For the same reason, it is only recently that higher acenes are being
successfully synthesized \cite{ToeBe.21.PushingLimitsAcene}. Whereas acenes as large as undecacene have been synthesized under cryogenic matrix
isolation conditions \cite{ShTaSa.18.EvolutionOpticalGap}, the largest acene that has been synthesized to date is dodecacene \cite{EiKueGa.20.DodecaceneGeneratedSurface}, obtained by OSS under ultra-high-vacuum conditions. 

In any case, as can be seen in \Figref{fig:acenes}a, acenes are a rare case in which the open-shell character sets in in spite of
gaining only one Clar sextet for each pair of radicals created. The radicals are largely delocalized over all the rings
spanned between the sextets, which lowers their energy and thereby justifies the anomalous threshold at which acenes
become open-shell. In addition, as mentioned above, it also relates to the continuously decreasing ratio of Clar
sextets (fixed to one) per carbon atom as the acenes grow longer, rendering the acenes less and less stable with
increasing size. For ratios beyond a lower limit of one Clar sextet for approximately 26 atoms (those forming the
``threshold'' value of six rings) the stabilization energy becomes too low and the generation of an additional Clar
sextet becomes favorable, along with the appearance of two radicals. 

Alternative factors that also lower the threshold of new Clar sextets per radical pair are topological defects, as will be discussed in \Secref{sec:defects}. 

\subsubsection{Associating spin-polarization to competing aromatic-quinoid conjugation patterns}
\label{sec:aromatic-quinoid}

\begin{figure*}
	\includegraphics[width=\textwidth]{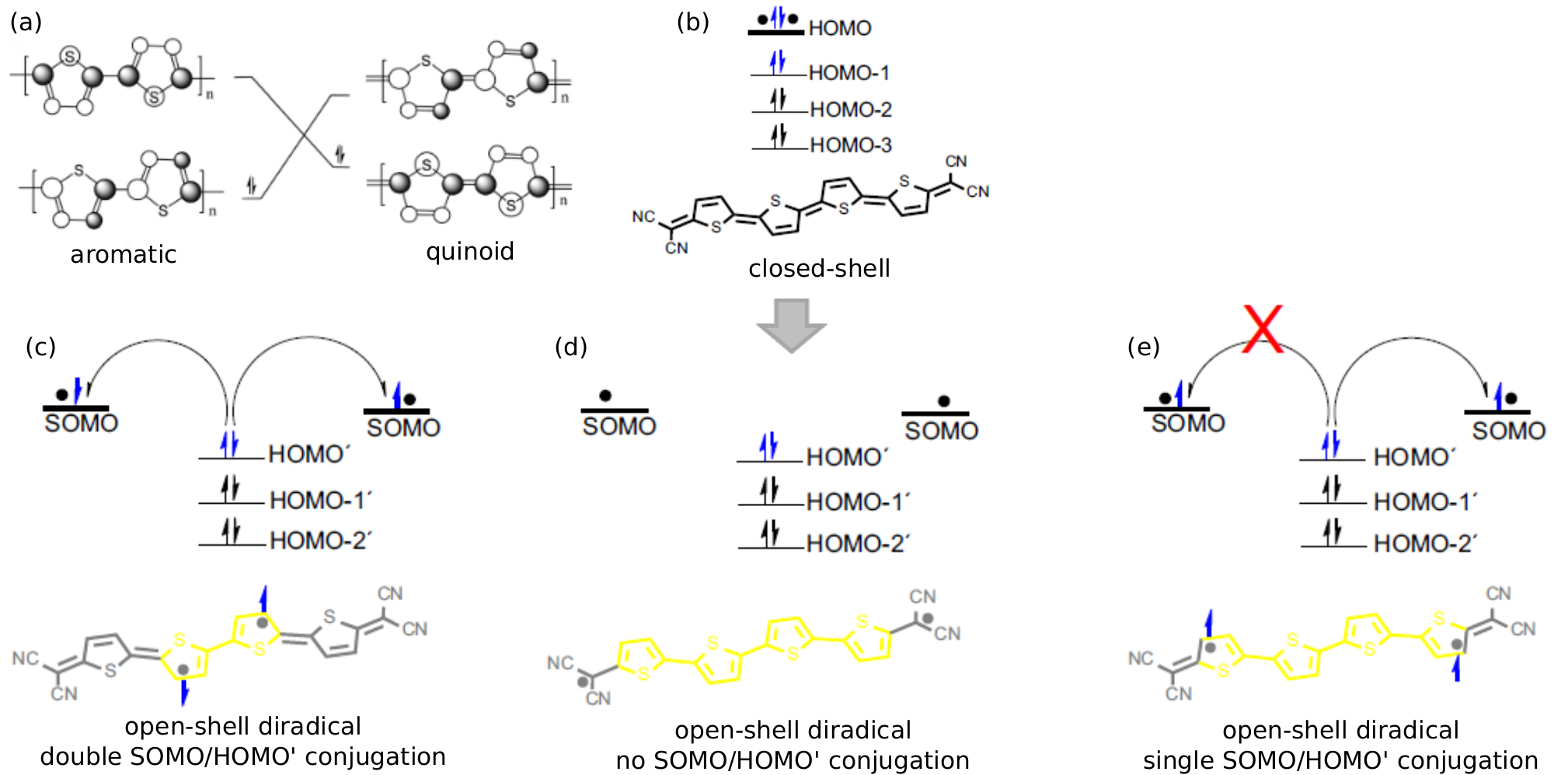}
	\caption{(a) Calculated wave functions at the onset of valence and conduction band for polythiophene in its aromatic and quinoid forms. Molecular structure drawings and associated orbitals with spin-up and spin-down electrons in closed-shell quinoid tetracyanoquaterthiophenequinodimethane (b), in its open-shell diradical configuration with no SOMO/HOMO$^\prime$ hybridization (d), in its open-shell singlet configuration with double SOMO/HOMO$^\prime$ hybridization (c) and in its open-shell triplet configuration with single SOMO/HOMO$^\prime$ hybridization. Note the differentiation between the HOMO and $^\prime$ orbitals, which do not correspond to each other. In fact, HOMO$^\prime$ in the open-shell relates to the HOMO--1 in the closed-shell configuration.
	Panel (a) republished with permission of the Royal Society of Chemistry, from \Ocite{GrPeTo.20.Quinonoidvsaromatic}; permission conveyed through Copyright Clearance Center, Inc.
	Panels b-e reprinted by permission from Springer Nature: Top. Curr. Chem. (Z). \Ocite{Ca.17.ParaQuinodimethanesUnified}. Copyright (2017).}
	\label{fig:conjugation}
\end{figure*}

Many of the cases undergoing spin-polarization of low-energy states can be understood from the point of view of a competition between aromaticity and $\pi$-electron delocalization, the latter being typically promoted by quinoid conjugation patterns \cite{ChLiWi.14.AromaticityDecreasesSingle}. In fact this applies to carbon-based materials beyond pure hydrocarbons and hexagonal patterns,
as for example shown in \Figref{fig:conjugation} with polythiophene. The change between aromatic and quinoid structures brings about a
reversal of HOMO and LUMO levels (\Figref{fig:conjugation}a) \cite{GrPeTo.20.Quinonoidvsaromatic}.
This reversal implies changes in its properties, particularly notable
when such structures are made finite. An example thereof is shown in the following with
tetracyanoquaterthiophenequinodimethane \cite{Ca.17.ParaQuinodimethanesUnified}.
Its closed-shell structure is displayed in \Figref{fig:conjugation}b, characterized by its
quinoid conjugation pattern. As the thiophene units change to their aromatic form, two radicals are concomitantly
generated at its ends (\Figref{fig:conjugation}d). The generation of $\pi$-radicals implies an energetic cost that may be lowered by
their hybridization with the electrons of the following doubly occupied orbitals (HOMO$^\prime$ in \Figref{fig:conjugation}c-e) \cite{Ca.17.ParaQuinodimethanesUnified}. The lowered
energy of such hybridized radicals facilitates its compensation by the aromatic stabilization, which may promote the
diradical character of the molecule. Several properties of the radicals can be explained by this hybridization
scenario. 

One is the usual antiferromagnetic alignment of the $\pi$-radicals in symmetric and linear molecules of this kind,
which renders their ground state a singlet. Under these conditions, the ``spin-up'' electron of the HOMO$^\prime$ can hybridize
with the ``spin-down'' singly-occupied molecular orbital (SOMO) electron and vice versa (\Figref{fig:conjugation}c), contributing to the delocalization and stabilization of
the radicals. Instead, if the two radicals align ferromagnetically into a triplet, only one of the $^\prime$ electrons
(spin-down) can hybridize with one of the two spin-up radicals, whereas the hybridization of the second (spin-up) HOMO$^\prime$
electron with the SOMO electrons is hindered by the Pauli exclusion principle (\Figref{fig:conjugation}e) \cite{Ca.17.ParaQuinodimethanesUnified}. The double
hybridization/stabilization of the singlet in comparison to the single hybridization/stabilization of the triplet cause
the former to be often energetically favored and thus commonly the ground state in linear diradical molecules. 

Another characteristic property of the radical states, namely their spatial distribution, can also be related to the
hybridization picture. The hybridization of the radical states with the extended doubly occupied orbitals allows for
their increased delocalization. Given that the hybridization increases the closer the in-gap radical states are to the
following doubly occupied orbitals, the radical delocalization scales inversely with the molecular band gaps \cite{CiSaTo.20.Tailoringtopologicalorder,WaSuGr.19.surfacesynthesischaracterization}.
As can be easily discerned from the wireframe chemical structure  drawings (\Figref{fig:conjugation}), the radical states appear at the ends
of the aromatic structure independently of whether this coincides with the molecular ends or with the junction to a
quinoid section. The delocalization of the radical states away from the edges thus implies a change of conjugation back
to the quinoidal form near the molecular ends (\Figref{fig:conjugation}c). Eventually, this may result in a maximum radical density
displaced away from the molecular ends, an effect that has indeed been characterized \cite{CiSaTo.20.Tailoringtopologicalorder} and directly observed in real
space by SPM on different diradical molecular structures \cite{CiSaTo.20.Tailoringtopologicalorder,LaBrBe.20.ProbingMagnetismTopological}.

\subsubsection{Analysis from a topology viewpoint}

As explained in \Secref{sec:top-band-analysis}, a useful alternative way to understand and predict the presence of radical states in certain molecular structures is topology. From the symmetries of the filled bands of infinite systems, or from the calculated Zak phase, the materials are classified as trivial or nontrivial. For the latter, zero energy modes appear at the structure's ends as they are made finite. 

\begin{figure*}
	\includegraphics[width=\textwidth]{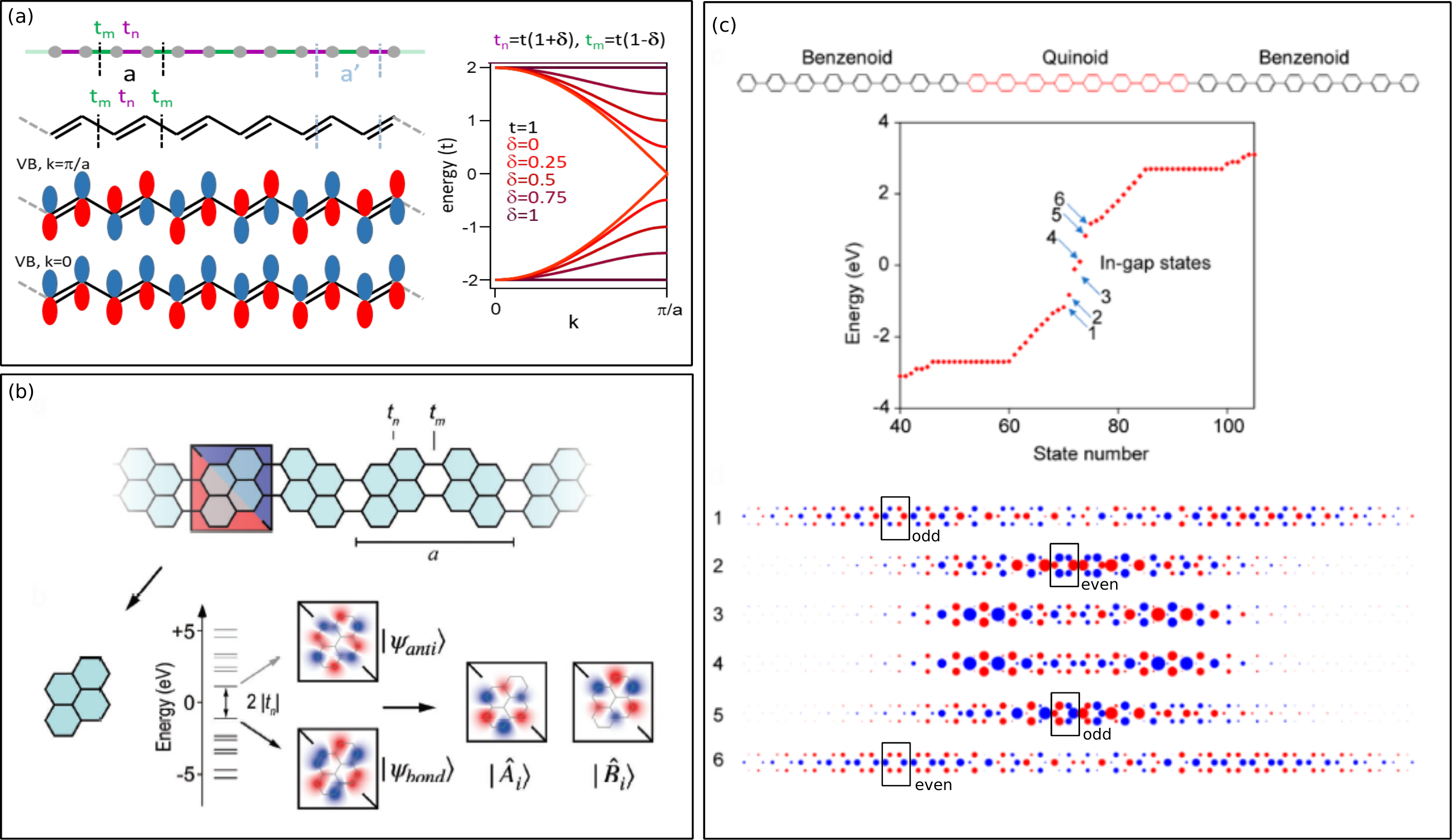}
	\caption{(a) Model of linear atoms with electrons coupled by alternating hopping constants $t_n$ and $t_m$ and two possible unit cell choices $a$ and $a^\prime$. Below is the chemical structure of polyacetylene, formed by carbon atoms with one $p_z$ electron each and alternating single and double bonds, that can be represented by this model. The calculated band structure is shown on the right for varying $t_n$ and $t_m$ values, and the wave functions of the valence band at the $\Gamma$ point and at the zone boundary are shown at the bottom. (b) Carbon backbone of a GNR formed by pyrene units and the modelling of its low-energy bands with the SSH model. The calculated HOMO and LUMO levels are represented as the bonding and antibonding combinations of the two orbitals that conform the SSH chain. The intra-unit cell hopping constant $t_n$ corresponds to half of the unit cell's HOMO--LUMO gap, whereas the inter-unit cell hopping constant depends on the inter-pyrene coupling motif and is obtained from fitting the SSH band-structure to that of the actual polymer. (c) Polyparaphenylene oligomer comprising benzenoid and aromatic sections, along with the energy spectrum for all of its electronic states. The six lowest energy states are labelled 1 to 6 and their calculated wave functions shown below. They can be assigned to the valence- and conduction-band onsets of the aromatic and quinoid sections (with accordingly reversed symmetries, as can be recognized from the marked unit cells in the boxes), plus two in-gap interface states.
	Panel b adapted with permission from  \Ocite{SuGrOv.20.MassiveDiracFermion}. Copyright (2020) WILEY-VCH Verlag GmbH \& Co. KGaA, Weinheim. Panel c adapted with permission from \Ocite{YuLiZh.20.ResolvingQuinoidStructure}. Copyright (2020) American Chemical Society.}
	\label{fig:ssh}
\end{figure*}

The simplest organic conjugated chain for which to calculate this is polyacetylene, which has only two $p_z$ electrons per
unit cell and thus only one filled $\pi$-band. It can be easily modelled with TB by a linear array of
electronic states coupled with alternating hopping constants $t_n$ and $t_m$ (\Figref{fig:ssh}a) in what is called the
Su-Schrieffer-Heeger (SSH) model \cite{SuScHe.79.Solitonsinpolyacetylene} The solutions to the Hamiltonian are given by 
\begin{equation}
E_{\pm}(k) = \pm \sqrt{t_n^2+t_m^2+2 t_n t_m \cos(k)}
\end{equation}
and the associated bands are plotted in \Figref{fig:ssh}a for a variety of values of ${\delta}=t_n-t_m$, showing how the band
width increases and the band gap shrinks with a decreasing ${\delta}$ parameter, eventually becoming metallic for
${\delta} = 0$. Double bonds display a larger hopping constant than single bonds, which in our diagram implies that 
$t_n > t_m$ and ${\delta}>0$. The wave functions are shown for the valence band states at the
${\Gamma}$ point and at the Brillouin zone boundary (\Figref{fig:ssh}a). What remains to be determined is the unit cell, which is
defined according to the finite chain's ends in such a way that the unit cell is commensurate with the chain's boundary
conditions. If the chain ends with a double bond (that is, the chain is ``cut'' through a single bond), the unit cell
will be given by $a$, marked with dashed black lines. On the contrary, if the chain ends with single bonds (that is, the
unit cell's ends and therefore the chain's ends ``cut'' through double bonds), the unit cell will be given by $a^\prime$,
marked with bright blue dashed lines. Whereas for $a$ the larger hopping constant is intra-unit cell, it is inter-unit
cell for $a^\prime$. 

Calculations of the Zak phase according to \Eqref{eq:zak} provide the same result as that extracted from the parity of the
Bloch wave functions according to \Eqref{eq:parity}. From \Figref{fig:ssh}a, we can observe that the parity of the valence band (VB) at $k = 0$ is even
($+1$) regardless of the unit cell choice. However, at the zone boundary the parity of the VB states is even ($+1$) for $a$ but
odd ($-1$) for $a^\prime$. Consequently also the  $\mathbb{Z}_2$  invariant changes from 0 (topologically trivial) for $a$ to $+1$
(topologically nontrivial) for $a^\prime$. The topologically non-trivial character of $a^\prime$ comes along with the appearance of
radical states at the chain's ends. An intuitive picture of these state's origin is that the chain's ends ``cut''
through what would have been a $\pi$-bond, leaving an unpaired electron behind.

The usefulness and simplicity of this SSH Hamiltonian is such that also other more complex polyaromatic structures have
been mapped onto this model under the assumption that the HOMO and LUMO orbitals of the unit cell can be viewed as the
bonding and antibonding dimer states of the SSH chain \cite{CiSaTo.20.Tailoringtopologicalorder,SuGrOv.20.MassiveDiracFermion}. This is exemplified in \Figref{fig:ssh}b with pyrene-based GNRs \cite{SuGrOv.20.MassiveDiracFermion}. The topological class of the structures is then simply determined from comparing the resulting intra
($t_n$) and inter-unit cell ($t_m$) hopping constants. If the inter-unit cell hopping constant is larger, the structure is
topologically non-trivial and will display end states. 

In fact, those radicals appear at any topological interface. Vacuum can be considered as a topologically trivial
material, and the ends of the chains discussed earlier represent an interface between a topologically non-trivial
polyacetylene chain and vacuum. However, substituting the vacuum by another topologically trivial carbon-based material
results in a similar radical state at the interface. For example, as discussed earlier, the change from aromatic to
quinoid structures brings about a reversal of HOMO and LUMO levels (\Figref{fig:conjugation}a). Such crossing is exactly what one expects
when changing the topological class of a given material, as can be immediately concluded also from analyzing their
respective symmetries. For the polythiophene pictured in \Figref{fig:conjugation}a the HOMO or VB onset wave function is characterized by an odd
parity, whereas it becomes even for the quinoid structure. Assuming that no other symmetry inversion occurs at the zone boundary or on other
bands, this implies a change in the topological class and therefore the presence or absence of radical end-states (\Figref{fig:conjugation}). In fact, \Figref{fig:conjugation}c also shows how the radical states appear at the junction between aromatic and quinoid sections
that come along with the hybridization of the end states with the fully occupied $^\prime$-orbitals.

Another instructive case is polyparaphenylene (PPP), which can also display aromatic and quinoid conjugation character,
with the associated cross-over of valence and conduction band onsets. Figure \ref{fig:ssh}c shows calculations of a finite PPP
chain with a quinoid section embedded between two aromatic (benzenoid) regions, each of them 8 unit cells long \cite{YuLiZh.20.ResolvingQuinoidStructure}. The
electronic states associated with this structure are calculated and the low-energy orbitals (labelled 1 to 6 in \Figref{fig:ssh}c) are analyzed. Aromatic structures normally display larger band gaps than their quinoid counterparts \cite{YuLiZh.20.ResolvingQuinoidStructure,MiOs.20.Smallbandgapquinoid}. It is
therefore not surprising to see that the third occupied and unoccupied orbitals (labelled as 1 and 6) correspond to
what can be seen as the valence and conduction band onsets of the aromatic segments (or rather the first confined state
of the bands in these finite aromatic segments). The second occupied and unoccupied orbitals (labelled as 2 and 5) in
turn correspond to the analogous valence and conduction band onsets of the quinoid segment. Looking at their respective
wave functions and the associated longitudinal symmetry within the unit cells (marked with squares for the aromatic and quinoid
segments separately), one can easily distinguish their reversal, confirming the different topological classes of either
conjugation pattern. Most interestingly, two in-gap states appear (labelled 3 and 4) around the aromatic-quinoid
junctions, although extending further into the quinoid segment that displays the lower band gap, exactly as expected
from topological interface states. Samples resembling the calculated structure have been realized experimentally on
Cu(111) surfaces and their characterization by SPM and STS has confirmed the presence of
those in-gap states \cite{YuLiZh.20.ResolvingQuinoidStructure}.

At this point it should be remarked that a topology analysis can be understood as if the bulk properties determine the properties at its edges or interfaces. Although it thus allows predicting the presence or absence of radical states in finite systems from calculations performed for infinite systems, it loses significance for systems of reduced dimensions. Nevertheless, many of the radical states studied in carbon-nanostructures can be understood from topological arguments, as are the end-states of different types of one-dimensional polymers \cite{CiSaTo.20.Tailoringtopologicalorder}, GNRs \cite{CaZhLo.17.TopologicalPhasesGraphene,WaTaPi.16.Giantedgestate, LaBrBe.20.ProbingMagnetismTopological}, interface states at GNR heterojunctions of varying width \cite{RiVeCa.18.Topologicalbandengineering}, or of varying backbones with, \eg, heteroatoms \cite{FrBrLi.20.MagnetismTopologicalBoundary}. However, it is also worth reminding that such radical states can be equally well explained with alternative arguments, as are for example the conjugation patterns discussed in the previous section, or counting rules for ``locally uncompensated'' lattice sites at the ends of long one-dimensional systems with two ends that are ``independent'' from one another.

\subsubsection{Controlling the aromatic-quinoid balance and topological class}

\begin{figure*}
 	\includegraphics[width=\textwidth]{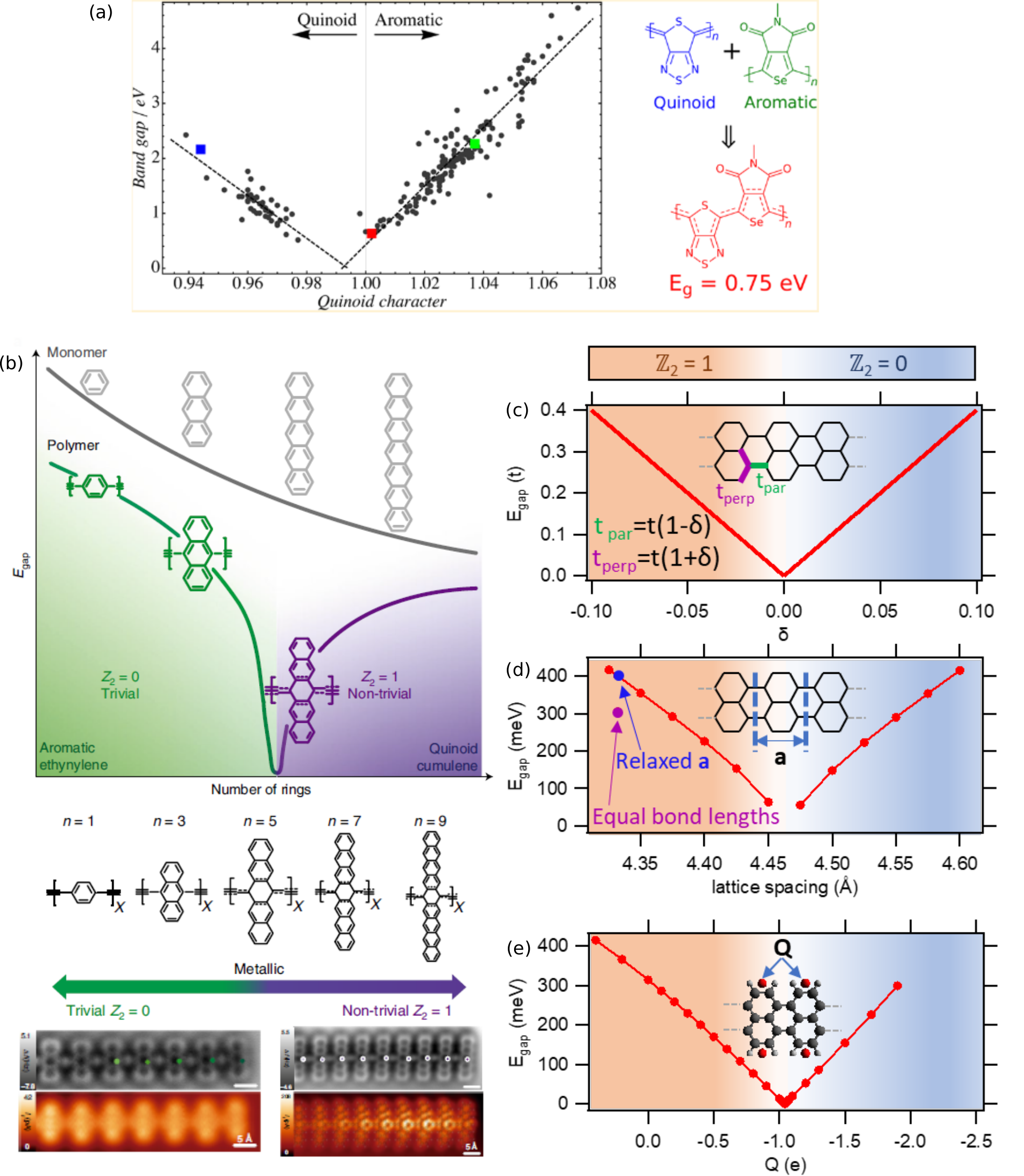}
 	\caption{(a) Calculated band gap of homopolymers and copolymers according to their quinoid or aromatic character, as represented by a parameter that indicates an increasing quinoid (< 1) or aromatic (> 1) character as the values depart from 1. Mixing aromatic (green) and quinoid (blue) units results in more balanced conjugation structures with consequently lower band gaps (red). (b) Schematic evolution of the band gap of acene monomers and acene-based polymers with increasing size of acene monomer. The latter case shows a phase transition between trivial and non-trivial topological classes accompanied by transformation of the $\pi$-conjugation of the polymer. At the bottom, noncontact-AFM and STM images of the ends of an anthracene-based polymer and a pentacene-based polymer are shown, revealing an end-state for the latter but not for the former. (c) Calculated band gap and topological class for 5-atom wide AGNRs as a function of the hopping constants in the bonds parallel to the ribbon's longitudinal axis ($t_{par}$) or perpendicular to it ($t_{perp}$). (d) Calculated band gap topological class for the same ribbons as a function of its lattice spacing $a$. (e) Calculated band gap topological class for the same ribbons as a function of the charge on the Yukawa potentials marked in the inset.
 	Panel a reprinted with permission from \Ocite{BeGaCo.13.LowBandGap}. Copyright (2013) American Chemical Society. Panel b reprinted by permission from Springer Nature: Nature Nanotechnology \Ocite{CiSaTo.20.Tailoringtopologicalorder}. Copyright (2020) 
 	Panels c-e reprinted with permission from \Ocite{LaBrBe.20.ProbingMagnetismTopological}. Copyright (2020) American Chemical Society.
 	}
 	\label{fig:topo}
\end{figure*}

The inspiring examples described above, along with many others, have motivated the rational design of molecular structures with premeditated
conjugation patterns to obtain materials according to our needs. Controlling the topological class determines the presence (or absence) of magnetic topological interface or end-states, which may be even rationally coupled by a tailored distance between them. In addition, creating materials that carefully
balance between their quinoid and aromatic structures typically have the lowest band gaps and also low reorganization
energies, thereby promoting high charge mobilities \cite{CiSaTo.20.Tailoringtopologicalorder,SuGrOv.20.MassiveDiracFermion,BeGaCo.13.LowBandGap}.
\Figref{fig:topo}a shows an example thereof, in which the band gap of a
variety of conjugated polymers is displayed as a function of their parametrized quinoid character \cite{BeGaCo.13.LowBandGap}.
The data display a
characteristic V-shape whose zero band gap value roughly coincides with a balanced aromatic-quinoid structure.
Interestingly, that balance (and therefore the polymer's band gap) can be controlled by an appropriate combination of
more aromatic and more quinoid moieties. Figure \ref{fig:topo}a illustrates a sample case with two constituents whose polymeric form
displays comparable band gaps but with quinoid (blue) or aromatic (green) character. Their combination in mixed polymers
results in structures with a relatively balanced aromatic/quinoid structure and a greatly diminished band gap \cite{BeGaCo.13.LowBandGap}.

Closely related to this, a recent work demonstrated how the use of differently sized units within acene-based polymers
linked by ethynyl bridges can be used to promote aromatic--ethynyl or cumulene--quinoid conjugation patterns (\Figref{fig:topo}b),
each belonging to a different topological class \cite{CiSaTo.20.Tailoringtopologicalorder,CoOt.20.Topologicalengineeringmetallic}. The pentacene case is closest to the cross-over point and is
therefore the polymer with lowest band gap, which amounts to only 0.35 eV as measured by STS on a Au(111) surface.
Whereas the anthracene-based polymers are topologically trivial, the pentacene-based polymers display a
topological nontrivial character, as predicted from calculations and confirmed experimentally with the absence and
presence of topological end-states, respectively (see bottom part of \Figref{fig:topo}b) \cite{CiSaTo.20.Tailoringtopologicalorder,CoOt.20.Topologicalengineeringmetallic}.
For the latter, the low band gap
promotes a significant hybridization of the end-states with the frontier orbitals, causing the end state's pronounced
delocalization and a maximum intensity shifted away from the actual polymer's ends, as discussed earlier in \Secref{sec:aromatic-quinoid}. The
associated rehybridization into a predominantly ethynyl-aromatic structure at the polymer's ends can be nicely resolved
from the bond-resolving nc-AFM measurements \cite{CiSaTo.20.Tailoringtopologicalorder}.

There are also alternative ways that have been proposed to move across the aromatic/quinoid or topological phase map
that do not even modify the backbone structure of the $p_z$ orbital network. Taking into account that topology can be
associated to orbital symmetries, it seems intuitively easy to modify those by changing the carbon backbone and thereby
the network of $p_z$ electrons. However, it is less intuitive to assume their modification for a fixed electron network. One of the various ways is through a polymer length control. 
By way of example, the previously described pentacene-based polymers undergo a topological transition and become trivial for lengths below $\sim$26 unit cells \cite{GoMeEd.21.AtomicScaleControl}. The reason underlying such length-dependent topological transition is a  pseudo Jahn--Teller effect driven by vibronic coupling of low-lying excited states 
to vibrational modes of specific symmetries, whereby a new ground state is established \cite{GoMeEd.21.AtomicScaleControl, Be.93.PseudoJahnTeller}. This mechanism may impose a similar length-dependence in the topological class also 
to other $\pi$-conjugated systems like graphene nanoribbons. 

Alternatively, TB simulations have shown that a key to move across the topological phase diagram is the control of the hopping constants between
lattice sites, in such a way that differences can be created for different types of bonds \cite{SoCoLo.06.EnergyGapsin,LaBrBe.20.ProbingMagnetismTopological}. For example, the
topological class of AGNRs with five atoms across its width (5-AGNRs) and ending with a row of
zigzag atoms changes from trivial to nontrivial if the $\pi$-bonds parallel to the ribbon's axis display a weaker
or stronger hopping constant than the transverse bonds (\Figref{fig:topo}c) \cite{LaBrBe.20.ProbingMagnetismTopological}. In fact, the same results were obtained if only the
hopping constant of the outermost parallel bonds between the singly hydrogenated carbon atoms at both sides of the ribbon
are made different from the rest. One way to controllably modify the hopping constants of parallel and transverse bonds
is by the application of uniaxial strain, as calculated in \Figref{fig:topo}d for varying 5-AGNR lattice constants \cite{LaBrBe.20.ProbingMagnetismTopological}. However, a
similar effect can be obtained if the electrons feel a modulated electrostatic potential on different bonds. This has
indeed been found to be the reason for 5-AGNRs to be topologically nontrivial. The positive partial charge on the
hydrogen atoms at the sides make the electrostatic potential for the electrons to be more favorable on the outer bonds
parallel to the ribbon, increasing the associated hopping constant and rendering 5-AGNRs nontrivial. This effect can,
however, be inverted by a modified electrostatic potential. This has been proved at the theoretical level by adding a
varying charge to Yukawa potential centers located in between the hydrogen atoms (\Figref{fig:topo}e) \cite{LaBrBe.20.ProbingMagnetismTopological}, but could eventually be
realized experimentally, \eg, with varying edge functionalization with polar groups \cite{CaHiVi.17.DopingGrapheneNanoribbons,LiBrVi.20.BandDepopulationGraphene} or electron-rich atoms like
fluorine instead of hydrogen.  

A somewhat related concept that has also been proposed at a theoretical level is based on a topological inversion
process driven by transverse electric fields. GNRs that are asymmetric across their width can be engineered so as to have the valence and conduction bands displaying
opposite energy shifts in response to transverse electric fields, potentially enabling a band inversion above a
critical electric field that would bring along a change in the topological class. A type of ribbons that has been
proposed to fulfil the requirements for transverse electric field-induced topological changes is 11-AGNRs with nitrogen
and boron heteroatom dimers on opposite sides of the ribbon \cite{ZhCaLo.21.ElectricFieldTunable}. Taking a step beyond merely moving across the
topological phase diagram for the whole ribbon, a superlattice electric field of modulated strength or direction (as
may, \eg, be created by a periodic array of parallel gates with alternating bias values), has been further proposed to
allow for the controlled generation of periodic topological interface states whose coupling can be programmed by the
spatial profile of the electric field that determines the interface state spacing \cite{ZhCaLo.21.ElectricFieldTunable}.

All in all, controlling the topological class of carbon nanostructures allows for the tailored generation of magnetic states at the ends, edges, or topological interfaces within their carbon backbones, which represents an extraordinary playground for the generation of custom-made materials.

\subsection{Topological defects}
\label{sec:defects}

\begin{figure}
	\includegraphics[width=\columnwidth]{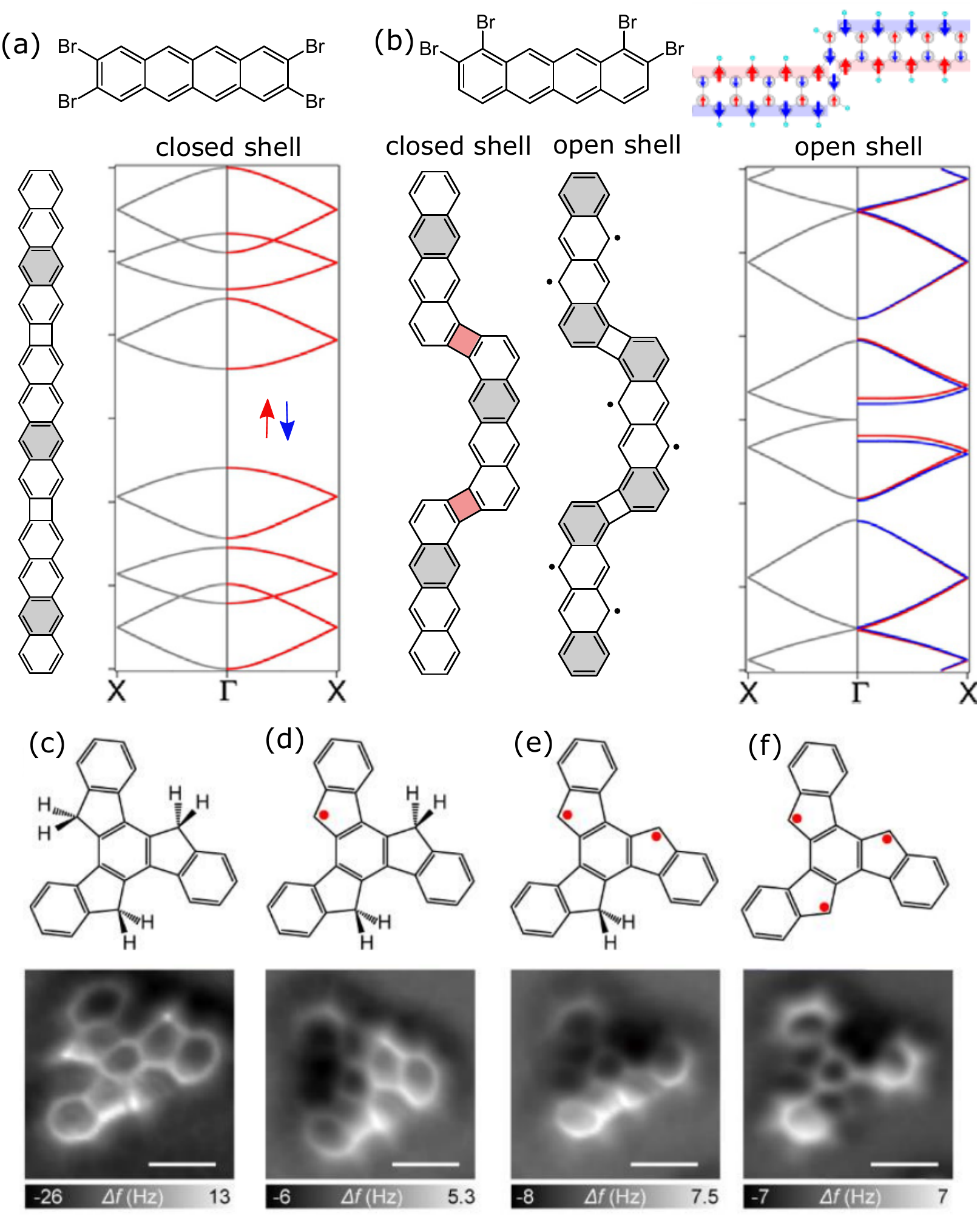}
	\caption{(a) Reactant and product structure, along with the calculated band structure for the linearly fused acenes displaying four-membered rings at the junctions. (b) Proposed reactant and expected product structure in its closed-shell and open-shell forms, along with the calculated band structure for the polymer comprising staggered acenes fused by four-membered rings. The top drawing depicts the schematic spin configurations with the carbon tetragon acting as a spin switch. In the molecular structures the Clar sextets are highlighted in grey and antiaromatic ciclobutadiene moieties in red. Chemical structure and associated nc-AFM images of truxene (c) and its derivatives after dehydrogenation of one (d), two (e) and three (f) of its $sp^3$ carbon atoms at the pentagonal rings. 
	Band structures in panel a and b reproduced with permission from \Ocite{SaDiNi.19.SurfaceSynthesisCharacterization}. Copyright (2019) Wiley-VCH Verlag GmbH \& Co. KGaA, Weinheim. Spin configuration in panel d reprinted with permission from \Ocite{CuZhZh.16.CarbonTetragonsas}. Copyright (2016) by the
American Physical Society. Panels c-f reprinted with permission from \Ocite{LiLiLi.22.TopologicalDefectsInduced}.}
	\label{fig:TopologicalDefects}
\end{figure}

Most of the structures discussed above are benzenoid, made up by hexagonal carbon rings. However, the presence of topological defects as given by carbon rings with a different number of atoms has a notable impact on the materials' properties, including their magnetism. For example we analyze in the following the effect of four-membered rings on two closely related linear conjugated structures. Both consist in
acene segments fused by four-membered rings. In one case the acene segments are fused in a linear configuration (\Figref{fig:TopologicalDefects}a), whereas the other displays a staggered configuration (\Figref{fig:TopologicalDefects}b). Interestingly, whereas the linear structure displays closed shell character \cite{SaDiNi.19.SurfaceSynthesisCharacterization}, the staggered
structure is predicted to be predominantly open-shell \cite{SaDiNi.19.SurfaceSynthesisCharacterization,CuZhZh.16.CarbonTetragonsas}. As explained in \Secref{sec:threshold exceptions}, acenes in their closed-shell structure
display one Clar sextet independently of their length and it is only for higher acenes that they display a notable
open-shell character. The structures shown in \Figref{fig:TopologicalDefects}a,b feature periodically fused tetracene segments, which on their
own are closed-shell structures. Each tetracene segment hosts one Clar sextet in its closed-shell form, which can be
located on any of its rings and can thus avoid being next to the four-membered rings. When the tetracene units are
fused in a linear way (\Figref{fig:TopologicalDefects}a), resonannce structures can be drawn such that no $\pi$-bond locates on the
four-membered ring. However, that is no longer possible when the tetracene units are fused in a staggered way (\Figref{fig:TopologicalDefects}b). In this configuration, $\pi$-bonds necessarily participate in the four-membered rings, which promotes their antiaromatic character (marked in red in \Figref{fig:TopologicalDefects}b) \cite{Pl.21.ExploitationBairdAromaticity} and is energetically unfavorable. However, this can be avoided in
the open-shell structure shown in \Figref{fig:TopologicalDefects}b. The drawn resonance structure has no $\pi$-bond on the four-membered
rings, avoiding their antiaromaticity \cite{ZeGuWa.22.ChemisorptionInducedFormation,KaTaIt.17.CompetingAnnuleneRadialene}. For the linear structure, an open-shell structure would thus gain one Clar
sextet per unit cell at the expense of two radicals, which is not energetically favored for short acene constituents (see \Secref{sec:Clar-sextets}, and \Secref{sec:Clar-threshold}).
Instead, for the staggered structure the open-shell configuration similarly gains only one Clar sextet, but
additionally limits the antiaromatic character of its four-membered rings. Under these circumstances, the open-shell
configuration is favored, representing a clear case in which the topological defect in form of four-membered rings lowers the threshold number of extra Clar sextets required to
favor open-shell structures. 

Interestingly, in addition to promote the open-shell character of these structures, the four membered rings act as
spin-switches \cite{CuZhZh.16.CarbonTetragonsas}. This can be understood from a conceptually very simple idea, namely that opposite spins in neighboring
C atoms allow for their $\pi$-bonding and therefore are stabilizing, whereas for equal spins the Pauli exclusion
principle prohibits their bonding and such configuration is therefore destabilizing \cite{RaGhGh.19.queststabletriplet}. Even-membered rings allow for
configurations in which the spins of all neighboring C atoms are antiferromagnetically oriented. The same applies to
the four-membered rings in the staggered structure of \Figref{fig:TopologicalDefects}b. However, the spin alternation causes a reversal of the
spin when comparing the same side of neighboring acene segments (\Figref{fig:TopologicalDefects}b). 

The scenario with odd-membered rings is different, since it does not allow for configurations in which all atoms display
an alternant spin alignment. This causes spin-frustration \cite{OrLaMe.16.Engineeringspinexchange}, which reduces the stability of such structures \cite{RaGhGh.19.queststabletriplet}, and is
typically mirrored in lower HOMO--LUMO gap values \cite{LaMoRe.21.ReassessingAlkyneCoupling}.  This is, \eg, illustrated in \Figref{fig:trends},  with structure \textbf{5} displaying a comparable gap to structure \textbf{3} in spite of its lower number of carbon atoms. These low HOMO--LUMO gaps in turn facilitate their
spin-polarization. Examples of structures synthesized and studied under vacuum in which odd-membered rings facilitate the spin-polarization are
polyindenoindene \cite{DiChUr.20.SurfaceSynthesisOligoindenoindene} or polyindenofluorene \cite{DiEiYa.19.SurfaceSynthesisAntiaromatic, DiFa.22.surfacesynthesisatomic}, in which their prevailing open-shell configurations gain only two Clar
sextets for every pair of radicals, both cases being therefore below the common ``three sextets per radical pair'' 
threshold described in \Secref{sec:Clar-sextets}.
Nanographenes with a single pentagonal ring \cite{LiSaCo.19.Singlespinlocalization, MiBeBe.20.TopologicalDefectInduced}
and GNRs with five-membered rings at the edges \cite{RiVeJi.20.Inducingmetallicitygraphene} have been characterized.

Besides facilitating the spin-polarization of low energy states by reducing the HOMO--LUMO gaps, topological defects in the form of odd-membered rings also have the peculiarity of potentially resulting in structures with an odd number of $p_z$ electrons, which automatically causes topological frustration and the concomitant appearance of radical states, as exemplified with structure \textbf{6} in \Figref{fig:trends}   \cite{MiBeBe.20.TopologicalDefectInduced,ZhLiZh.20.EngineeringMagneticCoupling}.

A nice example systematically following the magnetism of molecules arising from topological defects is based on truxene derivatives (\mbox{\Figref{fig:TopologicalDefects}c-f}) \cite{LiLiLi.22.TopologicalDefectsInduced,MiFaFe.22.NonbenzenoidHighSpin}.
The as-deposited truxene molecule displays three doubly hydrogenated $sp^3$ carbon atoms at the outer pentagon vertices (\Figref{fig:TopologicalDefects}c). As readily explained in previous sections, their dehydrogenation and $sp^3$ to $sp^2$ rehybridization can be triggered by tip-induced manipulations, controllably affecting one (\Figref{fig:TopologicalDefects}d), two (\Figref{fig:TopologicalDefects}e), or the three pentagonal rings (\Figref{fig:TopologicalDefects}f). Dehydrogenation of one ring results in an odd number of $p_z$ electrons and thus one radical and a spin state $S=1/2$. Dehydrogenantion of a second pentagonal ring makes the total number of $p_z$ electrons even again and the resonance structure can be drawn both as a closed- and open-shell. Interestingly, calculations predict the latter to dominate, with the two radical spins aligned ferromagnetically and a $S=1$ ground state. The same scenario holds when the third radical is generated, ultimately generating a high-spin quartet \cite{LiLiLi.22.TopologicalDefectsInduced,MiFaFe.22.NonbenzenoidHighSpin}. These topological defects thus facilitate the generation of high-spin molecules with smaller molecular weights as compared to, \eg, [n]-triangulenes, and predictions are such that the same ferromagnetic alignment would hold also for an increasing number of radicals in two-dimensional networks based on these truxene-derivatives \cite{LiLiLi.22.TopologicalDefectsInduced}.

\subsection{Heteroatoms}
\label{sec:heteroatoms}

\begin{figure*}
	\includegraphics[width=\textwidth]{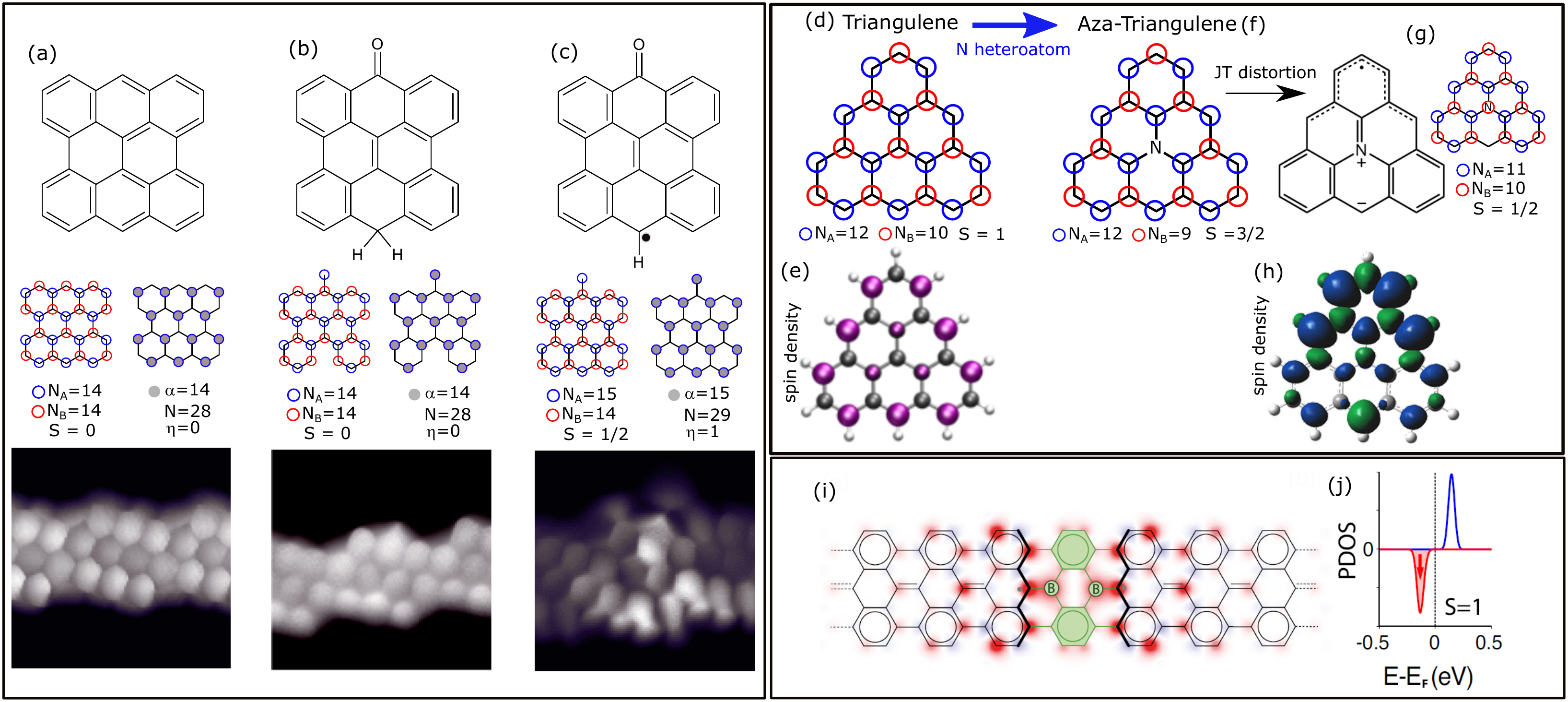}
	\caption{The unit cell, sublattice balance and $\eta$ analysis of (a) a chiral GNR with four zigzag lines across its width, (b) the same unit cell with the central edge atom functionalized on one side with a ketone and on the other side with an extra hydrogen, and (c) for the same unit cell with only one central edge atom functionalized with a ketone. Below are the corresponding bond-resolving STM images. (d) Counting rules applied to triangulene. (e) Spin density for triangulene. (f) Counting rules applied to aza-triangulene in its undistorted 3-fold symmetric form. (g) Best fitting resonance structure for the ground state of aza-triangulene undergoing a Jahn--Teller distortion and application of the counting rules. (h) Calculated spin density for aza-triangulene in its ground state. (i) Structure and spin density around a 2B substitution in 7-AGNRs. (j) Projected DOS (PDOS) on the structure displayed in panel (i).  
	Experimental data in panels (a-c) reprinted with permission from \Ocite{BeLaEd.21.ChemicalStability31}. Copyright (2021) American Chemical Society.
	Panel (e) reprinted by permission from Springer Nature: Nature Nanotechnology \Ocite{PaMiMa.17.Synthesischaracterizationtriangulene}, copyright (2017). 
	Panel (h) from \Ocite{WaBeFr.21.azatriangulene}.
	Panels (i,j) reprinted with permission from \Ocite{FrBrLi.20.MagnetismTopologicalBoundary}.
	\label{fig:heteroatoms}
	}
\end{figure*}

Although always in close relation to the previous sections, the origin of magnetism in carbon-based nanostructures (or alternatively a way to tune it) can also be sought in the presence of heteroatoms. Let us first focus on edge functionalization. Just as hydrogenation can cause an $sp^2$ to $sp^3$ rehybridization and thereby effectively remove a $p_z$ electron (and the process can be reversed by
dehydrogenation), different types of chemical functionalization can have the opposite effect. An example is pictured in
\Figref{fig:heteroatoms}a-c, which shows the unit cell of chiral GNRs that have been synthesized on surfaces under
vacuum \cite{OtDiGa.16.SubstrateIndependentGrowth,MeMoCa.20.Transferringaxialmolecular}. Sublattice imbalance and nullity of the pristine ribbons are both zero, displaying no magnetism (\Figref{fig:heteroatoms}d).
As will be discussed later on, these structures are highly unstable and easily react with gases like O$_2$ even at room
temperature (RT) \cite{BeLaEd.21.ChemicalStability31}. The most common reaction product is a ketone side-functionalization of the ribbon's edges. In the
oxidation process, the oxygen displaces the previously present hydrogen atom, whose migration to the opposite edge is
favorable and thus very common (\Figref{fig:heteroatoms}e) \cite{BeLaEd.21.ChemicalStability31}.
The oxygen displays an $sp^2$ configuration and therefore contributes with an
extra $p_z$ electron to the product structure, in particular to sublattice A, whereas the hydrogenation on the opposite
side removes a $p_z$ electron from that same sublattice. The sublattice balance and $\eta$ are consequently unaffected.
However, as the extra hydrogen is controllably removed by a scanning probe, the missing $p_z$ electron is recovered and,
with the extra $p_z$ electron of the ketone group, a sublattice imbalance is generated and $\eta$ becomes 1. That is, a
new radical is created, which experimentally becomes evident from the enhanced contrast in the low bias bond-resolving
STM images due to the associated Kondo resonance (\Figref{fig:heteroatoms}c) \cite{BeLaEd.21.ChemicalStability31}. 

A similar scenario occurs for chiral GNRs in which each unit cell is doubly functionalized with a ketone on either side \cite{LaBeEd.21.circumventingthestability}. Whereas the perfect structure shows sublattice balance and no magnetic states, defective unit cells with one missing ketone develop a radical state \cite{WaSaCa.22.MagneticInteractions}. These systems are very interesting to study the magnetic interactions between those states when more than one is present on the same ribbon at sufficiently close distances. A relevant conclusion from such studies is that Ovchinnikov's rule \cite{Ov.78.Multiplicitygroundstate} or Lieb's theorem \cite{Li.89.TwotheoremsHubbard} do not necessarily apply any longer as heteroatoms are involved. Along the same lines, the description of these heteroatoms as mere additional $p_z$ states in a TB approach turns out to be insufficient, and additional chemical information in the form of modified hopping constants and on-site energies are required for an adequate modelling \cite{WaSaCa.22.MagneticInteractions}.

In a similar way as edge functionalization with heteroatoms can bring about or modify the magnetism of carbon nanostructures, the same occurs within the backbone structure. By way of example, the addition of substitutional nitrogen is predicted to trigger ferromagnetism in graphene \cite{BlTuSo.17.Dopingwithgraphiticnitrogen,BaKa.19.Ferromagnetisminnitrogen}. Its effect on well-defined nanographene structures has been studied for example in aza-triangulene, which allows for an interesting comparison with the magnetic properties of conventional triangulene. As discussed in earlier sections and reminded in \Figref{fig:heteroatoms}d, the latter has a sublattice imbalance of 2 and thus displays a triplet ground state with $S=1$. The associated spin density displays a three-fold symmetric distribution as shown in \Figref{fig:heteroatoms}e \cite{PaMiMa.17.Synthesischaracterizationtriangulene}.  
If the central atom is exchanged by a nitrogen atom in aza-triangulene, naively one would assume a double occupancy of its $p_z$ orbital, which would in turn remove its contribution to the $N_B$ count in Ovchinnikov’s rule and result in a quartet ground state ($S=3/2$, \Figref{fig:heteroatoms}f). 
However, DFT calculations predict this molecule to have a doublet ground state ($S=1/2$). The odd number of $p_z$ electrons makes it favorable for the molecule to undergo a Jahn--Teller distortion  \cite{SaCaCa.19.Triangulargraphenenanofragments} and adopt a structure with lower symmetry that is best represented by the zwitterionic resonance form displayed in \Figref{fig:heteroatoms}g \cite{WaBeFr.21.azatriangulene}. As can be observed, the zwitterion displays only one radical that is delocalized around one of the molecule's vertices, in agreement with its predicted spin ($S=1/2$) and its spatial distribution (\Figref{fig:heteroatoms}h). Besides, it also reconciles the ground state $S=1/2$ with Ovchinnikov's rule, since the bonding nature of the N $p_z$ orbital justifies its counting towards $N_B$, whereas the carbon atom hosting the negative charge at the low side edge has its $p_z$ orbital doubly occupied and does not count towards $N_A$ ($N_A=11$, $N_B=10$, $S=1/2$, \Figref{fig:heteroatoms}g) \cite{WaBeFr.21.azatriangulene}. 

Another nice example of backbone heteroatom doping is the substitution of carbon by boron atoms. In contrast to N, B in its $sp^2$ configuration has its $p_z$ orbital empty. Naively this would equally remove it from its sublattice count and thereby cause a net spin according to Ovchinnikov’s rule, in a similar way as the $sp^2$ to $sp^3$ rehybridization by hydrogenation of graphene does \cite{GoGoMa.16.Atomicscalecontrol}. 
This is, however, not the case in graphene, where it merely acts as a point potential rather than rupturing the conjugated electron system \cite{WaSuRo.14.heteroatomdopedgraphene}. However, if two boron atoms are introduced into the backbone of GNRs instead, they can trigger a sufficiently strong modification of the conjugation pattern in their close proximity to modify the orbital's symmetry and thus their topology \cite{CaZhLo.17.TopologicalPhasesGraphene,FrBrLi.20.MagnetismTopologicalBoundary,ZhLiDo.22.pimagnetismspin}. 
As a result, topological interface states appear at the junctions to the undoped GNR regions. This has been realized for example with B atoms within 7-AGNRs as pictured in \Figref{fig:heteroatoms}i, which also represents the spatial distribution of the spin density \cite{FrBrLi.20.MagnetismTopologicalBoundary}. The latter looks reminiscent of the spin density associated to the end-states of 7-AGNRs (assuming a termination at the thicker solid lines), which distribute symmetrically to either side of the borylated section. Most interestingly, the two spins align ferromagnetically into a $S=1$ ground state (\Figref{fig:heteroatoms}i,j) in spite of their distribution on either side of the borylated section on opposite sublattices, which would normally favor an antiferromagnetic alignment. 
This surprising result has been ascribed to the strong barrier imposed by the borylated section, which prevents conjugated electrons from hopping across. The disconnection of the two boundary states at each side hinders the (antiparallel) kinetic exchange between them. The stabilization of the triplet configuration is then the result of the weak direct overlap between both spin-polarized boundary states through the 2B barrier, which, due to the tiny hopping between them, dominates the exchange interaction and induces the ferromagnetic alignment of the spins according to Hund's rule \cite{FrBrLi.20.MagnetismTopologicalBoundary}. 

It is worth noting that, apart from causing the appearance of magnetism \textit{via} one mechanism or another, the introduction of heteroatoms has also been found to allow for the stabilization of magnetic states and their electronic decoupling from the underlying metallic substrates merely by a modification of the molecular adsorbate's adsorption geometry \cite{BlZhBr.21.Spinsplittingdopant,GaSa.21.Cleversubstitutionsreveal}.

\subsection{Charge transfer}

All examples that have been reviewed so far correspond to carbon-based nanostructures that intrinsically display an
open-shell character. When molecules are adsorbed on solid substrates, another very important factor needs to be
considered. The effects of the molecule-substrate interactions on the magnetic properties of the molecular structures
can be manifold. They may quench the intrinsic magnetism of the molecules or instead endow magnetic properties to
carbon nanostructures that in a free-standing configuration would display a conventional closed-shell character. In this respect it is instructive to analyze the simple picture of the Anderson model for single magnetic states \cite{An.1961.LocalizedMagneticStates}.

\begin{figure}
	\centering
	\includegraphics[width=0.7\columnwidth]{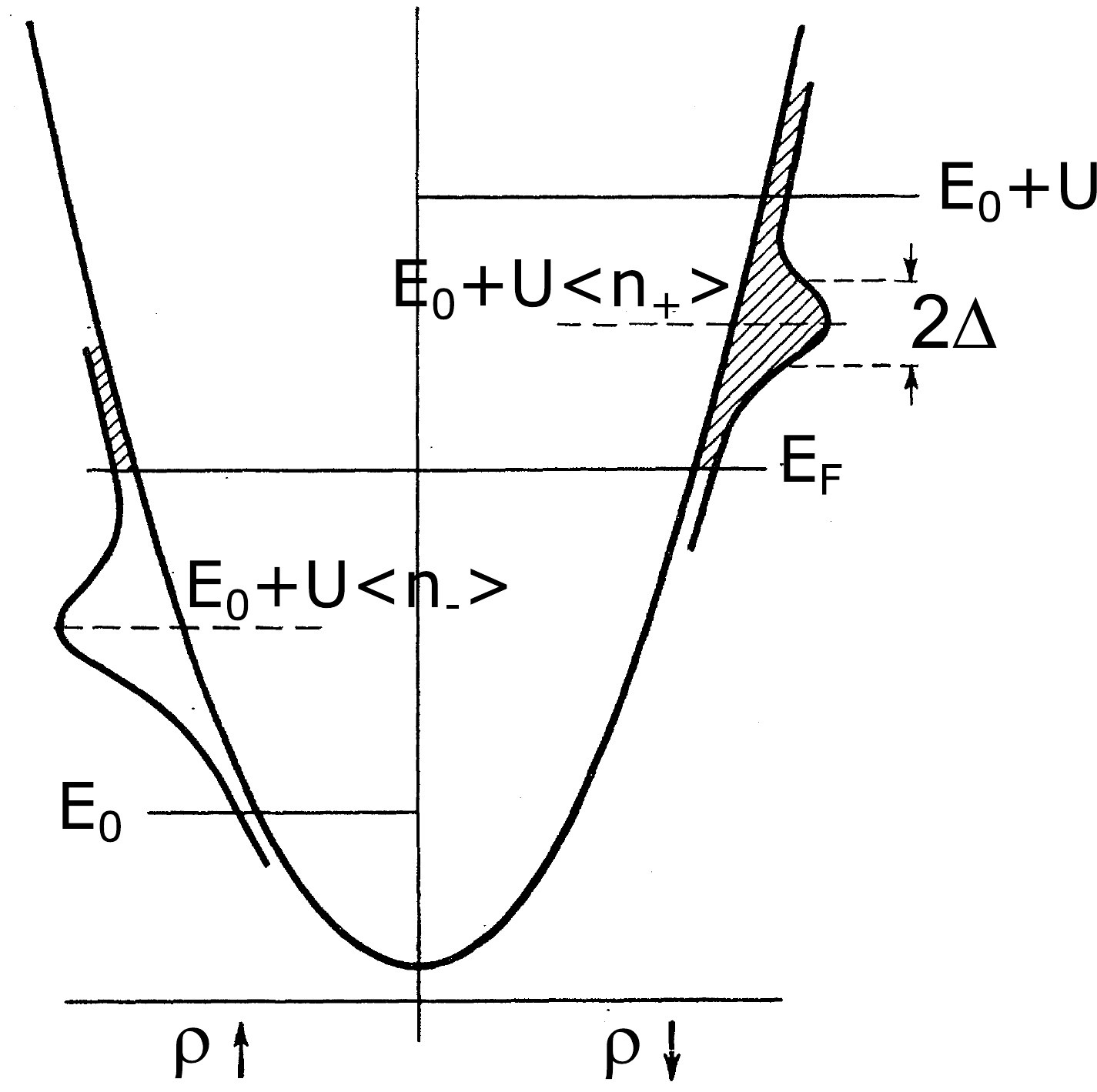}
	\caption{Density of state distribution for spin up (left) and spin down (right). The ``humps'' at $E_0+U\langle n_-\rangle$ and $E_0 + U \langle n_+\rangle$ are the virtual levels of the magnetic state, of width $2\Delta$, for up and down spins, respectively. The number of electrons $\langle n_-\rangle$ and $\langle n_+\rangle$ occupying them correspond to the integration of their unshaded region below the Fermi level and determine the Coulomb repulsion they feel from electrons of the opposite spin and thereby the energy deviation of the virtual states with respect to $E_0$ and $E_0+U$.
	Reprinted with permission from \Ocite{An.1961.LocalizedMagneticStates}.  Copyright (1961) by the American Physical Society. 
	\label{fig:anderson}
	}
\end{figure}

This model describes a magnetic state at energy $E_0$. Placing a second electron into that same state implies a Coulomb repulsion $U$ between the two electrons, thus requiring an addition energy of $E_0+U$. However, as the states interact with their environment (\eg, the substrate) and get a finite width, if their tails cross the Fermi level their occupation deviates from 1 and 0 and instead becomes $\langle n_+\rangle$ and $\langle n_-\rangle$ for spin up and spin down, respectively. The partial occupation $\langle n_-\rangle$ causes Coulomb repulsion on the spin up state and increases its energy to $E_0+U \langle n_-\rangle$, whereas the Coulomb repulsion on the spin down state is lowered to $E_0+U\langle n_+\rangle$. As may be directy inferred from this simple picture, the net spin density ($\langle n_+\rangle - \langle n_-\rangle$) is maximized for large $U$ values, low ${\Delta}$ values and for binding energies $E_0$ as close as possible to $U/2$. 

All of these relevant parameters can be affected by the substrate and the
supramolecular environment. By way of example, the binding energy $E_0$ of an electronic state will greatly vary with
substrates of different work function \cite{HoLueHu.17.ChargeTransferOrbital,BoElGo.14.SpectroscopicFingerprintsWork,GoBoEl.16.MultiComponentOrganic}. In addition, the state's width ${\Delta}$ is crucially dependent on the
specific substrate--adsorbate interactions and hybridization \cite{HoLueHu.17.ChargeTransferOrbital,GoBoEl.16.MultiComponentOrganic}. At the same time, the latter determines the
adsorption height that, along with the dielectric properties of the substrate, have an important effect on the Coulomb
energy $U$ \cite{WaTaPi.16.Giantedgestate,HoLueHu.17.ChargeTransferOrbital}. However, not only the substrate plays a role, but also the supramolecular environment. Besides the possibility of intermolecular charge transfer in so-called ``charge-transfer complexes'', different intermolecular
interactions can also bring along modified interaction strength and hybridization with the substrate \cite{GoBoEl.16.MultiComponentOrganic,GoMaEl.14.SelfAssemblyBicomponent,StLueWi.14.Unexpectedinterplaybonding, FrScHe.08.ReducingMolecule-SubstrateCoupling}, as
mirrored, \eg, in varying adsorption heights \cite{GoMaEl.14.SelfAssemblyBicomponent,StLueWi.14.Unexpectedinterplaybonding}. Also the energy level alignment may vary \cite{GoBoEl.16.MultiComponentOrganic,GoMaEl.14.SelfAssemblyBicomponent,StLueWi.14.Unexpectedinterplaybonding}, which
can eventually cause changes in the adsorbate's charging state \cite{BoElGo.14.SpectroscopicFingerprintsWork,FeKrSt.12.GatingChargeState}.

If a closed-shell adsorbate undergoes charge transfer to or from the substrate, it may display molecular
magnetism if the following conditions are additionally met. According to the Anderson model \cite{TeHeSc.09.SpectroscopicmanifestationsKondo,An.1961.LocalizedMagneticStates}, the Coulomb
interaction $U$ should be greater than ${\Delta}$ and the binding energy $E_0$ should be in the range of 0 to $U$, being most
favourable for magnetism if it is around $E_0=U/2$ \cite{An.1961.LocalizedMagneticStates}.
These conditions guarantee a close to integer charge state of the
adsorbate, with well-differentiated spin-up and spin-down electron population. A low ${\Delta}$ value, which translates into
a sufficiently small electronic coupling of adsorbate and substrate, is thus a key parameter to promote magnetism. 

\begin{figure*}
 	\includegraphics[width=\textwidth]{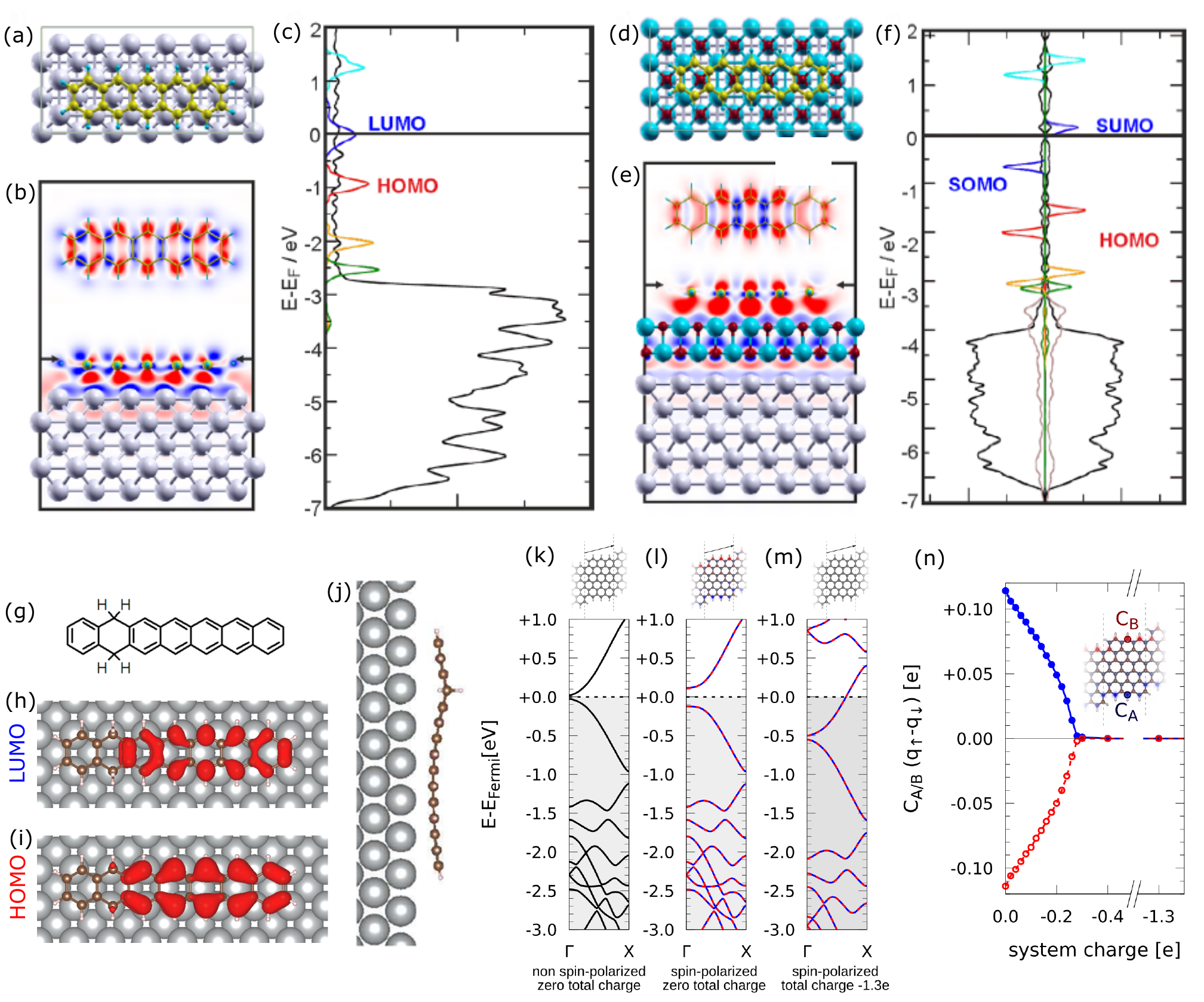}
 	\caption{(a) Top view of the most stable adsorption configuration of pentacene on Ag(001). (b) Horizontal (top) and vertical two-dimensional cuts showing the charge density differences induced by pentacene adsorption on the Ag(001) substrate. The arrows indicate where the horizontal cut has been made. Red shows accumulation and blue depletion of charge, respectively. (c) PDOS obtained for the pentacene/Ag(001) interface using the GGA functional. The DOS projected onto the Ag atoms (black) has been reduced for better visibility. In addition, the projections on the molecular orbitals are shown for LUMO+1 (cyan), LUMO (blue), HOMO (red), HOMO--1 (orange), and HOMO--2 (dark green). Equivalent calculations for the (d) adsorption configuration , (e) charge density differences and (f) spin-polarized PDOS for the pentacene/MgO/Ag(001) interface, the latter obtained with the HSE functional. (g) Chemical structure of 5,18-dihydroheptacene, calculated charge densities associated with (h) its LUMO and (i) its HOMO level, as well as (j) a lateral view of its adsorption configuration on Ag(001). (k) Relaxed atomic configuration (top) and electronic band structure (bottom) of chiral (3,1,6)-GNRs without spin‐polarization. (l) and (m) Spin density (top) and spin‐polarized band structure (bottom) for the neutral case and upon charge transfer of 1.3 electrons per unit cell, respectively. (n) Spin‐polarized electron density at the marked carbon atoms at the ribbon’s edges as a function of charge transfer. 
 	Panels (a-f) reprinted with permission from \Ocite{HoLueHu.17.ChargeTransferOrbital}.  and
 	Panels g-j reprinted with permission from \Ocite{MoCoRo.20.Electronicdecouplingpolyacenes}. http://creativecommons.org/licenses/by/4.0/.  
 	Panels k-n reprinted with permission from 
 	\Ocite{CoMePi.21.BandStructureEnergy}.
 	}
 	\label{fig:pentacene}
\end{figure*}

A good example is the case of pentacene on Ag(001) (\Figref{fig:pentacene}a,b). Experiment and calculations both reveal the transfer of
nearly a full electron from the silver substrate to pentacene's LUMO level \cite{HoLueHu.17.ChargeTransferOrbital}.
However, $E_0$ is roughly centered around
the Fermi level and the state's width ${\Delta}$ is much greater than $U$ (\Figref{fig:pentacene}c) \cite{HoLueHu.17.ChargeTransferOrbital}. The spin-up and spin-down
population on pentacene is thus similar and the result is a non-magnetic molecule-adsorbate system in spite of the
quasi integer charge transfer. However, as the pentacene molecule is decoupled from the Ag(001) by a MgO monolayer
(\Figref{fig:pentacene}d,e), the effect is two-fold. On the one hand, the screening from the substrate is significantly reduced, which
translates into a greatly enhanced $U$. On the other hand, the molecule's electronic coupling to the silver is much more
limited, which consequently reduces ${\Delta}$. In the end, two well-defined molecular resonances are observed below
and above the Fermi level that are associated to the SOMO and SUMO, respectively (\Figref{fig:pentacene}f) \cite{HoLueHu.17.ChargeTransferOrbital}. This situation exactly
fits the Anderson model applied to a magnetic molecule-substrate system. 

A closely related system with an intermediate molecule-substrate coupling between that of the systems described above
was studied with 5,18-dihydroheptacene directly on the Ag(001) surface \cite{MoCoRo.20.Electronicdecouplingpolyacenes}. This molecule shares the carbon backbone of
heptacene but displays doubly hydrogenated C atoms at its second ring (\Figref{fig:pentacene}g). The $sp^3$ hybridization of these atoms
breaks the conjugation along the molecule, for which the frontier electronic states are determined by the longest
conjugated segment within the molecule \cite{MoCoRo.20.Electronicdecouplingpolyacenes,CoMoDo.18.surfacesynthesisheptacene,ZuDoKr.17.NonaceneGeneratedSurface}, that is, a pentacene segment. This confers the 5,18-dihydroheptacene
molecule electronic properties that are strikingly similar to those of pentacene, as sustained, \eg, by the energy and
shape of its frontier orbitals (\Figref{fig:pentacene}h,i) \cite{MoCoRo.20.Electronicdecouplingpolyacenes,CoMoDo.18.surfacesynthesisheptacene}.
However, the $sp^3$ C atoms within the carbon backbone endow the
molecule with a non-planar structure that causes its self-decoupling from the underlying substrate (\Figref{fig:pentacene}j). Although
a less effective decoupling than that provided by insulating buffer layers like MgO, it is still sufficient for the
molecules to display in STM experiments charging peaks (with its associated vibronic satellites) that are associated to
a double tunnelling barrier and, most importantly, magnetism as proved by an associated Kondo resonance at
zero bias \cite{MoCoRo.20.Electronicdecouplingpolyacenes}. The Kondo resonance still implies a minor molecule-substrate coupling, since it relates to the
screening of the molecular spin by the substrate electrons. It is, however, small enough to stabilize a singly charged
molecule with net spin $S = 1/2$. 

Although exemplified above with pentacene or its derivatives on Ag substrates, the appearance of magnetism upon
adsorption in molecules that in their free-standing configuration are conventional closed-shell structures has been
observed also with other molecules and substrates \cite{FeFrPa.08.VibrationalKondoEffect,FeKrSt.12.GatingChargeState,GaStBa.13.Longrangemagnetic,SuMaRo.20.InducingOpenShell}.

Nevertheless, just as charge transfer may cause the appearance of magnetism in otherwise non-magnetic materials \cite{FeKrSt.12.GatingChargeState,MoCoRo.20.Electronicdecouplingpolyacenes}, 
it can equally well quench the magnetism of intrinsically magnetic carbon-based molecules \cite{LaBrBe.20.ProbingMagnetismTopological,WaBeFr.21.azatriangulene, ZhJiLi.20.PreciseControlpi,DuWa.12.TuningChargeSpin,PiKa.21.ImprintingTunablepi} or graphene
defects \cite{GoGoMa.16.Atomicscalecontrol}.
For the latter, this could be nicely tested by comparing the magnetism of defects in pristine and doped
graphene samples. Whereas indirect but unambiguous fingerprints of magnetism were observed with pristine graphene, the
use of $p$- or $n$-doped graphene removed the electron sustaining the unpaired spin moment or paired it up with an extra
electron, respectively, in either case resulting in non-magnetic systems \cite{GoGoMa.16.Atomicscalecontrol}.
Depending on the system, it may also be important to consider in which direction the electron transfer takes place. By way of example, aza-triangulene, which in its neutral form is predicted to display a ground state with $S=1/2$ \cite{SaCaCa.19.Triangulargraphenenanofragments,WaBeFr.21.azatriangulene}, presents a triplet ground state ($S=1$) when it donates an electron to the substrate (as occurs on Au(111)), but becomes a closed-shell molecule when it receives an extra electron from the underlying substrate (as occurs on Ag(111)) \cite{WaBeFr.21.azatriangulene}. 

In the case of carbon-based nanostructures that have been obtained by OSS, the most commonly utilized substrate
is Au(111). Its high work function endows this substrate a strong tendency to receive electrons from the
molecular adsorbates atop \cite{MeGaCa.17.WidthDependentBand}. The molecule's loss of spin-carrying electrons to the substrate can modify or even quench their
associated magnetism. This situation, in which the resonances of the magnetic states appear at energies above the Fermi
level and are thus unoccupied, has been observed with many disparate molecular systems, like the end-states of AGNRs \cite{WaTaPi.16.Giantedgestate,LaBrBe.20.ProbingMagnetismTopological,ChOtPe.13.TuningBandGap} or other linear polymers \cite{CiSaTo.20.Tailoringtopologicalorder}, a variety of topological interface states \cite{GrWaYa.18.Engineeringrobusttopological,RiVeCa.18.Topologicalbandengineering}, or the radical $\pi$-states of porphyrin derivatives \cite{ZhJiLi.20.PreciseControlpi}. For the latter, the charge transfer  of some specific derivatives could be controlled by modifying the adsorption geometry through scanning probe manipulations, which consequently allowed tuning its magnetic state accordingly \cite{ZhJiLi.20.PreciseControlpi}.

It should also be mentioned that even for systems in which the
magnetism is associated to more electrons, like in chiral or ZGNRs, charge transfer is predicted to similarly
affect the magnetism \cite{CaWaLe.14.Edgemagnetizationlocal,CoMePi.21.BandStructureEnergy,DuWa.12.TuningChargeSpin}.
By way of example, chiral (3,1)-GNRs with six carbon atoms across their width develop spin-polarized edge states, whereby their band structure widens its gap to avoid the instability associated with large density of states near the Fermi level (\Figref{fig:pentacene}k,l). However, a charge transfer of around one electron per unit cell, as experimentally observed on silver surfaces, quenches the magnetization and causes the band structure to become like that of non spin-polarized ribbons (\Figref{fig:pentacene}m) \cite{CoMePi.21.BandStructureEnergy}.
A more detailed analysis of the spin-polarization in this kind of ribbons as a function of charge transfer indeed reveals that an electron transfer of only 0.3 electrons per unit cell is already enough to fully quench any magnetization (\Figref{fig:pentacene}n) \cite{CoMePi.21.BandStructureEnergy}.

Another example are ZGNRs with six atoms across their width. An indirect fingerprint of their
magnetism (in particular the observation of a correlation gap) was only observed when adsorbed on a NaCl buffer layer,
but not directly on Au(111) \cite{RuWaYa.16.surfacesynthesisgraphene}. For this latter case, however, no evidence of the flat band was observed even above the
Fermi level. Therefore, besides charge transfer, an excessively large hybridization of the flat band with the
underlying substrate cannot be discarded either as the reason for the lacking magnetism, as indeed suggested by the appearance of those resonances in closely related ZGNRs with nitrogen dopants that allow for their electronic decoupling from the substrate by a modified adsorption geometry \cite{BlZhBr.21.Spinsplittingdopant,GaSa.21.Cleversubstitutionsreveal}. This same hybridization scenario may also be
the reason for the lacking magnetism in superheptazethrene on Au(111). Whereas in solution and in the solid state it
was found to have an open-shell character \cite{ZeSuHe.16.Superheptazethrene}, adsorbed on Au(111) it displays a closed-shell character in spite of no
obvious charge transfer evidences, with comparably shaped HOMO and LUMO resonances almost symmetrically positioned
around the Fermi level \cite{MiMeEi.20.surfacesynthesissuper}. Only its larger ``sister molecule'' supernonazethrene, with a consequently more robust magnetism, shows clear magnetic fingerprints on Au(111) \cite{TuMiMe.21.SynthesisCharacterization}. As discussed later on in the frame of the carbon nanostructure's reactivity, comparative studies on weakly and more strongly interacting surfaces also show how the magnetism of molecules like, \eg, [7]-triangulene may survive on the former but be quenched by the strong hybridization with the substrate on the latter \cite{MiXuEi.21.Synthesischaracterization7triangulene}.

Another important point to remind regarding charge transfer is that Ovchinnikov's rule and Lieb's theorem no longer apply for doped systems. Applied to GNRs this means that, for doping levels at which the magnetization is not yet quenched, ground states other than that predicted for neutral ribbons (whether chiral or zigzag edged) are allowed and may thus not correspond to the expected ferromagnetic alignment of the spins along each edge of the ribbons and an inter-edge antiferromagnetic alignment \cite{CaWaLe.14.Edgemagnetizationlocal,SaIsSa.09.PhaseControlGraphene,JuMa.09.CarrierDensityMagnetism}.

\section{Magnetic interactions}
\label{sec:interactions}

\begin{table*}
\caption{Examples of reported exchange-coupled electron spins in open-shell nanographenes.}
\label{tab:spin-interactions}
\begin{indented}
	\item[]
	\begin{tabular}{l  c  c  c}
	\br
	System & Substrate & $J$ (meV) & Reference \\
	\mr
    (3,1)-chiral GNR junctions  & Au(111) & 3 to 10 & \cite{LiSaCo.19.Singlespinlocalization}\\
    Clar's goblet & Au(111) & 23 & \cite{MiBeEi.20.Topologicalfrustrationinduces}\\
    \mbox{[3]}-Triangulene & gas phase & $-570$ $^a$ & \cite{SaCaCa.19.Triangulargraphenenanofragments}\\
    Extended triangulene (ETRI) & gas phase & $<-60$ $^b$ & \cite{LiSaCa.20.UncoveringTripletGround} \\
    Triangulene dimer & Au(111) & 14 & \cite{MiBeEi.20.CollectiveAllCarbon,MiCaWu.21.Observationfractionaledge}\\
    Triangulene nanostar & Au(111) & 18 & \cite{HiCaFr.21.SurfaceSynthesisCollective}\\
    Triangulene rings and chains & Au(111) & 14 $^c$ & \cite{MiCaWu.21.Observationfractionaledge}\\ 
    \mbox{[5]}-rhombene & Au(111) & 102 & \cite{MiYaCh.21.Largemagneticexchange}\\
    Super-nonazethrene & Au(111) & 51 &\cite{TuMiMe.21.SynthesisCharacterization}\\
    Ketone-functionalized chiral GNRs & Au(111) & $-14$ to 42 & \cite{WaSaCa.22.MagneticInteractions}\\
    Defective \mbox{[3]}-rhombene dimers & Au(111) & $-6.6$ to 42.9 & \cite{ZhLiXu.20.Designerspinorder}\\
    Peripentacene & Au(111) & 40.5 &\cite{SaUrVe.21.UnravellingOpenShell}\\
    Periheptacene & Au(111) & 49.2 &\cite{BiUrMu.22.SynthesisCharacterizationPeriHeptacene}\\
	\br
	\end{tabular}
\item[]$^a$ Calculation based on the restricted active space configuration interaction (RASCI) method.
\item[]$^b$ MFH calculations with $U>3$ eV.
\item[]$^c$ This work considers spin interactions beyond Heisenberg exchange within the bilinear-biquadratic (BLBQ) model.
\end{indented}
\end{table*}

\subsection{Electron spin pairs}
As we have seen in figures \ref{fig:goblet} and \ref{fig:methods} the emergence of two unpaired electrons in a graphene nanostructure can lead to a sizable spin interaction on the $J\sim 1$-100 meV scale, promising for potential applications in RT spin-logic operations \cite{MiBeEi.20.Topologicalfrustrationinduces}
Some characteristic observations of exchange-coupled spins are listed in \Tabref{tab:spin-interactions}. 
The current record of $J=102$ meV was observed in \mbox{[5]}-rhombene on Au(111) \cite{MiYaCh.21.Largemagneticexchange}.
Distance-dependent effects in the spin interaction have been reported \cite{MiBeEi.20.Topologicalfrustrationinduces,MiBeEi.20.CollectiveAllCarbon,WaSaCa.22.MagneticInteractions,SaUrCa.20.DiradicalOrganicOne},
including dimers of asymmetric rhombus-shaped nanographenes \cite{ZhLiZh.20.EngineeringMagneticCoupling,ZhLiXu.20.Designerspinorder}, in which the exchange coupling clearly correlates with the spacing between the rhombus sides displaying the largest spin density. 

A variety of analytical arguments and more complex calculations have been utilized to model and understand the interactions between the spins of different electrons in singly occupied molecular states \cite{OrFe.20.Probinglocalmoments,OrBoGa.19.ExchangeRulesDiradical}.
The dependence of exchange coupling $J$ on structural parameters have also been analyzed theoretically \cite{WaYaMe.09.TopologicalFrustrationGraphene,GeJiSh.18.Firstprinciplesstudy}.

Considering the simplest case of exchange coupling $J$ between only a pair of electron spins, we have seen that it can manifest itself in an antiferromagnetically (AFM) ordered ground state ($J>0$), exemplified by the Hubbard dimer (\Secref{sec:Hubbard-dimer}) and Clar's goblet (\Figref{fig:goblet}), or by a ferromagnetic (FM) ground state ($J<0$), exemplified by [3]-triangulene (\Figref{fig:sublattice_imbalance}).
It is worth to mention that ferromagnetic exchange $J$ is maximal for C$_3$ diradicals and that the symmetry also imposes that the lowest-energy excited state is a degenerate pair of closed-shell states \cite{OrBoGa.19.ExchangeRulesDiradical}
On the other hand, ferromagnetic exchange $J$ is minimal for $N_A=N_B$ diradicals, due to the disjointed nature of their zero modes \cite{RiVeJi.20.Inducingmetallicitygraphene,OrBoGa.19.ExchangeRulesDiradical}.

A typical fingerprint of the magnetic interactions may be seen in inelastic electron tunneling spectroscopy (IETS), \ie, symmetric features in $dI/dV$ at characteristic energies $|eV|=\Delta E_{ST}$ due to inelastic transitions between the singlet and triplet states. This is similar to spin IETS of single or few magnetic atoms with STM \cite{HeGuLu.04.Single-AtomSpin-FlipSpectroscopy,LoGa.09.EfficientSpinTransitions,Fe.09.TheoryofSingle-Spin,Fr.09.SpinInelasticElectron,Te.15.Spinexcitationsand, ChLoWi.19.ColloquiumAtomicspin}.
The possibility to split the triplet states by an external magnetic field allows to unambiguously confirm the spin origin of the IETS features. However, the Zeeman energy of $g\mu_B\sim 0.12$ meV/T is typically much smaller than the resolution in IETS (thermal broadening $5.4 k_BT\sim 2$ meV at liquid He temperatures, lock-in amplifier modulation voltage, finite lifetime of the spin states, etc.) making such detection difficult in practice.

We note here that the intrinsic singlet-triplet gap of a molecule is also known to be reduced on a surface due to renormalization of the energies by coupling with substrate electrons \cite{JaOrFe.21.Renormalizationspinexcitations,KoLoGa.12.Many-bodyeffectsin}.

Although IETS features for high-spin systems like [3]-triangulene or the related extended triangulene (ETRI) \cite{LiSaCa.20.UncoveringTripletGround} have not been reported experimentally, another manifestation of magnetism for these $S=1$ systems, namely the emergence of an underscreened Kondo resonance, have been reported (\mbox{\Figref{fig:spin-interactions}a-c}). This narrow resonance can further be split by an external magnetic field when the Zeeman energy exceeds the Kondo temperature.

\subsection{Interactions in few-spin systems}
Going beyond a pair of electron spins, IETS features have also been observed in open-shell nanographenes such as the triangulene dimer with four unpaired electrons (\Figref{fig:spin-interactions}d). Consistent with Lieb's theorem, the ground state is a singlet with the low-energy spectrum effectively described by an AFM ordering between two $S=1$ units (\Figref{fig:spin-interactions}e). The singlet-to-triplet excitation by IETS was observed around $V=14$ meV (\Figref{fig:spin-interactions}f). The direct transition to the quintet state ($S=2$) by single tunneling electrons is forbidden according to the selection rule $\Delta S = 0, \pm 1$ \cite{HeGuLu.04.Single-AtomSpin-FlipSpectroscopy,LoGa.09.EfficientSpinTransitions,Fe.09.TheoryofSingle-Spin,Fr.09.SpinInelasticElectron,Te.15.Spinexcitationsand, ChLoWi.19.ColloquiumAtomicspin}.

Very recently, [3]-triangulene has also been demonstrated as a suitable $S=1$ building block for realizing one-dimensional AFM systems through OSS strategies \cite{MiCaWu.21.Observationfractionaledge,HiCaFr.21.SurfaceSynthesisCollective}. As 
conjectured by Haldane \cite{ChLoWi.19.ColloquiumAtomicspin,Ha.83.NonlinearFieldTheory,Af.89.Quantumspinchains}, antiferromagnets for integer spins exhibit a finite excitation gap in the bulk, in contrast to the spin-half case.
Furthermore, open-ended $S=1$ chains are expected to host fractional $S=1/2$ (topological) edge states at the boundaries.
Figure \ref{fig:spin-interactions}g sketches the formation of coupled rings and chains of triangulene $S=1$ units, with the latter exhibiting fractional $S=1/2$ edge states.
The collective spin states in these systems were revealed by multiple IETS features (\mbox{\Figref{fig:spin-interactions}h-i}) well explained by the many-body spectrum obtained from exact diagonalization of the 1D Heisenberg model or its extension including a biquadratic coupling term as in the exactly solvable Affleck-Kennedy-Lieb-Tasaki (AKLT) model \cite{AfKeLi.87.Rigorousresultsvalence}. Notably, the fractional $S=1/2$ termini states expected for open chains were revealed by a Kondo resonance in tunnel spectroscopy (green spectra in \Figref{fig:spin-interactions}i).
Antiferromagnetically coupled $S=1/2$ and $S=1$ molecular 1D chains have also been reported with metal-free porphyrins \cite{WaZhJi.22.Realizationquantumnanomagnets}.

\subsection{Extended spin chains and associated band-engineering}
One-dimensional spin chains constructed with extended GNRs have also been proposed through topological band engineering with heterostructures or heteroatoms \cite{GrWaYa.18.Engineeringrobusttopological,RiVeJi.20.Inducingmetallicitygraphene,JoJaBo.19.CorrelatedTopologicalStates,CaZhLo.17.TopologicalPhasesGraphene,LiCh.18.TopologicalPropertiesGapped,RiVeCa.18.Topologicalbandengineering,FrBrLi.20.MagnetismTopologicalBoundary,PiKa.21.ImprintingTunablepi,SuYaGr.20.CoupledSpinStates,SuYaYa.21.EvolutionoftheTopological,JiLo.21.TopologyClassificationusing}. 
By rational design of precursor molecules and their mixtures, it is possible to 
vary the inter-radical spacing (and therefore their coupling) to engineer regular or disordered spin arrangements.

\begin{figure*}
 	\includegraphics[width=\textwidth]{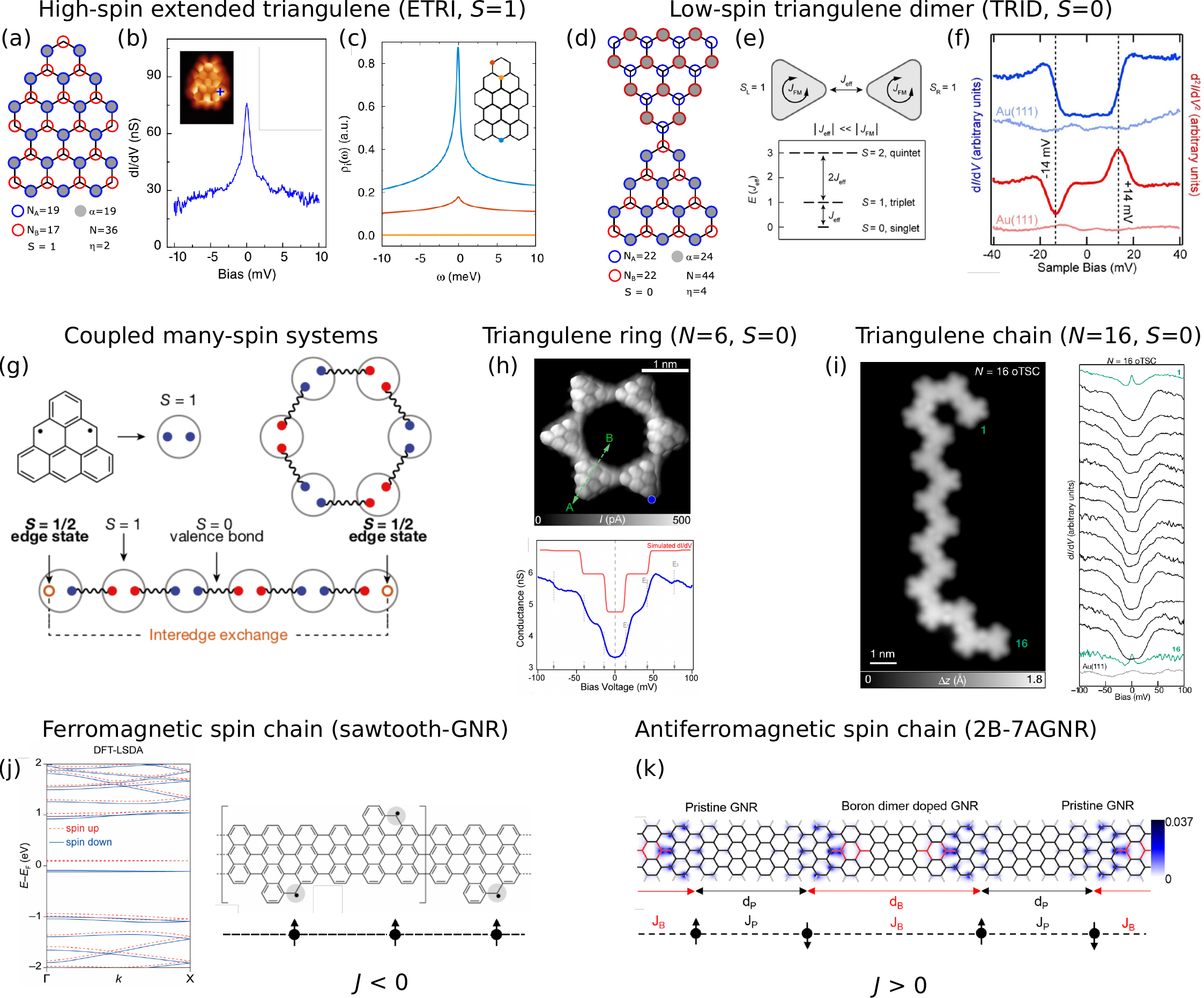}
 	\caption{Interacting spins in graphene nanostructures and nanoribbons.
 	(a) Structure of extended triangulene (ETRI) with indication of the sublattice imbalance.
 	(b) Experimental signature of a zero-bias $S=1$ Kondo resonance related to the triplet ground state of ETRI.
 	(c) Spectral functions computed for ETRI coupled to the conducting electrons in the substrate, confirming the $S=1$ Kondo effect.
 	 (d) Structure of a triangulene dimer (TRID) with sublattice balance.
 	 (e) Heisenberg dimer model and level diagram for the TRID with intra-triangulene ferromagnetic exchange and inter-triangulene antiferromagnetic exchange.
 	 (f) Experimental $dI/dV$ (blue curve) and IETS (red curve) spectra acquired on TRID, revealing
 	 inelastic the singlet-triplet excitation at 14 meV.
 	 (g) Heisenberg spin chains and rings of antiferromagnetically coupled $S=1$ triangulenes.
 	 (h) Experimental bond-resolving STM image of a triangulene nanostar composed of $N=6$ units and
 	 corresponding experimental and theoretical $dI/dV$ spectra.
 	 (i) Experimental STM image of a triangulene chain of $N=16$ units and corresponding 
 	 experimental $dI/dV$ spectra recorded over each unit. The zero-bias feature observed at the ends of the
 	 chain are Kondo resonances associated with the $S=1/2$ edge states illustrated in panel g.
 	 (j) Spin-polarized electronic band structure predicted from LSDA-DFT calculations for a sawtooth-GNR with radicals at each zigzag segment at the edges of a 7-AGNR backbone. The obtained ferromagnetic ordering ($J<0$) is in accordance with Lieb's theorem due to sublattice imbalance. 
 	 (k) Theoretical prediction of antiferromagnetic ordering ($J>0$) between periodic interfaces of pristine
 	 and borylated 7-AGNR units.
 	 Panel b reprinted with permission from \Ocite{LiSaCa.20.UncoveringTripletGround}. Copyright (2020) by the American Physical Society.
 	 Panel c reprinted with permission from  \Ocite{JaOrFe.21.Renormalizationspinexcitations}. Copyright (2021) by the American Physical Society. 
 	 Panels e-f reprinted with permission from \Ocite{MiBeEi.20.CollectiveAllCarbon}.
 	 Panels g,i reprinted by permission from Springer Nature: Nature \Ocite{MiCaWu.21.Observationfractionaledge}, copyright (2021).  
 	 Panel h reprinted with permission from \Ocite{HiCaFr.21.SurfaceSynthesisCollective}.
 	 Panel j from \Ocite{RiVeJi.20.Inducingmetallicitygraphene}. Reprinted with permission from AAAS. 
 	 Panel k reprinted with permission from \Ocite{CaZhLo.17.TopologicalPhasesGraphene}. Copyright (2017) by the American Physical Society. 
 	}
 	\label{fig:spin-interactions}
\end{figure*}

In \Figref{fig:spin-interactions}j we show an example of a 7AGNR decorated with sawtooth-like, short zigzag edges that host radical states. Since the location of the radicals are on the same sublattice sites, it follows from Lieb's theorem that the ground state is a FM order of the spins, also confirmed by DFT calculations in the local spin density approximation (LSDA) as seen in the corresponding spin-polarized band structure \cite{RiVeJi.20.Inducingmetallicitygraphene}.
This GNR structure was experimentally realized in references~\cite{RiVeJi.20.Inducingmetallicitygraphene, SuYaGr.20.CoupledSpinStates}. A metallic behavior was reported on the Au(111) substrate, instead of the gapped FM phase predicted for the freestanding ribbon, a difference that was explained as due to a combination of $p$-doping and surface electric fields induced by the substrate \cite{RiVeJi.20.Inducingmetallicitygraphene}.

Another class of GNR spin chains is that of AGNRs with borylated segments as shown in \Figref{fig:spin-interactions}k \cite{CaZhLo.17.TopologicalPhasesGraphene,CaGaCo.18.ElectronicPropertiesSubstitutionally}. As discussed in \Secref{sec:heteroatoms} the combination of pristine 7AGNR and 2B-7AGNR segments results in topological interface states that, depending on their spatial separation, may either hybridize as filled closed-shell states or exhibit spin polarization at the ends of the pristine segments. We note that in the shown example the exchange couplings $J_P>0$ and $J_B>0$ lead to an AFM ordering. However, for a single 2B-AGNR unit between pristine segments, DFT calculations have shown a FM ordering \cite{FrBrLi.20.MagnetismTopologicalBoundary}.

\section{Reactivity}
\label{sec:reactivity}

\subsection{Reactivity under UHV}

\begin{figure*}
	\includegraphics[width=\textwidth]{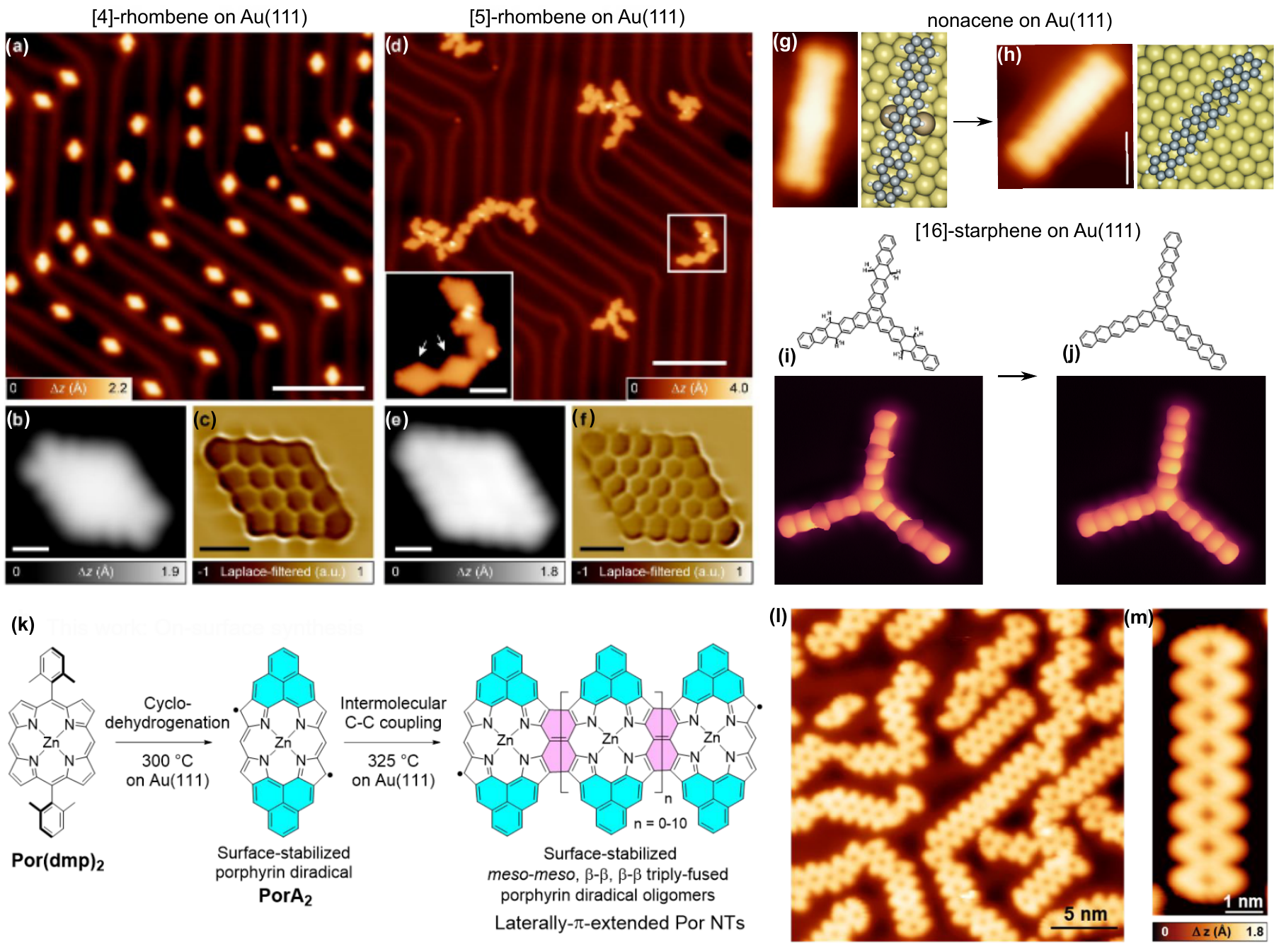}
	\caption{(a) Overview STM image of the sample after annealing the [4]-rhombene precursor on a Au(111) surface. (b) STM and (c) bond-resolving STM image of a single [4]-rhombene product. (d) Overview STM image of the sample after annealing the [5]-rhombene precursor on a Au(111) surface. (e) STM and (f) bond-resolving STM image of a single [5]-rhombene product. (g) STM image and associated model structure of nonacene on Au(111) interacting with Au adatoms. (h) STM image and associated model structure of nonacene on Au(111) in the absence of Au adatoms. (i) Bond-resolving STM image of [16]-starphene on Au(111) hydrogenated at its most reactive rings and (j) the same molecule after tip-induced dehydrogenation. (k) Reactant and synthetic route for porphyrine nanotapes. (l) Overview and (m) zoom-in image of the porphyrine nanotapes on Au(111).
	Panels a-f reprinted by permission from Springer Nature: Nature Chemistry, \Ocite{MiYaCh.21.Largemagneticexchange} Copyright (2020).  
	Panels g and h reprinted with permission from \Ocite{UrMiHa.19.surfacelightinduced}. http://creativecommons.org/licenses/by/4.0/. 
	Panels i and j reprinted with permission from \Ocite{HoCoLa.21.LargeStarpheneComprising}. Copyright (2021) Wiley-VCH GmbH. 
	Panels k-m reprinted with permission from \Ocite{SuMaRo.21.BottomFabricationAtomic}.}
	\label{fig:reactivity}
\end{figure*}

Radical states are intrinsically associated with a pronounced reactivity, although the latter may be reduced by steric protection of the radical sites, by electronic stabilization or by a combination of these two strategies \cite{StChZe.19.DoDiradicalsBehave,ArShSh.21.SynthesisandIsolation,VaMaKa.22.TrimesityltrianguleneaPersistent, BeArCe.16.ChemistryEdgeGraphene}. One of
the ways in which electronic stabilization can be promoted is by radical delocalization, which as discussed
earlier can be enhanced by the hybridization of the radical states with the following doubly occupied orbitals.
Such delocalization reduces the spin density at the radical site, which, although not the only one, is the main
determining factor in the reactivity \cite{StChZe.19.DoDiradicalsBehave}. For that reason, molecular structures with large spin density like Clar's 
goblet \cite{MiBeEi.20.Topologicalfrustrationinduces} or triangulenes \cite{MiXuEi.21.Synthesischaracterization7triangulene} show a remarkable tendency to react and polymerize on their supporting surfaces, lowering
the yield of the isolated target products. It is particularly instructive to compare the reactivity of closely related
molecules with and without spin polarization, as is the case of extended triangulene and double triangulene \cite{LiSaCa.20.UncoveringTripletGround}. The
latter shows a prevalent closed-shell character, whereas the former is an open-shell structure with notable spin
polarization. As a result, the sample preparation of extended triangulene by OSS on Au(111) results
mostly in dimerized units, with monomers amounting to only about 12.5 \% \cite{LiSaCa.20.UncoveringTripletGround}.
In contrast, in the synthesis of double
triangulene nearly all molecules remain as monomers. A similar scenario is found comparing rhombene molecules of
different size, namely [4]-rhombene and [5]-rhombene, which display a prevailing closed-shell and open-shell character,
respectively \cite{MiYaCh.21.Largemagneticexchange}. The former is obtained by OSS on Au(111) in quasi full yield, whereas the latter is
only rarely found as a monomer (\Figref{fig:reactivity}a-f) \cite{MiYaCh.21.Largemagneticexchange}. Similar conclusions were reached from theoretical calculations comparing zigzag- and armchair-edged GNRs \cite{JiSuDa.07.UniqueChemicalReactivity}.

The high reactivity of radical states has indeed also been used as an indirect hint or support for the partial
open-shell character of molecular structures on which no other magnetic fingerprint was experimentally measured. For
example, this is the case of higher acenes. Heptacene, and to a greater extent nonacene, show a tendency to coordinate
to metal adatoms at their most reactive central ring \cite{ZaBe.12.Reactivityacenesmechanisms,ZaZaRe.11.ProductsMechanismAcene,UrMiHa.19.surfacelightinduced} when adsorbed on Au(111) (\Figref{fig:reactivity}g), although the adatoms
can be controllably be removed by scanning probe manipulation (\Figref{fig:reactivity}h) \cite{UrMiHa.19.surfacelightinduced}. The fact that shorter acenes remain stable and uncoordinated
under the same conditions has been thus correlated with the closed-shell character of short acenes and the increasing
open-shell character of acenes as they grow longer (in particular above hexacene), as supported also by
calculations \cite{UrMiHa.19.surfacelightinduced}. A similar reasoning has been applied to the case of [16]-starphene on Au(111), whose notable tendency
to become hydrogenated at the central rings of each of its arms (\Figref{fig:reactivity}i), although easily reversible by scanning probe
manipulation (\Figref{fig:reactivity}j), has also been associated to its partial open-shell character \cite{HoCoLa.21.LargeStarpheneComprising}. 

Although the examples outlined above present the increased reactivity of open-shell molecules as a downside, it may also
be used constructively. A beautiful example is the formation of porphyrine nanotapes, in which an initial
cyclodehydrogenation step first forms surface-supported porphyrine diradical molecules (\Figref{fig:reactivity}.k) \cite{SuMaRo.21.BottomFabricationAtomic}. The reactivity of
the associated radicals is particularly high at specific molecular positions, ultimately driving a relatively
well-defined and selective polymerization process towards atomically precise porphyrine nanotapes (\Figref{fig:reactivity}.k-m) \cite{SuMaRo.21.BottomFabricationAtomic}.  

A very important parameter with regard to the reactivity of the molecules is the supporting substrate, which can affect
it in different ways. For example, a strong molecule-substrate interaction may quench the radical character of the
molecules and thereby their reactivity, although at the expense of an equally quenched magnetism. However, even if the
molecules retain their radical character, strong molecule-substrate interactions may also be associated to strongly
corrugated adsorption energies that limit the molecule's diffusion and thereby the potential polymerization events.
Either way, the substrate can limit the molecule's reactivity and thereby increase the yield and selectivity during
their synthesis. By way of example, whereas isolated [7]-triangulene could not be obtained on the weakly interacting
Au(111) because its pronounced reactivity caused the molecule's covalent coupling into molecular clusters, on Cu(111)
the yield of isolated [7]-triangulene was dramatically improved \cite{MiXuEi.21.Synthesischaracterization7triangulene}.
However, the desired open-shell septet ($S = 3$)
anticipated from counting rules and confirmed by gas-phase calculations turned out to become a trivial closed-shell
electronic structure on Cu(111) \cite{MiXuEi.21.Synthesischaracterization7triangulene}. Similar findings were observed also for other molecules like [5]-rhombene, which in
its isolated form is only a minority product on Au(111) because its pronounced radical character and high diffusion
promote the molecule's random polymerization (\Figref{fig:reactivity}d), while on the more interactive Cu(110) its yield is orders of
magnitude higher \cite{MiYaCh.21.Largemagneticexchange}.

\subsection{Reactivity in controlled atmospheres and ambient conditions}

The previous section discussed the reactivity of open-shell carbon-nanostructures under ultra-high-vacuum, which, as
already seen, is of key importance for their synthesis. However, most of the potential applications envisioned for
these materials requires further processing \cite{BePeMa.13.Bottomgraphenenanoribbon,LlFaBo.17.Shortchannelfield,MoViKr.18.Bottomsynthesismultifunctional,FaSaLa.17.Highvacuumsynthesis,MuLlJa.21.TransferFreeSynthesis}.
By way of example, the metallic substrates commonly used in their
synthesis is detrimental or directly impedes many optoelectronic applications, therefore requiring the material's
transfer to other functional substrates. Besides, scalable device applications need to be applicable out of the vacuum
environment. Whereas closed-shell structures with armchair edges have successfully survived transfer processes and
ultimate device utilization that included exposure to a multitude of environments like different solvents or
air \cite{BePeMa.13.Bottomgraphenenanoribbon,LlFaBo.17.Shortchannelfield,MoViKr.18.Bottomsynthesismultifunctional,FaSaLa.17.Highvacuumsynthesis,MuLlJa.21.TransferFreeSynthesis}, the same scenario poses serious challenges for open-shell carbon-based nanostructures. 

\begin{figure}
 	\includegraphics[width=\columnwidth]{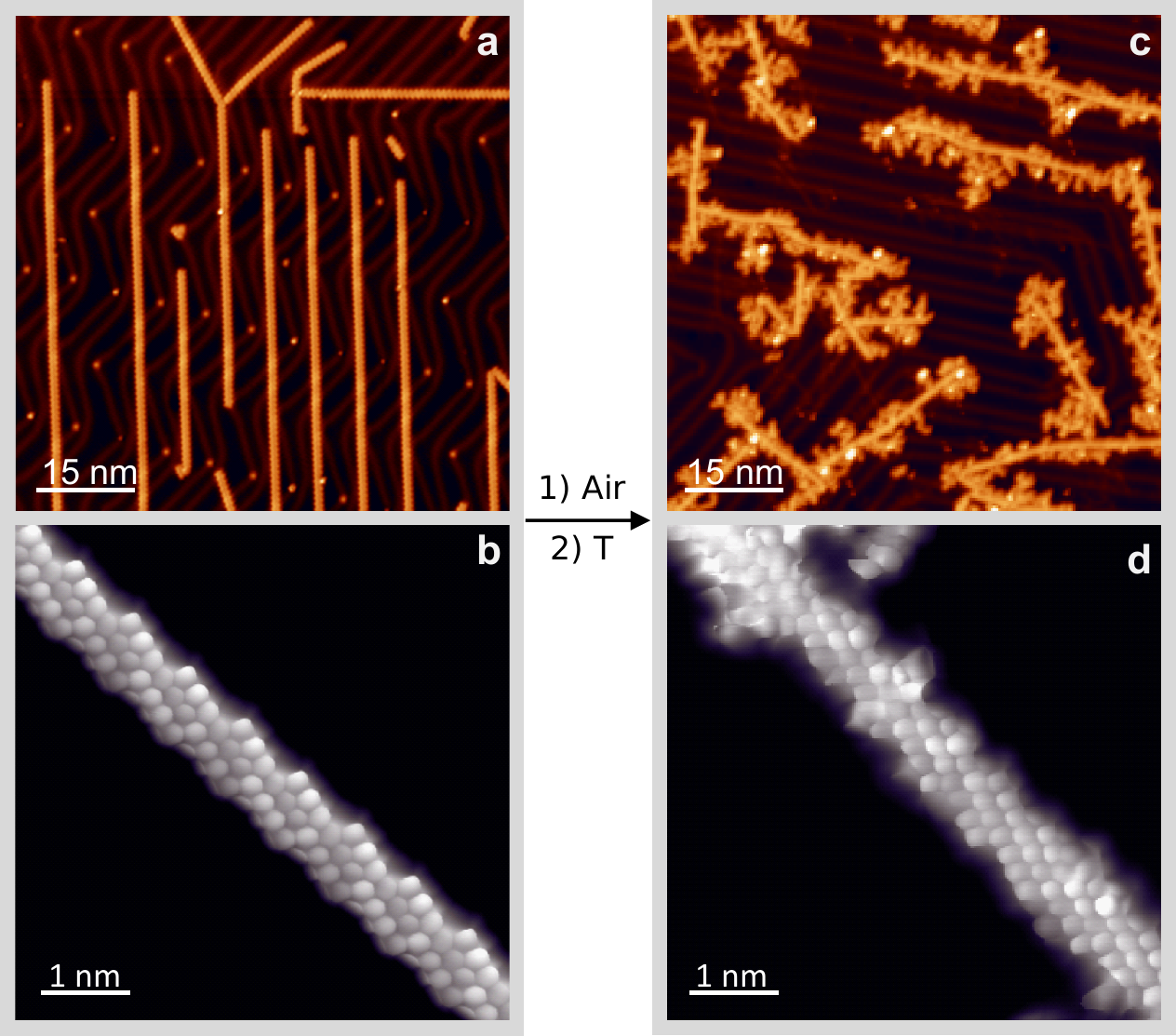}
 	\caption{Overview STM images (a, c) and bond-resolving STM images (b, d) of the chiral GNRs as grown on Au(111) under UHV (a, b), and after exposure to air for 3 days, followed by an annealing at 200 °C for 15 h (c, d).
 	Reprinted with permission from \Ocite{BeLaEd.21.ChemicalStability31}. Copyright (2021) American Chemical Society.}
 	\label{fig:cgnr}
\end{figure}

By way of example, it has been recently shown that narrow (3,1)-chiral GNRs, which display an
alternating edge structure composed by three zigzag and one armchair units (\Figref{fig:cgnr} and \Figref{fig:sextets}h-j), are dramatically
affected by oxidizing environments in spite of their predominantly closed-shell character \cite{BeLaEd.21.ChemicalStability31}. This can be observed in
\Figref{fig:cgnr}, which shows overview and bond-resolving STM images of the GNRs before and after exposure to air. The overview
images show the long and straight shape of the as-synthesized ribbons (\Figref{fig:cgnr}a), which turns curvy after air exposure
(\Figref{fig:cgnr}b), already hinting at their chemical modification. The bond-resolving images confirm the degradation scenario,
on which the protruding signals from many of the C atoms at the GNR edges, as well as modification of the shapes, sizes
and brightness of the hexagons that form the carbon backbone (\Figref{fig:cgnr}d), all relate to the chemical alteration after the
air exposure \cite{BeLaEd.21.ChemicalStability31}. The characterization of the air-exposed sample required an additional annealing treatment at 200 $^\circ$C to desorb the largest part of contaminants that came with the air exposure (although as may be
observed in \Figref{fig:cgnr}c, many remained still adsorbed around the GNRs in spite of the treatment). This posed the question,
whether the degradation had been thermally activated by the annealing \cite{MaXiPu.18.Oxidizationstabilityatomically} or was already occurring at RT. More
controlled experiments with exposure to low pressure ($3 \times 10^{-5}$ mbar) of pure oxygen gas at RT confirmed
that a multitude of oxidation products appear also under those more gentle conditions, further revealing the most
reactive sites and the nature of the most prevalent reaction products \cite{BeLaEd.21.ChemicalStability31}. The rationalization of the remarkable
reactivity invoked the small but apparently already sufficient open-shell character of the structure. 

Similar findings with regard to the lack of stability and of the most common reaction product, were found also for the
zigzag ends of 5-AGNRs \cite{LaBrBe.20.ProbingMagnetismTopological}. Exposure to low pressures ($2 \times 10^{-5}$ mbar) of pure oxygen
resulted in ketone-functionalized ribbon ends independently of the ribbon length that determines its varying open- to
closed-shell character \cite{LaBrBe.20.ProbingMagnetismTopological}.
Along the same lines, a low band gap pyrene-based GNR that also displays
periodic zigzag-oriented edge sites (\Figref{fig:cgnr}b) \cite{SuGrOv.20.MassiveDiracFermion,BaSmCa.21.Morphologicalcharacterizationelectronic}, has been equally shown to oxidize already at RT with a 20 minute
exposure to only $10^{-6}$ mbar of pure oxygen \cite{BaSmCa.21.Morphologicalcharacterizationelectronic}.

At this point it is important to remark various issues. First, that the examples discussed above show a predominantly
closed-shell character and that the lack of stability will be even worse in structures with increasing open-shell
character. Second, that all these cases already suffered irreversible damage upon exposure to low pressures of pure
oxygen (still in the high-vacuum range), which is beyond doubt a lower limit regarding the harshness of the treatment
to which the carbon-based nanostructures may be exposed, if compared to a conventional transfer process or device
implementation in solvents or air. Altogether, this brings up the need to revise the approaches and devise new
strategies for the actual implementation of magnetic carbon-based nanostructures in scalable devices. 
Although still far from allowing for this ultimate goal, the development of chemical protection and deprotection strategies may be a first step in that direction \cite{LaBeEd.21.circumventingthestability}.

\section{Outlook}
\label{sec:outlook}

As stressed throughout the first sections of this review, the rapid progress made in the development of magnetic carbon-based materials is intimately linked to the advances in the field of OSS. For that reason, it also shares many of its challenges. The latter have been analyzed and listed in several OSS reviews \cite{ClOt.19.ControllingChemicalCoupling,WaZh.19.Confinedsurfaceorganic,HeFuSt.17.CovalentBondFormation,DoLiLi.15.SurfaceActivatedCoupling,LiFe.20.SyntheticTailoringGraphene,ShGaFu.17.FrontiersOnSurfaceSynthesis,HoRiSt.21.Atomicallyprecisegraphene}, and range from the development of purification methods to get rid of all the byproducts of non-selective reactions, to the need of non-metallic substrates \cite{KoStIz.20.Rationalsynthesisatomically} for many technological applications (or even just for a better characterization), which thus requires either new synthetic approaches or efficient, non-invasive transfer methods. The latter are particularly challenging when dealing with open-shell nanostructures, given their unstable and reactive character. The development of chemical protection and deprotection methods seem promising in this respect \cite{LaBeEd.21.circumventingthestability}, but would require further optimization and its combination with other forms of protection like, \eg, capping layers to allow for the eventual implementation of such open-shell carbon nanomaterials in actual devices. 
Another challenge faced by the field is the ability to bridge from the nanoscale functionality at the molecular level to the mesoscale for actual device engineering, maintaining the atomic precision to avoid the problems associated with defects \cite{PiBaCe.21.EdgeDisorderBottom}, and all of it in a scalable manner.  

Besides these more generic challenges, there are many others specifically for magnetic carbon nanostructures. One of them is, \eg, the direct measurement of their magnetic properties by spin-polarized STM and electron-spin-resonance STM, the latter even allowing for the manipulation and characterization of the dynamic response of the spins.
On the theory side, there are corresponding challenges to develop complete quantum descriptions of the $\pi$-electron spin physics of nanographenes taking explicitly into account
the various environmental interactions (\eg, electronic coupling to substrates/electrodes and hyperfine coupling to nuclear spins) and the spin dynamics that can be probed or induced by time-dependent external drives (nonequilibrium).
Electrical spin manipulation is one example of the latter \cite{OrGaLa.18.Electricalspinmanipulation}.

When it comes to actual applications, open-shell carbon nanostructures have been proposed as a promising platform for many disparate uses. 
By way of example, radical molecules may act as particularly efficient photon emitters, whose doublet-spin nature avoids the formation of triplet excitons that limit the electroluminescence efficiency of non-radical emitters \cite{ZhLaRo.18.SinglePhotonEmission,AbHeGu.20.Understandingluminescentnature,AiEvDo.18.Efficientradicalbased,PeObZh.15.OrganicLightEmitting}. 
This may be of good use in conventional applications like organic light-emitting diodes \cite{AbHeGu.20.Understandingluminescentnature, AiEvDo.18.Efficientradicalbased,PeObZh.15.OrganicLightEmitting}, but also in more advanced quantum applications like the generation of efficient single-photon emitters \cite{ZhLaRo.18.SinglePhotonEmission}.
In fact, bright, narrow-band, and tunable light emission from individual GNR junctions has already been demonstrated \cite{ChAfSc.18.BrightElectroluminescenceSingle}.

Graphene-based nanostructures have been also proposed for a variety of other spintronics \cite{SoTuDu.15.Spintransporthydrogenated} or quantum applications, including spin and valley filters \cite{RyTwBe.07.Valleyfilterand, WiAdBe.08.SpinCurrentsRough}, spin rectification \cite{NiHuLi.22.PerfectSpinSeebeckEffect}, 
logic gates \cite{WaYaMe.09.TopologicalFrustrationGraphene,GeJiSh.18.Firstprinciplesstudy, BuGiOw.15.ImprovedAllCarbon}, or spin qubits hosted by structures that range from zero to two dimensions \cite{TrBuLo.07.Spinqubitsgraphene, PeFlPe.08.GrapheneAntidotLattices:}.
Another area of application is for electron quantum optics, in which GNR-based beam splitters have been proposed as a platform for controlled motion of charge/spin, interferometry, and entanglement \cite{BoCrRo.11.QuantumTransportin, LiHePi.16.5050electronicbeam, BrEnPa.17.tunableelectronicbeam, SaBrGi.20.Crossedgraphenenanoribbons, Sanz2022}.
Graphene nanostructures thus seem promising for use in the emergent field of quantum-coherent nanoscience \cite{HeOlVa.21.Quantumcoherentnanoscience}.
Furthermore, in analogy with nitrogen-vacancy (NV) color centres in diamond \cite{RoTeHi.14.Magnetometrynitrogenvacancy},
they could also become components in quantum sensors for various metrology applications.

Most of these applications, however, have been only proposed at the theoretical level and their experimental realization still needs to be demonstrated in coming years. 
With many groups worldwide working in the field, we will surely witness important advances in a near future addressing many of the above mentioned challenges and creating new ones aiming at novel directions.

\section*{Acknowledgments}
The authors acknowledge fruitful discussions with Alejandro Berdonces-Layunta, Pedro Brandimarte, Jan Patrick Calupitan, David Casanova, Martina Corso, David Ecija, Ferdinand Evers, Roman Fasel, Niklas Friedrich, Carlos Garcia, Aran Garcia-Lekue, Geza Giedke, Jingcheng Li, Ricardo Ortiz, Jose Ignacio Pascual, Diego Peña, Francisco Romero Lara, Daniel Sanchez-Portal, Soﬁa Sanz, Shiyong Wang and Tao Wang.
This work was funded by the Spanish MCIN/AEI/ 10.13039/501100011033 (PID2020-115406GB-I00 and PID2019-107338RB-C63), the Basque Department of Education (PIBA-2020-1-0014), and the European Union’s Horizon 2020 (FET-Open project SPRING Grant No.~863098).

\section*{References}
\providecommand{\newblock}{}

\end{document}